\documentclass[a4paper,oneside,11pt]{article}
\usepackage{cprform}
\usepackage{amsmath,amstext,amsfonts,amsbsy,amssymb,amscd,bbm,epsfig,lscape}
\usepackage{appendix}
\usepackage{hyperref}

\usepackage{subfigure}
\usepackage{times}
\usepackage{float}

\newcommand{\ci}[1]{\boldsymbol{#1}}

\newcommand{\Dslash}{\relax{\kern+.25em / \kern-.70em D}}

\newcommand{\Real}{\relax{\mathsf{\Gamma\kern-.35em R}}}
\newcommand{\Int}{\relax{\mathsf{Z\kern-.40em Z}}}

\newcommand{\obar}[1]{\kern3pt\overline{\kern-2pt #1\kern-0pt}\kern1pt}

\newcommand{\corrbar}[1]{\kern3pt\overline{\kern-2pt #1\kern-0pt}\kern1pt}

\newcommand{\oVApAVren}[1]{\kern3pt\overline{\kern-2pt #1\kern-0pt}\kern1pt_{\rm\scriptscriptstyle VA+AV;s}}

\newcommand{\zbar}{\kern3pt\overline{\kern-2pt Z\kern-0pt}\kern1pt}

\newcommand{\zbarVApAV}[1]{\kern3pt\overline{\kern-2pt Z\kern-0pt}\kern1pt_{\rm\scriptscriptstyle VA+AV #1}}

\newcommand{\LSB}{\raisebox{-0.3ex}{\mbox{\LARGE$\left[\right.$}}}
\newcommand{\RSB}{\raisebox{-0.3ex}{\mbox{\LARGE$\left.\right]$}}}

\def\gsim{\mathrel{\raise2pt\hbox to 8pt{\raise -5pt\hbox{$\sim$}\hss{$>$}}}}
\def\rsim{\mathrel{\raise2pt\hbox to 8pt{\raise -5pt\hbox{$\sim$}\hss{$>$}}}}
\def\lsim{\mathrel{\raise2pt\hbox to 8pt{\raise -5pt\hbox{$\sim$}\hss{$<$}}}}
\newcommand{\be}{\begin{equation}}
\newcommand{\ee}{\end{equation}}

\begin{document}

\bibliographystyle{mybibstyle}


\begin{titlepage}

\vspace*{-50truemm}
\begin{flushright}
\hspace*{-1cm} ICCUB-11-204, IFIC/12-48, RM3-TH/12-13, ROM2F/2012/05
 
\end{flushright}\vspace{5truemm}

\centerline{\Large \bf  Kaon Mixing Beyond the SM from $\mathbf{N_f=2}$ tmQCD} 
\centerline{\Large \bf and model independent constraints from the UTA}
\vskip 5 true mm
\centerline{\bigrm V.~Bertone$^{(a)}$, N.~Carrasco$^{(b)}$, M.~Ciuchini$^{(c)}$, 
P.~Dimopoulos$^{(d)}$, R.~Frezzotti$^{(d,e)}$, }
\centerline{\bigrm  V.~Gim\'enez$^{(b)}$, V.~Lubicz$^{(f,c)}$, G.~Martinelli$^{(g,h)}$, F.~Mescia$^{(i)}$,  
M.~Papinutto$^{(j,k)}$\footnote{On leave of absence from Dipartimento di Fisica,
Universit\`a di Roma "La Sapienza" Piazzale A. Moro, I-00185 Rome, Italy.},}
\centerline{\bigrm    G.C.~Rossi$^{(d,e)}$, L.~Silvestrini$^{(h)}$,  S.~Simula$^{(c)}$, C.~Tarantino$^{(f,c)}$, 
 A.~Vladikas$^{(e)}$ }

\vspace*{1truemm}
\begin{figure}[!h]
  \begin{center}
    \includegraphics[scale=0.70]{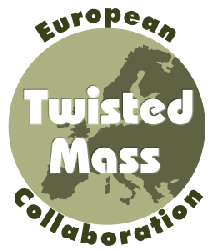}
 \end{center}
\end{figure}
\vspace*{2truemm}

\vskip 1 true mm
\vskip 0 true mm
\centerline{\it $^{(a)}$ Physikalisches Institut, Albert-Ludwigs-Universit\"at Freiburg,}
\centerline{\it Hermann-Herder-Straße 3, D-79104 Freiburg i. B., Germany}
\vskip 2 true mm
\centerline{\it $^{(b)}$  Departament de F\'{\i}sica Te\`orica and IFIC, Univ. de Val\`encia-CSIC}
\centerline{\it Dr.~Moliner 50, E-46100 Val\`encia, Spain}
\vskip 2 true mm
\centerline{\it $^{(c)}$ INFN, Sezione di Roma Tre}
\centerline{\it c/o Dipartimento di Fisica, Universit\`a  Roma Tre}
\centerline{\it Via della Vasca Navale 84, I-00146 Rome, Italy}
\vskip 2 true mm
\centerline{\it $^{(d)}$ Dipartimento di Fisica, Universit\`a di Roma ``Tor Vergata''}
\centerline{\it Via della Ricerca Scientifica 1, I-00133 Rome, Italy}
\vskip 2 true mm
\centerline{\it $^{(e)}$ INFN, Sezione di ``Tor Vergata"}
\centerline{\it c/o Dipartimento di Fisica, Universit\`a di Roma ``Tor Vergata''}
\centerline{\it Via della Ricerca Scientifica 1, I-00133 Rome, Italy}
\vskip 2 true mm
\centerline{\it $^{(f)}$ Dipartimento di Fisica, Universit\`a  Roma Tre}
\centerline{\it Via della Vasca Navale 84, I-00146 Rome, Italy}
\vskip 2 true mm
\centerline{\it $^{(g)}$ SISSA - Via Bonomea 265 - 34136, Trieste - Italy}
\vskip 2 true mm
\centerline{\it $^{(h)}$ INFN, Sezione di Roma, Piazzale A. Moro, I-00185 Rome, Italy}
\vskip 2 true mm
\centerline{\it $^{(i)}$ Departament d'Estructura i Constituents de la Mat\`eria}
\centerline{\it and Institut de Ci\`encies del Cosmos}
\centerline{\it Universitat de Barcelona, 6a planta, Diagonal 647}
\centerline{\it E-08028 Barcelona, Spain}
\vskip 2 true mm
\centerline{\it $^{(j)}$ Laboratoire de Physique Subatomique et de Cosmologie}
\centerline{\it UJF/CNRS-IN2P3/INPG}
\centerline{\it 53 rue des Martyrs, 38026 Grenoble, France}
\vskip 2 true mm
\centerline{\it $^{(k)}$ Dpto. de F\'{\i}sica Te\'orica and Instituto de F\'{\i}sica Te\'orica UAM/CSIC}
\centerline{\it Universidad Aut\'onoma de Madrid Cantoblanco E-28049 Madrid, Spain}
\vskip 2 true mm

\vskip 20 true mm


\thicktablerule
\vskip 3 true mm
\noindent{\tenbf Abstract}
\vskip 1 true mm
\noindent
{\tenrm  
We present  the first unquenched, continuum limit, lattice QCD results for the matrix elements of the operators 
describing neutral kaon oscillations in
extensions of the Standard Model. Owing to the accuracy of our  calculation on
$\Delta S=2$ weak Hamiltonian matrix elements, 
we are able to provide a refined Unitarity Triangle analysis improving the bounds coming from model 
independent constraints on New Physics.
In our non-perturbative computation we use a combination of $N_f=2$ maximally twisted sea
quarks and Osterwalder-Seiler valence quarks in order to achieve both
O($a$)-improvement and continuum-like renormalization properties for the relevant
four-fermion operators. The calculation of the
renormalization constants has been performed non-perturbatively in the RI-MOM scheme.
Based on  simulations at four  values of the lattice spacing and  a number of quark masses we 
have extrapolated/interpolated our results to the continuum
limit and physical light/strange quark masses. 
}
\vskip 3 true mm
\thicktablerule
\eject
\end{titlepage}

\section{Introduction}
\label{sec:introI}

The fundamental target of present-day research activity in Particle Physics is the search for New Physics (NP) effects 
beyond the Standard Model (SM) predictions. Two main routes are followed, one based on the study of processes in 
which the direct production of NP particles at high energy colliders like the LHC takes place, and a second 
one based on the indirect investigation of NP effects coming from the exchange of virtual NP particles.

In the so-called indirect approach a crucial role is played by Flavor Physics processes that are sensitive 
to NP through loop effects. These processes  vanish at tree level in the SM, and some of them are theoretically 
very clean, despite the fact that they are loop mediated, and in some cases also CKM or helicity suppressed. 
Among them, $\Delta F=2$ transitions have always provided some of the most stringent constraints on NP. 
For instance, the constraints from $K^0-\bar K^0$ oscillations are particularly stringent for NP models 
that generate transitions between quarks of different 
chiralities~\cite{Gabbiani:1996hi}-\cite{Gabrielli:1995bd}.  
Therefore, an accurate determination of the $\Delta S=2$ bag parameters ($B$-parameters) is crucial 
to the improvement of NP constraints.

In the present work, we provide  the first accurate lattice determination of the $\Delta S=2$ $B$-parameters relevant 
for physics beyond the SM, calculated in the continuum limit and  using data from unquenched, $N_f=2$, 
dynamical quark simulations\footnote{While finalizing our paper we became aware of the work of Ref.~\cite{Boyle:2012temp}
where $B$-parameters and ratios of four-fermion matrix elements 
have been computed at one lattice spacing with $N_f=2+1$ dynamical quarks.}.
Our results represent a significant 
improvement with respect to the (quenched) input values of Refs.~\cite{Donini:1999nn} and~\cite{Babich:2006bh} 
used so far in phenomenological analyses. 
The calculation of the $B_{K}$ parameter that is relevant
for the $K^{0}-\bar{K}^{0}$ mixing in the SM has been presented in $\cite{Constantinou:2010qv}$ using three values of the lattice 
spacing. In the present work we update that value by adding a fourth (finer) lattice spacing. 
We note that the difference between the two 
results is about half standard deviation.

The outline of the paper is as follows. Section~\ref{sec:introII} contains a brief description of the $\Delta S=2$ 
matrix elements of the effective weak Hamiltonian describing the most general pattern of $K^0-\bar{K}^0$ 
oscillations. In Section~\ref{sec:PhenAnal},  based on the results of this work for the $\Delta S=2$ $B$-parameters, 
we discuss the implications for NP of our updated Unitarity Triangle (UT) analysis~\cite{Bona:2007vi}. 
In Section~\ref{sec:tmQCD-gen} we illustrate the main theoretical features of the 
lattice setup employed in our simulations
(twisted mass lattice QCD~\cite{Frezzotti:2000nk},~\cite{FrezzoRoss1})  
and we describe  the strategy for obtaining accurate numerical 
estimates of the $B$-parameters as well as  ratios of kaon four-fermion matrix elements. 
In Section~\ref{sec:latres} we collect  our numerical results. In Section~\ref{sec:concl} we give  
our estimates of the various $B$-parameters and matrix elements ratios and  compare the present 
results with the  previous  determinations existing in the literature.  
In five Appendices we discuss a number of technicalities:  i) the renormalization properties
of the four-fermion operators in our ``mixed action" setup \cite{Frezzotti:2004wz};
ii-iii) the RI-MOM computation of renormalization constants (RCs) and corresponding results respectively;
iv) tables of lattice data on pseudoscalar meson masses, 
decay constants and bare four-fermion matrix elements;
v) complete results on renormalized four-fermion matrix elements obtained 
by using various formulae for the chiral extrapolation and two alternative procedures for the RI-MOM determination of RCs.

\section{$\mathbf{\Delta S=2}$ effective weak Hamiltonian}
\label{sec:introII}

The general form of the $\Delta S=2$ effective weak Hamiltonian is 
\begin{equation}
{\cal H}_{\rm{eff}}^{\Delta S=2} = \frac{1}{4}\sum_{i=1}^{5} C_i {\cal O}_i 
+ \frac{1}{4}\sum_{i=1}^{3} \tilde{C}_i \tilde{{\cal O}}_i \, ,\label{Heff}
\end{equation} 
where in the so-called SUSY basis~(\cite{Gabbiani:1996hi},~\cite{Bagger:1997gg}) 
the four-fermion operators ${\cal O}_i$ and $\tilde{{\cal O}}_i$ have the form
\begin{eqnarray}\label{def_Oi}
{\cal O}_1 &=& [\bar{s}^\alpha \gamma_\mu (1-\gamma_5)d^\alpha][\bar{s}^\beta \gamma_\mu (1-\gamma_5)d^\beta] \nonumber \\
{\cal O}_2 &=& [\bar{s}^\alpha (1-\gamma_5)d^\alpha][\bar{s}^\beta  (1-\gamma_5)d^\beta] \nonumber \\
{\cal O}_3 &=& [\bar{s}^\alpha (1-\gamma_5)d^\beta][\bar{s}^\beta  (1-\gamma_5)d^\alpha]  \\
{\cal O}_4 &=& [\bar{s}^\alpha (1-\gamma_5)d^\alpha][\bar{s}^\beta  (1+\gamma_5)d^\beta] \nonumber \\
{\cal O}_5 &=& [\bar{s}^\alpha (1-\gamma_5)d^\beta][\bar{s}^\beta  (1+\gamma_5)d^\alpha] \nonumber
\end{eqnarray}
\begin{eqnarray}
\tilde{{\cal O}}_1 &=& [\bar{s}^\alpha \gamma_\mu (1+\gamma_5)d^\alpha][\bar{s}^\beta \gamma_\mu (1+\gamma_5)d^\beta] \nonumber \\
\tilde{{\cal O}}_2 &=& [\bar{s}^\alpha (1+\gamma_5)d^\alpha][\bar{s}^\beta  (1+\gamma_5)d^\beta]  \\
\tilde{{\cal O}}_3 &=& [\bar{s}^\alpha (1+\gamma_5)d^\beta][\bar{s}^\beta  (1+\gamma_5)d^\alpha] \nonumber 
\end{eqnarray}
with $\alpha$ and $\beta$ denoting color indices. Spin indices are implicitly contracted within square brackets. 
The Wilson coefficients $C_i$ and $\tilde{C}_i$  have an implicit renormalization  scale dependence which 
is compensated by the  scale dependence of  the renormalization constants of the corresponding  
operators\footnote{We thank Robert Ziegler for pointing out the numerical factors of  $1/4$ in Eq.~(\ref{Heff}), 
which were missing in the previous version of the paper.}. 
 
Notice that the parity-even parts of the operators ${\cal O}_i$ and $\tilde{{\cal O}}_i$ are equal. 
From now on and for notational simplicity we will denote by $O_i$ ($i=1, \ldots, 5$) the parity-even 
components of the operators~(\ref{def_Oi}).  
Due to parity conservation in strong interactions, in the study of $\bar K^0-K^0$ oscillations it is then sufficient 
to consider only the matrix elements $\langle \bar{K}^0 | O_i | K^0 \rangle$. 
We recall that in the SM only the kaon matrix element of the operator ${\cal O}_1$ comes into play.

The bag parameters, $B_i$ ($i=1, \ldots, 5$), provide the value of four-fermion matrix elements in units 
of the magnitude of their vacuum saturation approximation. More explicitly they are defined 
by the equations~\cite{Allton:1998sm}
\begin{eqnarray}
 \langle \bar{K}^{0} | O_1(\mu) | K^{0} \rangle &=& \xi_1\, B_1(\mu) ~ m_{K}^{2} f_{K}^{2}  \label{B1} \\
 \langle \bar{K}^{0} | O_i(\mu) | K^{0} \rangle &=& \xi_i\, B_i(\mu) ~ 
\LSB \dfrac{ m_{K}^{2} f_{K}}{ m_s (\mu) + m_d(\mu)} \RSB^2  ~~~ {\rm for} ~~ i=2,\ldots, 5, 
\label{Bi} 
\end{eqnarray}
with $\xi_i=(8/3,\, -5/3,\, 1/3,\, 2,\, 2/3)$.
In the above relations one recognizes $B_1$ as the familiar $B_K$. 
We recall that, as suggested by the parametrization adopted in the r.h.s.\ of 
Eqs.~(\ref{B1}) and~(\ref{Bi}), the $\langle \bar{K}^{0}|O_1(\mu)|K^{0}\rangle$ matrix element 
is expected to vanish in the chiral limit unlike the other four ones.

An alternative way which has the merit of allowing a more accurate evaluation of matrix elements, 
is to consider 
the  matrix elements ratios
$R_i = \langle \bar{K}^0|O_i|K^0\rangle/\langle \bar{K}^0|O_1|K^0\rangle$, $i=2,\ldots,5$, 
as first proposed in Ref.~\cite{Donini:1999nn}.
For details on our lattice implementation see Section~\ref{sec:tmQCD-gen} and in particular Eq.~(\ref{REN_Riratio}).

For the reader's convenience we here anticipate our final continuum results for $B_i$ and $R_i$ 
in the $\overline{\rm{MS}}$  scheme of Buras {\it et al.}, defined in Ref.~\cite{mu:4ferm-nlo}, 
and  the RI-MOM scheme~\footnote{Actually, instead of the standard version of the RI-MOM scheme defined in~\cite{renorm_mom:paper1} 
we employ the RI'-MOM scheme~\cite{Franco:1998bm}.  The prime on RI is to remind 
about the specific definition we adopted for the quark field RC, $Z_q\, =\, \Sigma_1$, where $\Sigma_1$ 
is the quark propagator form factor defined in Eq.~(3.5) 
of Ref.~\cite{Constantinou:2010gr}.} at 2~GeV,
see Tables~\ref{results_MSbar_intro} and~\ref{results_RIMOM_intro} respectively. Details on the 
calculation and uncertainty estimates are given in Section~\ref{sec:latres}. 
In Appendix~\ref{App_CL} we also give the final continuum results for $B_i$ and $R_i$ in the $\overline{\rm{MS}}$ 
and  the RI-MOM scheme at 3~GeV.

\begin{table}[!h]
\begin{center}
\begin{tabular}{|c|c|c|c|c|}
\hline 
\multicolumn{5}{|c|}{$\overline {\rm{MS}}$ (2 GeV)}\tabularnewline
\hline
\hline 
$B_{1}$ & $B_{2}$ & $B_{3}$ & $B_{4}$ & $B_{5}$\tabularnewline
\hline
\hline 
0.53(2) & 0.52(2) & 0.89(5) & 0.78(3) & 0.57(4)\tabularnewline
\hline
\multicolumn{1}{c}{}  \tabularnewline
\hline 
$R_{1}$ & $R_{2}$ & $R_{3}$ & $R_{4}$ & $R_{5}$\tabularnewline
\hline
\hline 
1 & -14.0(5) & 4.8(3) & 24.2(8) & 5.9(4)\tabularnewline
\hline
\end{tabular}
\caption{Continuum limit results for $B_i$ and $R_i$, renormalized in the $\overline{\rm{MS}}$ scheme of Ref.~\cite{mu:4ferm-nlo}
at 2~GeV.}
\label{results_MSbar_intro}
\end{center}
\end{table}

\begin{table}[!h]
\begin{center}
\begin{tabular}{|c|c|c|c|c|}
\hline 
\multicolumn{5}{|c|}{RI-MOM (2 GeV)}\tabularnewline
\hline
\hline
$B_{1}$ & $B_{2}$ & $B_{3}$ & $B_{4}$ & $B_{5}$\tabularnewline
\hline
\hline 
0.52(2) & 0.70(2) & 1.22(7) & 1.00(4) & 0.69(5)\tabularnewline
\hline
\multicolumn{1}{c} {}\tabularnewline
\hline 
$R_{1}$ & $R_{2}$ & $R_{3}$ & $R_{4}$ & $R_{5}$\tabularnewline
\hline
\hline 
1 & -12.9(4) & 4.5(2) & 21.2(7) & 4.7(3)\tabularnewline
\hline
\end{tabular}
\caption{Continuum limit results for $B_i$ and $R_i$, renormalized in the RI-MOM scheme at 2~GeV.}
\label{results_RIMOM_intro}
\end{center}
\end{table}

\section{Model-independent constraints on $\mathbf{ \Delta S=2}$ operators and
  New Physics scale from the Unitarity Triangle analysis}
\label{sec:PhenAnal}

$\Delta F=2$ processes provide some of the most stringent constraints
on NP generalizations of the SM. Several phenomenological analyses of
$\Delta F=2$ processes have been performed in the last years, both for
specific NP models and in model-independent frameworks. 
A generalization of the UT analysis, which allows for NP effects by including
the most significant flavour constraints on NP available at the time was
performed in Ref.~\cite{Bona:2007vi}.
The result was a simultaneous determination of the CKM
parameters and the size of NP contributions to $\Delta F=2$ processes
in the neutral kaon and $B_{d,s}$ meson sectors.

The NP generalization of the UT analysis consists in including in the
theoretical param-etrization of the various observables  the matrix
elements of operators which, though absent in the SM, may appear in
some of its extensions. The analysis shows that the constraints coming
from $K^0-\bar K^0$ matrix elements are the most stringent ones, in
particular for models that generate transitions between quarks of
different chiralities 
(see Refs.~\cite{Gabbiani:1996hi} - \cite{Gabrielli:1995bd}). 
Thus an accurate determination
of the $\Delta S=2$ $B$-parameters is crucial to the improvement of
the NP constraints.

The results for the $\Delta S=2$ $B$-parameters obtained in the
present work come from unquenched $N_f=2$ lattice QCD 
data carefully extrapolated to the continuum limit. They hence 
represent a significant progress with respect to the input values used
in the UT analysis performed in Ref.~\cite{Bona:2007vi}, where
quenched lattice numbers (without a systematic continuum limit extrapolation analysis) 
computed more than five years ago in
Refs.~\cite{Donini:1999nn} and~\cite{Babich:2006bh} were employed.
For this reason, we present here an update of the analysis of
Ref.~\cite{Bona:2007vi} based on  our new  values of the $\Delta S=2$
$B$-parameters. The new
ingredients entering the analysis are collected in 
Tables~\ref{results_MSbar_intro} and~\ref{results_RIMOM_intro}. 
For all the other input data we use the numbers quoted in
Ref.~\cite{utfitwebpage} in the Summer 2012 analysis. 

In the present NP-oriented analysis, the relations
among experimental observables and the CKM matrix elements are extended 
by taking into consideration the most general form of the $\Delta S=2$
effective weak Hamilonian (see Eq.~(\ref{Heff})). 
The effective weak Hamilonian is parameterized
by Wilson coefficients of the form
\begin{equation}
  C_i (\Lambda) = \frac{F_i L_i}{\Lambda^2}\, ,\qquad i=2,\ldots,5\, ,
  \label{eq:cgenstruct}
\end{equation}
where $F_i$ is the (generally complex) relevant NP flavor coupling,
$L_i$ is a (loop) factor which depends on the interactions that
generate $C_i(\Lambda)$, and $\Lambda$ is the scale of NP, i.e.\ the
typical mass of new particles mediating $\Delta S=2$ transitions. For
a generic strongly interacting theory with an unconstrained flavor
structure, one expects $F_i \sim L_i \sim 1$, so that the
phenomenologically allowed range for each of the Wilson coefficients
can be immediately translated into a lower bound on
$\Lambda$. Specific assumptions on the flavor structure of NP
correspond to special choices of the $F_i$ functions. For example
Minimal Flavor Violation
(MFV) models~\cite{Misiak:1997ei}-\cite{D'Ambrosio:2002ex} 
correspond to $F_1=F_\mathrm{SM}$ and $F_{i \neq 1}=0$.

Following Ref.~\cite{Bona:2007vi}, in deriving the lower bounds on
the NP scale $\Lambda$, we assume $L_i = 1$, that corresponds to
strongly-interacting and/or tree-level coupled NP. Two other
interesting possibilities are given by loop-mediated NP contributions
proportional to either $\alpha_s^2$ or $\alpha_W^2$. The first case
corresponds for example to gluino exchange in the minimal
supersymmetric SM. The second case applies to all models with SM-like
loop-mediated weak interactions. To obtain the lower bound on
$\Lambda$ entailed by loop-mediated contributions, one simply has to
multiply the bounds we quote in the following by
$\alpha_s(\Lambda)\sim 0.1$ or $\alpha_W \sim 0.03$.

In agreement with Ref.~\cite{Bona:2007vi}, we find that in the $K^0$
sector, due to the non-vanishing chiral limit (chiral enhancement) of
their matrix elements, all bounds coming from the contributions of
non-standard operators (i.e.\ from the operators $O_i$ with $i\neq 1$)
are more than one order of magnitude stronger than the bound from the
SM $O_1$ operator. 

\begin{table}[!h]
\begin{center}
\begin{tabular}{|@{}ccc|}
\hline\hline
 & $95\%$ allowed range  &
Lower limit on $\Lambda$ \\
&(GeV$^{-2}$) &
 (TeV)\\
\hline
\phantom{A}Im\,$C^K_1$ & $[-2.1,3.4] \cdot 10^{-15}$ & $1.7 \cdot 10^{4}$ \\
\phantom{A}Im\,$C^K_2$ & $[-2.1,1.4] \cdot 10^{-17}$ & $22 \cdot 10^{4}$  \\
\phantom{A}Im\,$C^K_3$ & $[-5.1,7.8] \cdot 10^{-17}$ & $11 \cdot 10^{4}$  \\
\phantom{A}Im\,$C^K_4$ & $[-3.0,4.7] \cdot 10^{-18}$ & $46 \cdot 10^{4}$  \\
\phantom{A}Im\,$C^K_5$ & $[-0.9,1.4] \cdot 10^{-17}$ & $27 \cdot 10^{4}$ \\
\hline
\hline
\phantom{A}Im\,$C^K_1$ & $[-4.4,2.8] \cdot 10^{-15}$ & $1.5 \cdot 10^{4}$ \\
\phantom{A}Im\,$C^K_2$ & $[-5.1,9.3] \cdot 10^{-17}$ & $10 \cdot 10^{4}$  \\
\phantom{A}Im\,$C^K_3$ & $[-3.1,1.7] \cdot 10^{-16}$ & $5.7 \cdot 10^{4}$ \\
\phantom{A}Im\,$C^K_4$ & $[-1.8,0.9] \cdot 10^{-17}$ & $24 \cdot 10^{4}$  \\
\phantom{A}Im\,$C^K_5$ & $[-5.2,2.8] \cdot 10^{-17}$ & $14 \cdot 10^{4}$  \\
\hline
\hline
\end{tabular}
\end{center}
\caption {$95\%$ probability range for the Im\,$C_K^i$ coefficients
  and the corresponding lower bounds on the NP scale, $\Lambda$, for
  a generic strongly interacting NP with generic flavor structure ($L_i=F_i=1)$. In the
  lower panel the results of~\cite{Bona:2007vi} are
  displayed for comparison.} 
\label{tab:all}
\end{table}

\begin{figure}[tb]
\begin{center}
\includegraphics[scale=1.22,angle=0]{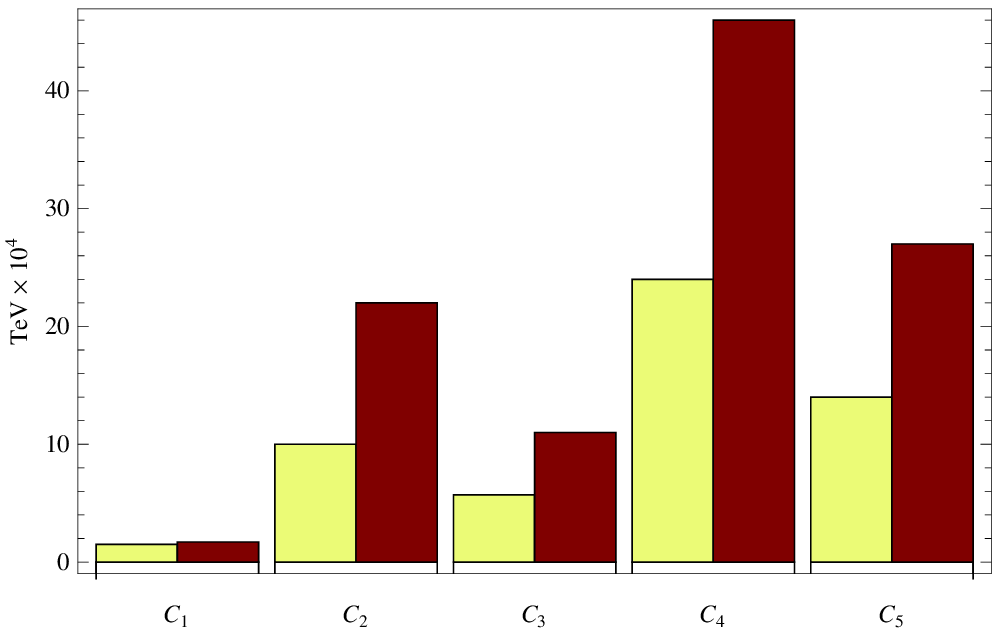} 
\caption{\sl The lower bounds on the NP scale, provided by the constraints on  
Im\,$C^K_i$ ($i=1,\ldots,5$) for generic NP flavor structure, are
shown as brown bars. For comparison, we plot  the bounds
of Ref.~\cite{Bona:2007vi} as yellow bars.  
\label{fig:all}}
\end{center}
\end{figure}

The results for the upper bounds on the Im$\,C^K_i$ coefficients and
the corresponding lower bounds on the NP scale $\Lambda$ are collected
in Table~\ref{tab:all} where they are compared
to the previous results
of Ref.~\cite{Bona:2007vi}. The superscript $K$ is to recall that we
are reporting the bounds coming from the kaon sector we are here
analyzing. Although several input parameters have been updated with
respect to Ref.~\cite{Bona:2007vi} (see Ref.~\cite{utfitwebpage}), the
more stringent constraints on the Wilson coefficients of the
non-standard operators and, consequently, on the NP scale, mainly come
from the improved accuracy achieved in the values of the $\Delta S=2$
$B$-parameters obtained in the present work. This can be realized by
comparing the small improvement of the bound coming from Im$\,C^K_1$,
obtained using a value of the $B_1$-parameter very close to the one
taken in Ref.~\cite{Bona:2007vi}, with those coming from the other
coefficients using the new $B$-parameters.  We observe that the
analysis is performed (as in~\cite{Bona:2007vi}) by switching on one
coefficient at the time in each sector, thus excluding the possibility
of having accidental cancellations among the contributions of
different operators.  Therefore, the reader should keep in mind that
the bounds may be weakened if, instead, some accidental cancellation
occurs.

In Fig.~\ref{fig:all} we show the comparison between the lower bounds
on the NP scale obtained for the case of a generic strongly
interacting NP with generic flavor structure by the constraints on the
Im$\,C^K_i$ coefficients coming from the present generalized UT
analysis, and the previous results of Ref.~\cite{Bona:2007vi}.

As a specific example of NP models  we consider the warped five-dimensional
extensions of the SM discussed in Ref.~\cite{Csaki:2008zd}, where the
origin of hierarchies in quark masses and mixings is explained via the
localization properties of quark wave functions in the fifth dimension. In
particular, in the Randall-Sundrum (RS) scenario one has
\begin{equation}
  \label{eq:RS}
  L_4 = (g_s^*)^2, \quad F_4 = \frac{2 m_d m_s}{Y_*^2
    v^2}\,, \quad \Lambda = M_{G}\,,
\end{equation}
where $M_G$ and $g_s^* \sim 6$ are the mass and coupling of
Kaluza-Klein excitations of the gluon, $Y_* \sim 3$ is the
five-dimensional Yukawa coupling (whose flavour structure is assumed
to be anarchic), $m_{d} \sim 3$ MeV and $m_s \sim 50$ MeV are
$\overline{\mathrm{MS}}$ quark masses at the high scale and $v = 246$
GeV is the Higgs vev. Running from a reference scale of $5$ TeV, we
obtain at $95\%$ probability Im\,$C^K_4 \in [-4.7,10.6] \cdot
10^{-18}$, from which we get
\begin{equation}
  \label{eq:RSbound}
  M_G > 43\, \mathrm{TeV}.
\end{equation}
Considering instead gauge-Higgs unification (GHU) models, from
Ref.~\cite{Csaki:2008zd} we have
\begin{equation}
  \label{eq:GHU}
  L_4 = (g_s^*)^2\,, \quad F_4 \sim \frac{8 m_d m_s}{g_*^2
    v^2}\,, \quad \Lambda = M_{G}\,,
\end{equation}
where in this case $g_* \sim 4$ is the five-dimensional gauge coupling in
units of the radius of the compact dimension. We obtain the bound
\begin{equation}
  \label{eq:GHUbound}
  M_G > 65\, \mathrm{TeV}.
\end{equation}

\section{Non-perturbative lattice computation of the $\mathbf{\Delta S=2}$ matrix elements}
\label{sec:tmQCD-gen}

Lattice QCD provides an ideal first principle framework in which non-perturbative computation of hadronic matrix elements can be 
performed with controlled systematic uncertainties. In the last few years a number of lattice determinations of $B_K$ of  
increasing precision have appeared in the literature based on a variety of lattice regularizations with $N_f=2$ or $N_f=3$ 
dynamical 
fermions~\cite{Constantinou:2010qv}, \cite{Aubin:2009jh}-\cite{Bae:2011ff}. 
For recent reviews see 
Refs.~\cite{Colangelo:2010et}-\cite{Laiho:2011rb}.  

Very little has been done in the literature concerning the calculation of the physical matrix elements of the full 
operator basis of Eq.~(\ref{def_Oi}). 
Calculations of the whole set of $\Delta S=2$ renormalized operators have been carried out, 
in the quenched approximation, 
using improved Wilson fermions (see Refs.~\cite{Donini:1999nn} and~\cite{Allton:1998sm}) or the chirality conserving 
overlap and domain-wall fermions (see Refs.~\cite{Babich:2006bh} and~\cite{Nakamura:2006eq}, respectively). Very recently a $N_f=2+1$ 
dynamical quark calculation appeared \cite{Boyle:2012temp} that uses domain-wall fermions at one value of the lattice 
spacing\footnote{ A preliminary computation of the bare matrix elements 
using unquenched $N_f=2+1$ domain--wall dynamical fermions was presented in Ref.~\cite{Wennekers:2008sg}.}. 

The first calculation of the kaon matrix elements of the whole operator basis was performed employing Clover improved 
Wilson fermions. Since the Clover term coefficient was set to its tree-level value, matrix elements were 
affected by O($g_0^2 a$) discretization errors. Simulations were carried out at two values of the gauge coupling 
corresponding to lattice spacings $\sim 0.07$ and $\sim 0.09$~fm~(\cite{Donini:1999nn},~\cite{Allton:1998sm}). 
The major source of systematic errors was, however, the uncertainty related to the construction of the multiplicatively renormalizable lattice operators $O_i$. 
In fact, owing to the breaking of the chiral symmetry intrinsic in the Wilson fermion action, the bare
counterparts of each of them mix with all the other operators of equal dimension including those 
with ``wrong'' chiral transformation properties~\cite{Bochicchio:1985xa}. 
All the mixing coefficients and the overall RC were computed 
in the non-perturbative RI-MOM scheme~\cite{renorm_mom:paper1}.
 
Similar quenched computations were carried out using overlap 
and domain-wall fermions (see Refs.~\cite{Babich:2006bh} 
and~\cite{Nakamura:2006eq}, respectively). Though performed at pretty coarse 
lattice spacings (namely $a\sim 0.09$ and $\sim 0.13$~fm in 
the case of overlap fermions and $a\sim 0.1$~fm in the case of domain-wall fermions), 
these simulations have the advantage 
that the renormalization properties of the operators entering the four-fermion basis 
 are  as in the continuum, and lattice artifacts are O($a^2$).  Also in this case the non-perturbative 
RI-MOM scheme was used in the computation of the various RCs.

In the following sections we present a new unquenched computation of the kaon matrix elements of the full $\Delta S=2$ 
four-fermion operator basis employing lattice data from simulations with $N_f=2$ dynamical fermions, performed at four 
rather fine values of the lattice spacing in the interval [0.05, 0.1]~fm. We are thus able to safely extrapolate 
the lattice estimators of all the relevant  matrix elements to the continuum limit (CL). 

We use a mixed fermion action setup where we adopt different regularizations for sea and valence quarks. In particular we 
introduce maximally twisted (Mtm) sea quarks~\cite{FrezzoRoss1} that we take in combination with Osterwalder--Seiler 
(OS)~\cite{Osterwalder:1977pc} valence quarks. 
This strategy has been suggested in Ref.~\cite{Frezzotti:2004wz} as a way 
of setting up a computational 
framework allowing for a calculation of $\Delta S=2$ four-fermion matrix elements that is both 
automatically O($a$) improved and free of wrong chirality mixing effects. 
A proof of the latter point is given in Appendix~\ref{APP_Op_Bas} and its validity is numerically verified in Appendix~\ref{App_RIMOM}, while O$(a)$ improvement of physical quantities 
is a genuine property of the setup of Ref.~\cite{Frezzotti:2004wz}. As a consequence
unitarity violations due to different sea and 
valence quark regularization yield only O($a^2$) artifacts, provided renormalized sea and valence 
quark masses are matched. In our case, the matching of the renormalized quark masses is obtained by simply taking identical 
values for the corresponding sea and valence bare mass parameters. 

The interesting lattice setup briefly described above has already been successfully tested in 
$B_K$ computations both in the quenched approximation~\cite{Dimopoulos:2009es}
and on ensembles with $N_f=2$~\cite{Constantinou:2010qv} and $N_f=2+1+1$ dynamical quarks~\cite{Carrasco:2011gr}, 
as well as in unquenched ($N_f=2+1+1$) studies of meson masses and
decay constants~\cite{Farchioni:2010tb} and nucleon sigma terms~\cite{Dinter:2012tt}.

\subsection{Sea and valence quark regularization}
\label{sec:seavalreg}

The Mtm-LQCD action of the light quark flavor doublet can be written in 
the so-called ``physical basis" in the form~\cite{FrezzoRoss1}
\begin{equation}
S^{\rm Mtm}_{sea} =  a^4 \sum_x \bar \psi (x) \Big \{ \frac{1}{2} \sum_\mu \gamma_\mu(\nabla_\mu 
+ \nabla^\ast_\mu ) - i \gamma_5 \tau^3 r_{\rm{sea}} \big [ M_{\rm cr} - 
\dfrac{a}{2} 
\sum_\mu \nabla^\ast_\mu \nabla_\mu \big ] + \mu_{sea} \Big \} \psi(x)\, .\label{SPHYS}
\end{equation}
The subscript $sea$ is to remind us that this action will be used to generate unquenched gauge configurations. 
The field $\psi$ describes a mass degenerate up and down doublet with  bare (twisted) mass $\mu_{sea}$. 
The parameter $M_{\rm cr}$ is the critical mass that one has to fix non-perturbatively at its optimal value 
(as proposed in Refs.~\cite{Frezzotti:2005gi}-\cite{Aoki:2004ta}  
and implemented in Refs.~\cite{Boucaud:2007uk} and~\cite{Boucaud:2008xu}) 
to guarantee the O($a$)-improvement of physical observables and get rid of all the unwanted 
leading chirally enhanced cutoff effects. In the gauge sector the tree-level improved action 
proposed in Ref.~\cite{Weisz:1982zw} has been used.  

For valence quarks we use the OS regularization~\cite{Osterwalder:1977pc}. 
The full valence action is given by the sum of the contributions of each {\it individual} 
valence flavour $q_f$ and reads~\cite{Frezzotti:2004wz}
\begin{equation}
S^{\rm OS}_{val} =  a^4 \sum_{x,f} \bar q_f (x) \Big \{ \frac{1}{2} \sum_\mu \gamma_\mu(\nabla_\mu 
+ \nabla^\ast_\mu ) - i \gamma_5 r_f \big [ M_{\rm cr} - 
\frac{a}{2} \sum_\mu \nabla^\ast_\mu \nabla_\mu \big ] + \mu_f \Big \} q_f(x)\, ,
\label{OSvalact}
\end{equation}
where the index $f$ labels the valence flavors and $M_{\rm cr}$ is the same critical mass parameter 
which appears in Eq.~(\ref{SPHYS}). We denote by $r_f$ and $\mu_f$ the values of the Wilson parameter 
and the twisted quark mass of each valence flavor. 

\subsection{Lattice operators and correlation functions} 
\label{sec:LATOP}

In the strategy proposed in Ref.~\cite{Frezzotti:2004wz}, which we follow here, 
four species of OS valence quark flavors ($q_f$, $f=1, \ldots, 4$) are introduced, 
two of which ($q_1$ and $q_3$) will represent the valence strange quark with masses 
$\mu_1=\mu_3 \equiv \mu_{``s"}$, while the other two ($q_2$ and $q_4$) will be 
identified with the light up/down quarks having masses $\mu_2=\mu_4 \equiv \mu_\ell$. 
The corresponding $r_f$ Wilson parameters must obey the relation
\begin{equation}
r_1 = r_2 = r_3 = -r_4\, .
\label{RPAR}
\end{equation}
In the numerical computations reported in the present work we 
have averaged over the two cases $r_1=\pm1$, holding $r_2$, $r_3$ and $r_4$ related to $r_1$ as in Eq.~(\ref{RPAR}).

As in the case of the computation of $B_{K}(\equiv B_1)$, in the calculation of the bag parameters $B_i$ ($i=2, \ldots, 5$), we need to consider the axial currents
\begin{equation}
A_\mu^{12} \,= \, \bar q_1 \gamma_\mu \gamma_5 q_2 \qquad \qquad
A_\mu^{34} \, = \, \bar q_3 \gamma_\mu \gamma_5 q_4 . \label{ADEF}
\end{equation}
and the pseudoscalar quark densities 
\begin{equation}
P^{12} \,= \, \bar q_1 \gamma_5 q_2 \qquad \qquad
P^{34} \, = \, \bar q_3 \gamma_5 q_4 \, . \label{PDEF}
\end{equation}
In addition, we need to consider the following set of four-fermion operators 
\begin{eqnarray}
&&O^{MA}_{1[\pm]}= 2\big{\{}\big{(}[\bar q_1^\alpha\gamma_\mu q_2^\alpha][\bar q_3^\beta\gamma_\mu q_4^\beta]+[\bar q_1^\alpha\gamma_\mu \gamma_5 q_2^\alpha][\bar q_3^\beta\gamma_\mu \gamma_5 q_4^\beta]\big{)}\pm \big{(}2\leftrightarrow 4\big{)}\big{\}}\nonumber \\
&&O^{MA}_{2[\pm]}=2\big{\{}\big{(}[\bar q_1^\alpha q_2^\alpha][\bar q_3^\beta q_4^\beta]+[\bar q_1^\alpha\gamma_5 q_2^\alpha][\bar q_3^\beta\gamma_5 q_4^\beta]\big{)}\pm \big{(}2\leftrightarrow 4\big{)}\big{\}} \nonumber \\
&&O^{MA}_{3[\pm]}= 2\big{\{}\big{(}[\bar q_1^\alpha q_2^\beta][\bar q_3^\beta q_4^\alpha]+[\bar q_1^\alpha\gamma_5 q_2^\beta][\bar q_3^\beta\gamma_5 q_4^\alpha]\big{)}\pm \big{(}2\leftrightarrow 4\big{)}\big{\}}\nonumber \\
&&O^{MA}_{4[\pm]}= 2\big{\{}\big{(}[\bar q_1^\alpha q_2^\alpha][\bar q_3^\beta q_4^\beta]-[\bar q_1^\alpha\gamma_5 q_2^\alpha][\bar q_3^\beta\gamma_5 q_4^\beta]\big{)}\pm \big{(}2\leftrightarrow 4\big{)}\big{\}}\nonumber \\
&&O^{MA}_{5[\pm]}=2\big{\{}\big{(}[\bar q_1^\alpha q_2^\beta][\bar q_3^\beta q_4^\alpha]-[\bar q_1^\alpha\gamma_5 q_2^\beta][\bar q_3^\beta\gamma_5 q_4^\alpha]\big{)}\pm \big{(}2\leftrightarrow 4\big{)}\big{\}}\, ,
\label{OMAPM_v2}
\end{eqnarray}
where square parentheses denote spin invariants and $\alpha$ and $\beta$ 
  are color indices. 

In the mixed action (MA) approach defined above the following properties can be proved 
(see Appendix~A and Ref.~\cite{Frezzotti:2004wz}). 
\begin{enumerate}
\item The operators $O^{MA}_{i[+]}$ defined in Eq.~(\ref{OMAPM_v2}) 
enjoy continuum-like renormalization properties,
\begin{equation}
\label{renorm_pattern_main}
\left( \begin{array}{c}
 O_{1[+]}^{MA} \\
 O_{2[+]}^{MA} \\
 O_{3[+]}^{MA} \\
 O_{4[+]}^{MA} \\
 O_{5[+]}^{MA} \end{array} \right)_{\rm ren} = 
\left( \begin{array}{ccccc}
Z_{11} & 0 & 0 & 0 & 0 \\
0 & Z_{22} & Z_{23} & 0 & 0 \\
0 & Z_{32} & Z_{33} & 0 & 0 \\
0 & 0 & 0 & Z_{44} & Z_{45} \\
0 & 0 & 0 & Z_{54} & Z_{55} \\
\end{array} \right)
\left( \begin{array}{c}
O_{1[+]}^{MA}\\
O_{2[+]}^{MA} \\
O_{3[+]}^{MA} \\
O_{4[+]}^{MA} \\
O_{5[+]}^{MA} \end{array} \right)
\end{equation}
where the matrix $Z_{ij}$ is defined in 
Eqs.(\ref{renorm_pattern}),~(\ref{RENZ}) and~(\ref{ZQ}).   

\item The axial currents and pseudoscalar quark densities, defined in 
Eqs.~(\ref{ADEF})  and~(\ref{PDEF}), are renormalized according to the 
formulae~(\cite{Frezzotti:2004wz},~\cite{Constantinou:2010gr}) 
\begin{eqnarray}
&&[A_\mu^{12}]_{\rm ren} \,= Z_A \, A_\mu^{12}  \qquad \qquad
[A_\mu^{34}]_{\rm ren} \, = Z_V \, A_\mu^{34} \, ,\label{APREN}\\
&&[P^{12}]_{\rm ren} \,= Z_S \, P^{12}  \qquad \qquad
[P^{34}]_{\rm ren} \, = Z_P \, P^{34} \, .\label{PREN}
\end{eqnarray} 
\item 
The matrix elements $\langle P^{43}|O^{MA}_{i[+]}|P^{12}\rangle_{\rm ren}$, obtained from correlation
functions of the renormalized operators in Eqs.~(\ref{renorm_pattern_main})--(\ref{PREN}), 
with the identification $\mu_1=\mu_3=\mu_s$ (= bare strange quark mass) 
and $\mu_2=\mu_4=\mu_\ell$ (= bare up-down quark mass), will tend in the limit $a\to 0$ 
to the continuum matrix elements $\langle \bar K^0| O_{i}|K^0\rangle$ 
of the (parity-even parts of the) operators of Eq.~(\ref{def_Oi}) with mere O($a^2$) discretization errors. 

\end{enumerate}
The key statement (iii) follows by noting that the set of fermionic Wick
contractions contributing to the three-point correlation functions 
(see Eq.~(\ref{PQPi-correl}) below) from which the 
matrix elements $\langle P^{43}|O^{MA}_{i[+]}|P^{12}\rangle_{\rm ren}$
are extracted coincides with the set of Wick contractions in the
the three-point continuum QCD correlator relevant for the computation
of the matrix elements $\langle \bar K^0| O_{i}|K^0\rangle$ and by exploiting
the general renormalizability and O($a$) improvement properties of our
MA lattice setup (spelled out in Ref.~\cite{Frezzotti:2004wz}), 
as well as the renormalization 
properties~(\ref{renorm_pattern_main}) of the operators $O^{MA}_{i[+]}$.
Concerning the issue of O($a$) improvement we recall here that \\
\begin{itemize}
\item the bare matrix elements of $O_{i[+]}^{MA}$'s are free from O($a$) cutoff effects;
\item the relevant renormalization constants in the RI-MOM scheme can also be evaluated with 
      no O($a$) artefacts.
\end{itemize} 
The first of these properties was derived in Ref.~\cite{Frezzotti:2004wz} 
for generic hadron masses and matrix elements~\footnote{
A simpler derivation might also be given along the lines of App.~A of 
Ref.~\cite{Frezzotti:2005gi} exploiting the symmetries of the MA lattice setup of 
Ref.~\cite{Frezzotti:2004wz}.                          }.
The second property above follows from the remark that the 
one-particle-irreducible vertices entering the RI-MOM
renormalization conditions are O($a$) improved in our MA setup, 
just because such vertices turn out to be
invariant under parity transformations of their external momenta.
The argument here is closely analogous to the one presented for the renormalization
constants of quark bilinear operators in the Appendix of Ref.~\cite{Constantinou:2010gr}.

In the construction of correlation functions we follow the general procedure outlined 
in Ref.~\cite{Constantinou:2010qv}. We use periodic boundary conditions in every direction for all fields, 
except for the quark fields on which we 
impose anti-periodic boundary conditions in the time direction. At time slices $y_0$ and $y_0+T/2$ ``wall" 
operators with $K^0$-meson quantum numbers are inserted. 
The first operator is constructed in terms of $\bar q_2$ and $q_1$ quark fields and the second in terms 
of $\bar q_4$ and $q_3$ quark fields. Explicitly they have the 
expressions 
\begin{eqnarray}{\cal P}^{21}_{y_0} &=&   \Big{(}\dfrac{a}{L}\Big{)}^3\, \sum_{\vec y} \bar q_2(\vec y, y_0) 
\gamma_5 q_1(\vec y , y_0) \nonumber \\ 
{\cal P}^{43}_{y_0+\frac{T}{2}} &=&   \Big{(}\dfrac{a}{L}\Big{)}^3\, \sum_{\vec y} \bar q_4(\vec y, y_0+T/2) 
\gamma_5 q_3(\vec y , y_0+T/2) \label{K-WALL}\end{eqnarray}
The correlators we then need to compute are~\footnote{For the special case of $i=1$, the  evaluation of 
$B_1=B_{\rm{K}}$ (see Ref.~\cite{Constantinou:2010qv}) requires 
the use of the two-point correlation functions that involve the axial current and have the form 
$C_{A_0P}(x_0) = \Big{(}{a}/{L}\Big{)}^3\sum_{\vec x} \langle A_0^{12}(\vec x,x_0) \,{\cal P}^{21}_{y_0}\rangle$ and 
$C_{PA_0}^\prime(x_0) = \Big{(}{a}/{L}\Big{)}^3 \sum_{\vec x} \langle {\cal P}^{43}_{y_0 + \frac{T}{2}}  
\,A_0^{34}(\vec x,x_0) \rangle$.} 
\begin{eqnarray}
\hspace{-1.5cm}&&C_i(x_0) = \Big{(}\dfrac{a}{L}\Big{)}^3\sum_{\vec x} \langle 
{\cal P}^{43}_{y_0 + \frac{T}{2}} \,O_{i[+]}^{MA} (\vec x,x_0) \, {\cal P}^{21}_{y_0} \rangle\, , \quad i=1, \ldots, 
5\, ,
\label{PQPi-correl}\\
\hspace{-1.5cm}&&C_{PP}(x_0) = \Big{(}\dfrac{a}{L}\Big{)}^3\sum_{\vec x} 
\langle P^{12}(\vec x,x_0) \,{\cal P}^{21}_{y_0} \rangle\, ,
\label{PP-12}\\
\hspace{-1.5cm}&&C_{PP}^\prime(x_0) = \Big{(}\dfrac{a}{L}\Big{)}^3 \sum_{\vec x} 
\langle {\cal P}^{43}_{y_0 + \frac{T}{2}}  \,P^{34}(\vec x,x_0) \rangle \, .
\label{PP-34}
\end{eqnarray}
To improve the signal-to-noise ratio a sum has been performed over the spatial position of the four-fermion operator, 
and for each gauge configuration the time slice $y_0$ is randomly chosen. An important contribution to the reduction of 
statistical fluctuations comes also from summing over the spatial position of 
both kaon interpolating fields in Eq.~(\ref{PQPi-correl}).
After summing over the the spatial position of the four-fermion operator, which is known from experience to be crucial for the 
signal, to project on zero 3-momentum states, just one further
spatial sum is needed.
The second spatial sum, which we are able to do, gives a further signal improvement. 
These spatial sums were implemented and carried out at a reasonably low computational price by 
means of the stochastic technique discussed in sect. 2.2 of Ref.~\cite{Constantinou:2010qv}.

For large time separation $y_0 \ll x_0 \ll y_0+T/2$ the (plateau of the) following ratio estimate
\begin{equation}
E[B_{i}^{(b)}](x_0) \,= \, \dfrac{C_{i}(x_0)}{C_{PP}(x_0) \,\, C_{PP}^\prime(x_0)}, \quad  i=2, \ldots, 5 
\label{Biratio}
\end{equation}
provides an estimate of the  
$B_i^{(b)}$ ($i=2, \ldots, 5$) 
bag parameter~\footnote{In the following the superscript $(b)$ denotes bare quantities.} since\begin{equation}
E[B_{i}^{(b)}](x_0) \, \mathop {\xrightarrow{\hspace*{1.2cm}}} \limits^{y_0 \ll x_0 \ll y_0+T/2} \, \left.  \dfrac{\langle \bar{K}^0 | O_{i[+]}^{MA} | K^0 \rangle}{\langle \bar{K}^0 | P^{12} | 0 \rangle \, 
\langle 0 | P^{34} | K^0 \rangle} \right|^{(b)} = \xi_i\, B_i^{(b)}, \quad i=2, \ldots, 5\, ,
\label{bareBi}
\end{equation} 
with $(\xi_2,\, \xi_3,\, \xi_4,\, \xi_5)=(-5/3,\, 1/3,\, 2,\, 2/3)$. \\
We also compute the ratios 
\begin{equation} E[R_{i}^{(b)}](x_0) \,= \, \dfrac{C_{i}(x_0)}{C_{1}(x_0)},  \qquad i=2, \ldots, 5 \, ,
\label{ROiratio} \end{equation} 
which for large time separations yield the ratios of the corresponding kaon four-fermion matrix elements, according to the formula 
\begin{equation}
E[R_{i}^{(b)}](x_0) \, \mathop {\xrightarrow{\hspace*{1.2cm}}} \limits^{y_0 \ll x_0 \ll y_0+T/2} \, 
\left. \dfrac{\langle \bar{K}^0 | O_{i[+]}^{MA} | K^0 \rangle}
{\langle \bar{K}^0 | O_{1[+]}^{MA} | K^0 \rangle} \right|^{(b)}
\,=\,R_i^{(b)},  \qquad i=2, \ldots, 5 \, . 
\label{OioverO1}
\end{equation}  

\subsection{Estimates of renormalized quantities} 
\label{sec:LATBK}

Recalling Eqs.~(\ref{renorm_pattern_main}) and (\ref{PREN}), the renormalized values of the bag parameters 
will be given by the formula
\begin{equation}
B_{i} = \dfrac{Z_{ij}}{Z_S \,Z_P} \, \, B_{j}^{(b)}, \quad i,\, j = 2, \ldots, 5 \, .
\label{REN_RBi}
\end{equation}
where a sum over the repeated index $j$ is understood.
As for the renormalized expression of the four-fermion operator ratios of Eq.~(\ref{OioverO1}), 
we choose to evaluate the rescaled quantity
\begin{equation}
\tilde R_{i} \,= \, \Big(\dfrac{f_K}{m_K}\Big)^2_{\rm{expt.}}\, \LSB \dfrac{M^{12} M^{34}}{F^{12}F^{34}}\, 
\dfrac{Z_{ij}}{Z_{11}} \, \, R_{j}^{(b)} \RSB_{\rm{Lat.}}, \quad i,\, j = 2, \ldots, 5 \, ,
\label{REN_Riratio}
\end{equation} 
where $M^{lk}$ and $F^{lk}$ are the mass and decay constant of the pseudoscalar meson made 
of the $(\bar q_l\,q_k)$ quark pair. 
In order to compensate the chiral vanishing of the 
$\langle\bar{K}^0|O_1|K^0\rangle$ matrix element we have multiplied the ratios $R_{i}^{(b)}$ 
by the factor $M^{12} M^{34}/F^{12}F^{34}$; its form has been chosen in such a way to  
partially compensate the  lattice artifacts affecting the different lattice 
discretizations of kaon mesons (resulting from different choices of the OS $r_f$-parameters) we use. 
Furthermore, we have remultiplied the quantity in the square bracket by the ratio of the experimental 
values of the kaon decay constant ($f_K^{\rm{exp}}=156.1$~MeV) and its mass ($m_K^{\rm{exp}}=494.4$~MeV). 
The definition we get in this way is in line with the one proposed in Ref.~\cite{Babich:2006bh}.
Based on the discussion of Section 4.2 (in  particular item iii), we note 
that in the continuum limit and at the physical values of the $u/d$ and $s$ quark masses 
the quantity $\tilde R_{i}$ of Eq.~(\ref{REN_Riratio}) 
provides the right  estimate for the ratio of the renormalized 
matrix elements of interest, i.e.
\begin{equation}
R_i = \dfrac{\langle \bar{K}^0 | O_i | K^0 \rangle}{\langle \bar{K}^0 | O_1 | K^0 \rangle}
\label{REN_Riratio_V2}
\end{equation}

We end this section by recalling that in the twisted mass mixed action 
setup of Ref.~\cite{Frezzotti:2004wz} we are using, lattice estimators of physical 
quantities are  only affected by O($a^2$) lattice artifacts. 
This is true in particular for the kaon mass and decay constant as well as
for the kaon four-fermion matrix elements.

\section{Simulations, data analysis and results}
\label{sec:latres}

The ETM Collaboration has generated $N_f=2$ configuration ensembles at four values of the inverse bare gauge coupling,
 $\beta$ and at a number of light quark masses, $\mu_{sea}$. 
The values of the simulated lattice spacings lie in the interval [0.05, 0.1] fm. Bare quark mass parameters are 
chosen so as to have light pseudoscalar mesons (``pions") in the range $280 \leq m_{\rm PS} \leq 500$~MeV and 
heavy-light pseudoscalar mesons (``kaons") in the range $450 \leq m_{PS} \leq 650$~MeV. Simulation details are given in Table~\ref{runs}.

The value of the light $u/d$ quark mass parameter, $a\mu_\ell$, is common to sea and valence quarks, while the 
heavier quark (the would-be strange quark that we denote by $``s"$, see Table~\ref{runs}) is quenched. As discussed 
in Section~\ref{sec:CBHAT}, we will get to the physical kaon mass by suitably interpolating (extrapolating) data 
in $\mu_{``s"}$ ($\mu_\ell$) to the ``physical" value $\mu_s$ ($\mu_{u/d}$), while simultaneously taking the 
continuum limit. The ``physical" values $\mu_{u/d}$ and $\mu_s$ of the quark masses are known and can be found 
in Ref.~\cite{Blossier:2010cr}. The quark bilinear RCs, $Z_P$ and $Z_S$, 
have been computed in the non-perturbative 
RI-MOM scheme in Ref.~\cite{Constantinou:2010gr}.
A RC computation for the full basis of the four-fermion operators using RI-MOM techniques is presented in 
Appendix~\ref{App_RIMOM}. 
In Appendix~\ref{App_RCS} we collect the values of  the four-fermion RCs that are used in this work,
 as well as $Z_P$ and $Z_S$. 

\begin{table}[!h]
\begin{center}
\begin{tabular}{|c|c|c|c|}
\hline
\multicolumn{4}{|c|}{\textbf{$\beta=3.80$}, $a \sim 0.10$ fm}\tabularnewline
\hline
\hline
$a\mu_{\ell}=a\mu_{sea}$ & $a^{-4}(L^{3}\times T)$ & $a\mu_{``s"}$ & $N_{stat}$\tabularnewline
\hline
0.0080 & $24^{3}\times48$ & 0.0165, 0.0200, 0.0250 & 170\tabularnewline
\hline
0.0110 & {}`` & {}`` & 180\tabularnewline
\hline
\hline
\multicolumn{4}{|c|}{$\beta=3.90$, $a \sim 0.09$ fm}\tabularnewline
\hline
\hline
0.0040 & $24^{3}\times48$ & 0.0150, 0.0220, 0.0270 & 400\tabularnewline
\hline
0.0064 & {}`` & {}`` & 200\tabularnewline
\hline
0.0085 & {}`` & {}`` & 200\tabularnewline
\hline
0.0100 & {}`` & {}`` & 160\tabularnewline
\hline
0.0030 & $32^{3}\times64$ & {}`` & 300\tabularnewline
\hline
0.0040 & {}`` & {}`` & 160\tabularnewline
\hline
\hline
\multicolumn{4}{|c|}{$\beta=4.05$, $a \sim 0.07$ fm}\tabularnewline
\hline
\hline
0.0030 & $32^{3}\times64$ & 0.0120, 0.0150, 0.0180 & 190\tabularnewline
\hline
0.0060 & {}`` & {}`` & 150\tabularnewline
\hline
0.0080 & {}`` & {}`` & 220\tabularnewline
\hline
\hline
\multicolumn{4}{|c|}{$\beta=4.20$, $a \sim 0.05$ fm}\tabularnewline
\hline
\hline
0.0020 & $48^{3}\times96$ & 0.016, 0.0129, 0.0142 & 96\tabularnewline
\hline
0.0065 & $32^{3}\times64$ & {}`` & 144\tabularnewline
\hline
\end{tabular}
\caption{Details of simulation runs at $\beta=$3.80, 3.90, 4.05 and 4.20.}
\label{runs}
\end{center}
\end{table}

\subsection{Extracting bare estimates from lattice data}
\label{sec:ERLD}

Bare results for the ratio of the four-fermion  matrix elements $R_i^{(b)}$ (c.f. Eq.~(\ref{OioverO1}))
with $i=2, \ldots, 5$ at the four $\beta$ 
values and combinations of quark masses are listed in Tables~\ref{Rbare_b380} - \ref{Rbare_b420}
of Appendix~\ref{APP_RESULTS}. In   Tables~\ref{Bbare_b380} - \ref{Bbare_b420}
of the same Appendix we collect the results for the bare quantities $\xi_i\, B_i^{(b)}$ ($i=2, \ldots, 5$) 
at each value of $\beta$. 

\begin{figure}[!ht]
\subfigure[]{\includegraphics[scale=0.50,angle=-0]{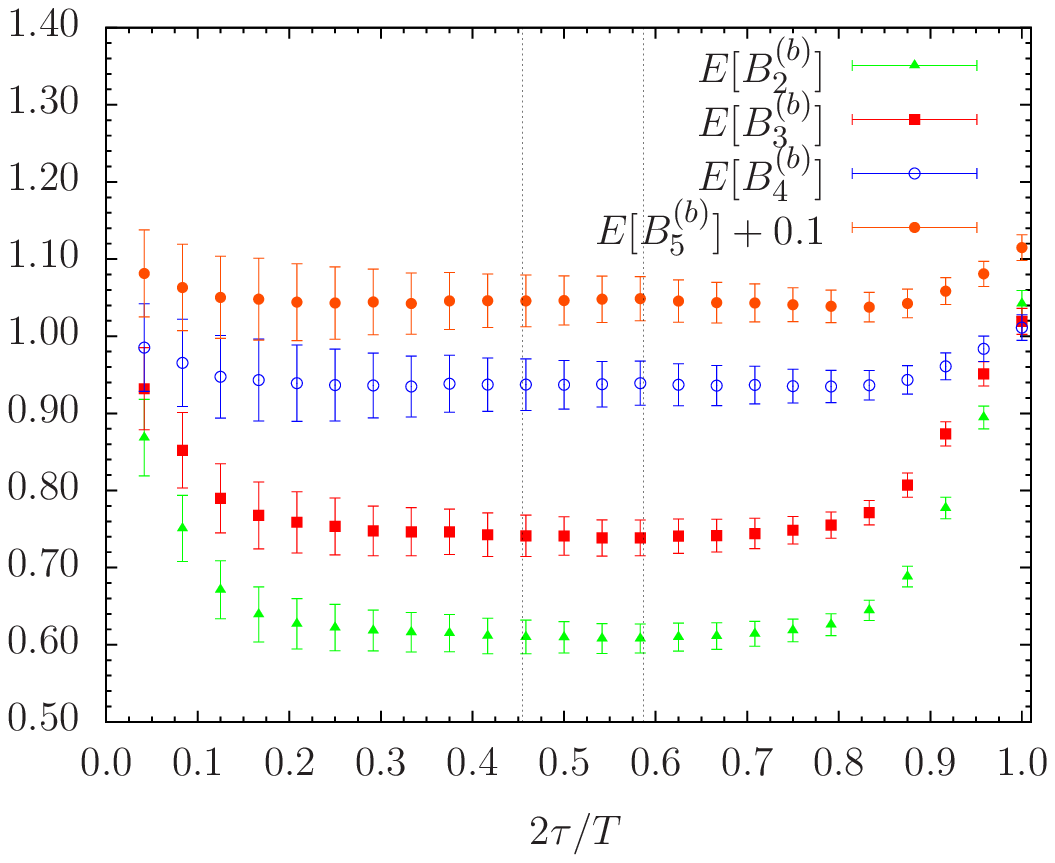}}
\subfigure[]{\includegraphics[scale=0.50,angle=-0]{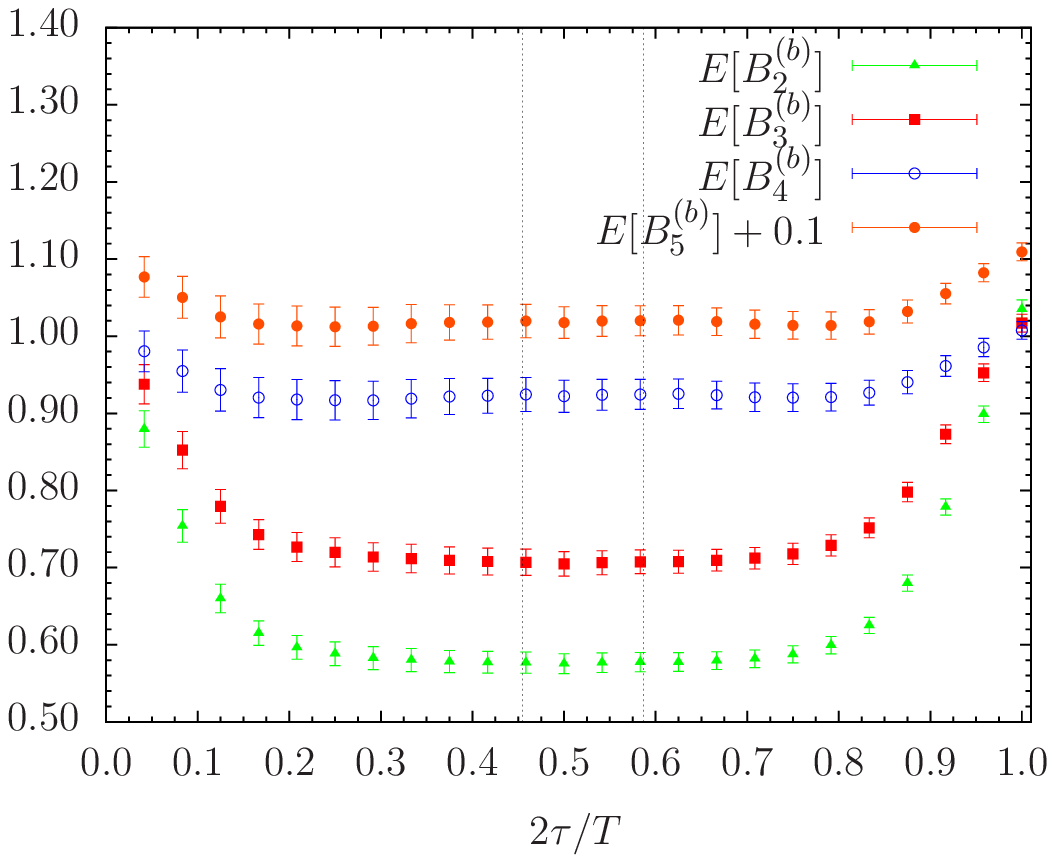}}
\subfigure[]{\includegraphics[scale=0.50,angle=-0]{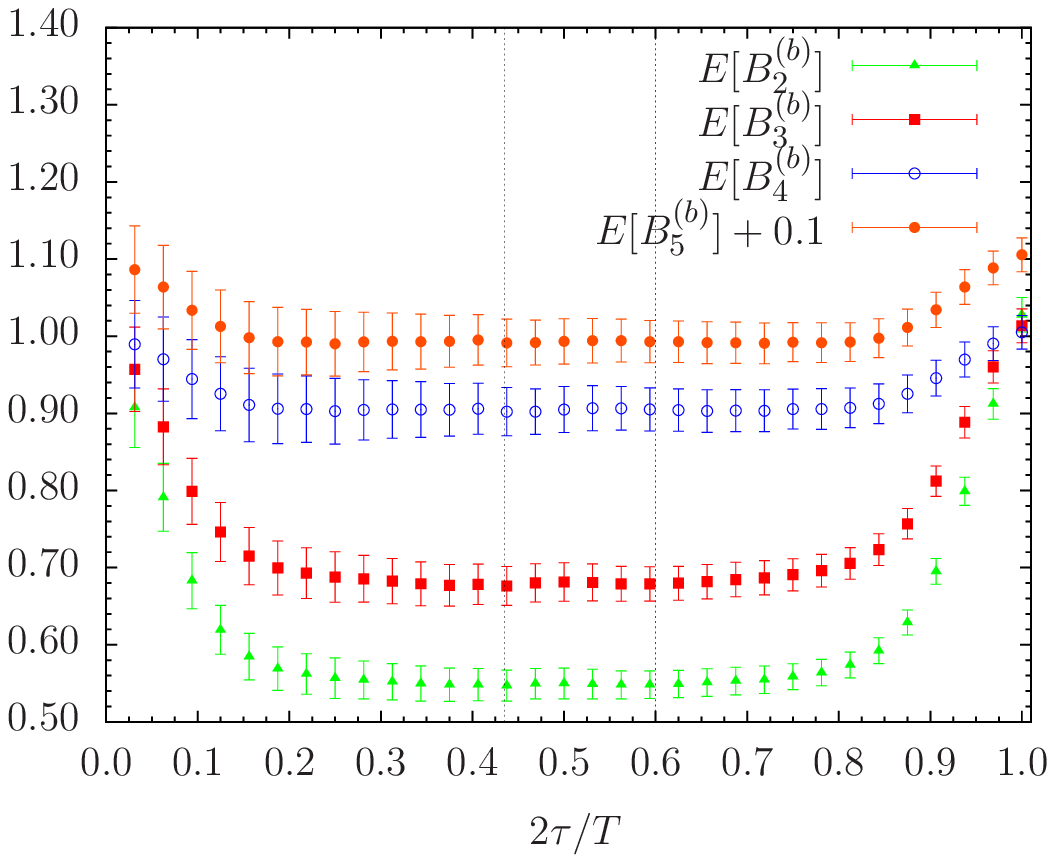}}
\hspace*{3.2cm} \subfigure[]{\includegraphics[scale=0.50,angle=-0]{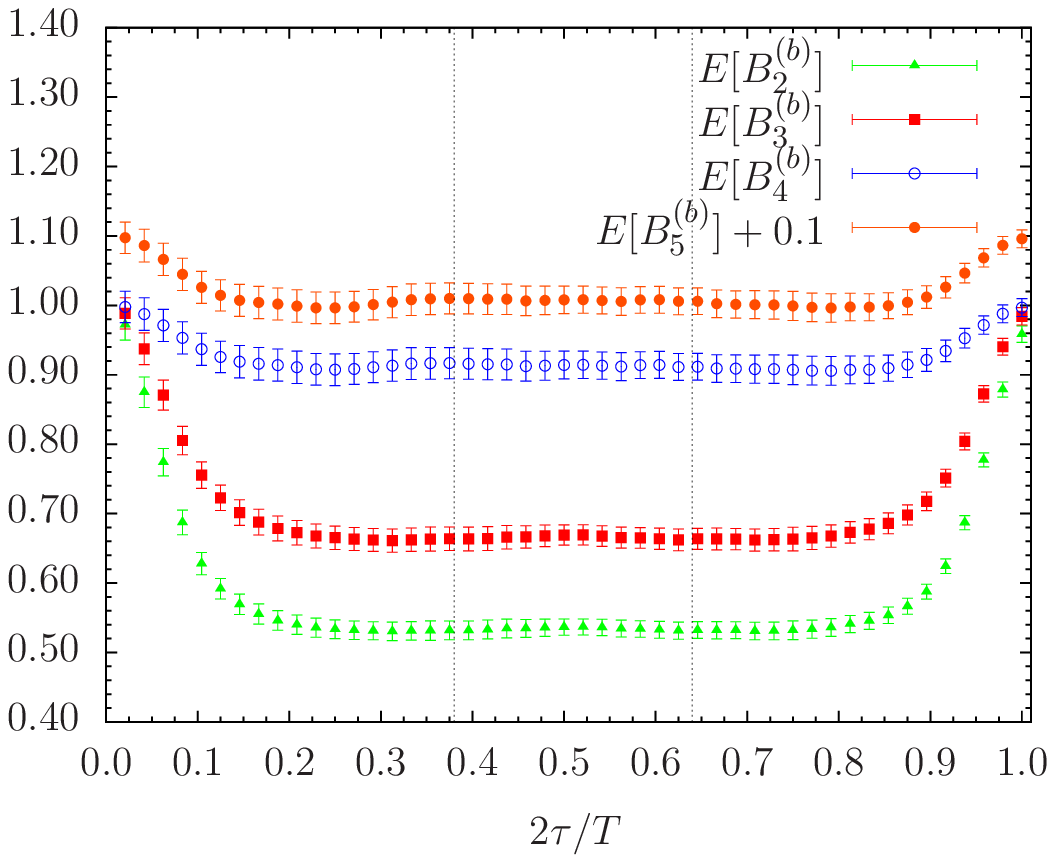}}
\vskip -0.5cm
\begin{center}
\caption{\sl Data and time-plateaux for $E[B_i^{(b)}](\tau)$ ($i=2, \dots, 5$) 
(see Eq.~(\ref{Biratio})) plotted vs. $2 \tau/T \equiv 2(x_0-y_0)/T$. 
In panel (a) we show data for $\beta = 3.80$ and $(a\mu_\ell, a\mu_{``s"}) = (0.0080, 0.0165)$  
on a $24^3 \times 48$ lattice; in panel (b)
for $\beta = 3.90$, $(a\mu_\ell, a\mu_{``s"}) = (0.0040, 0.0150)$ on a 
$24^3 \times 48$ lattice; in panel (c) for $\beta = 4.05$ and $(a\mu_\ell, a\mu_{``s"}) = (0.0030, 0.0120)$ 
on a $32^3 \times 64$ lattice; 
in panel (d) for $\beta = 4.20$ and $(a\mu_\ell, a\mu_{``s"}) = (0.0020, 0.0129)$ on a $48^3 \times 96$ lattice. 
Vertical dotted lines delimit the plateau region. For clarity data for $E[B_5^{(b)}](\tau)$ have been slightly shifted. 
}
\label{fig:RBi}
\end{center}
\end{figure}

\begin{figure}[!ht]
\subfigure[]{\includegraphics[scale=0.50,angle=-0]{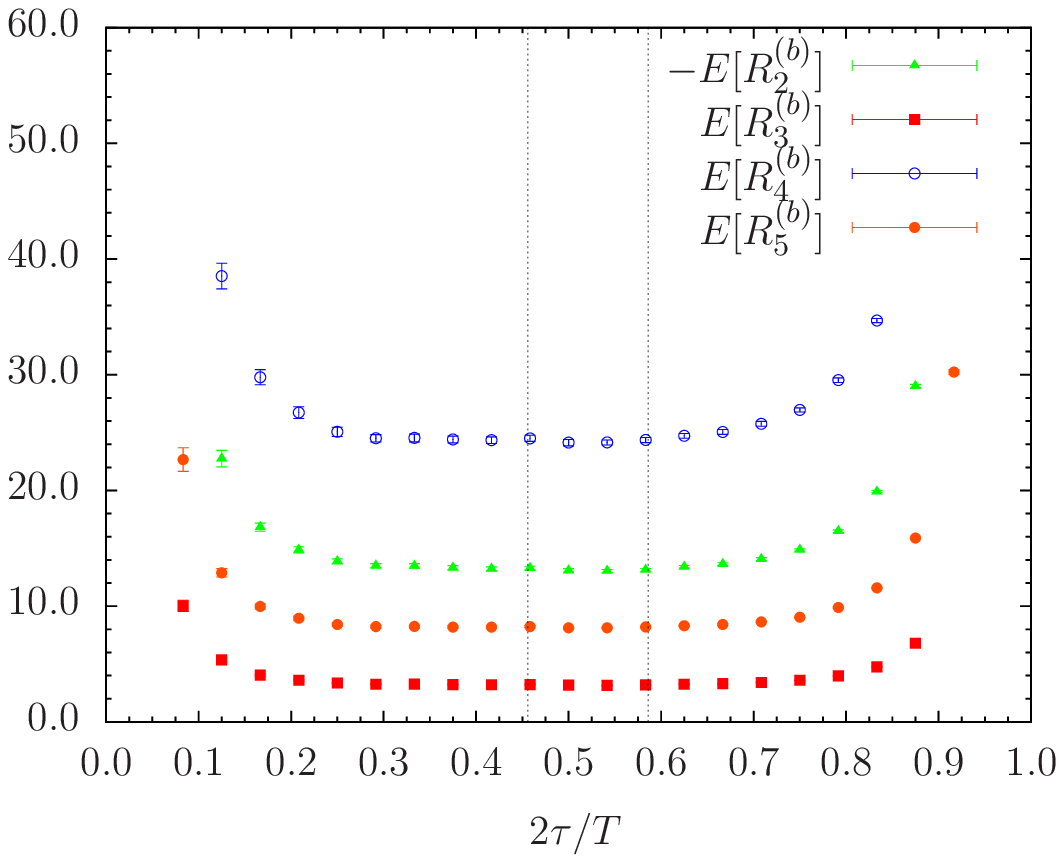}}
\subfigure[]{\includegraphics[scale=0.50,angle=-0]{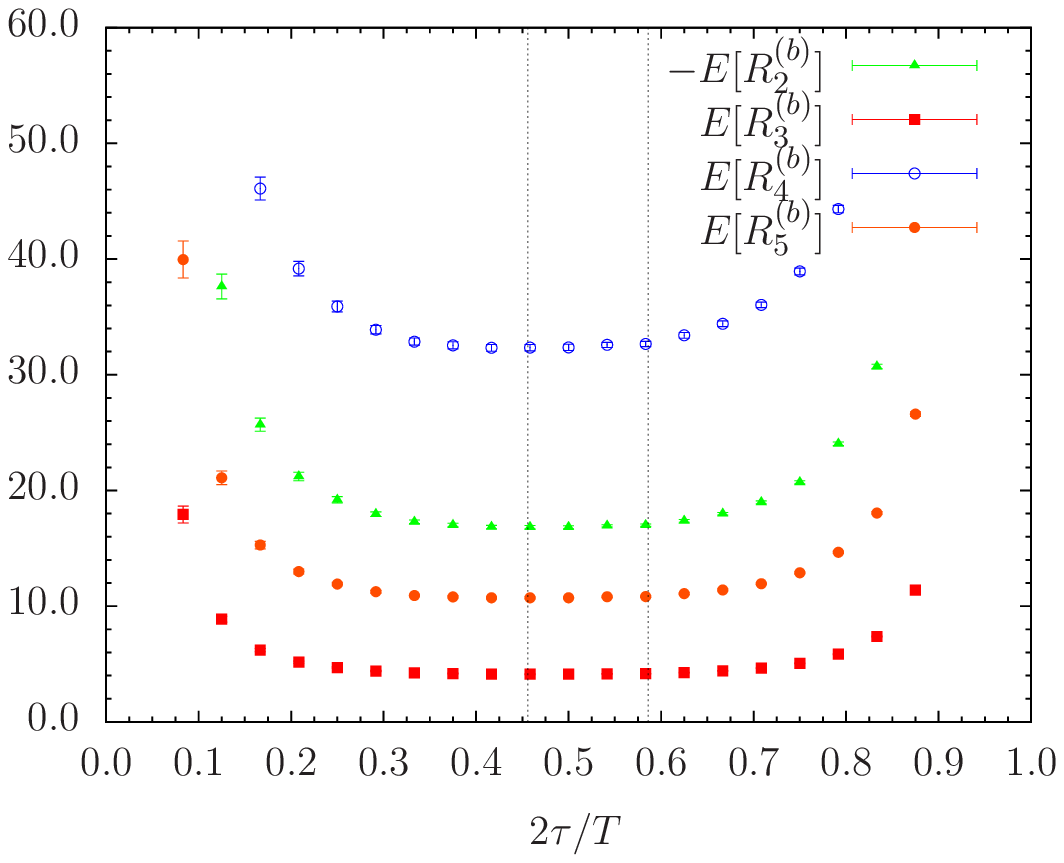}}
\subfigure[]{\includegraphics[scale=0.50,angle=-0]{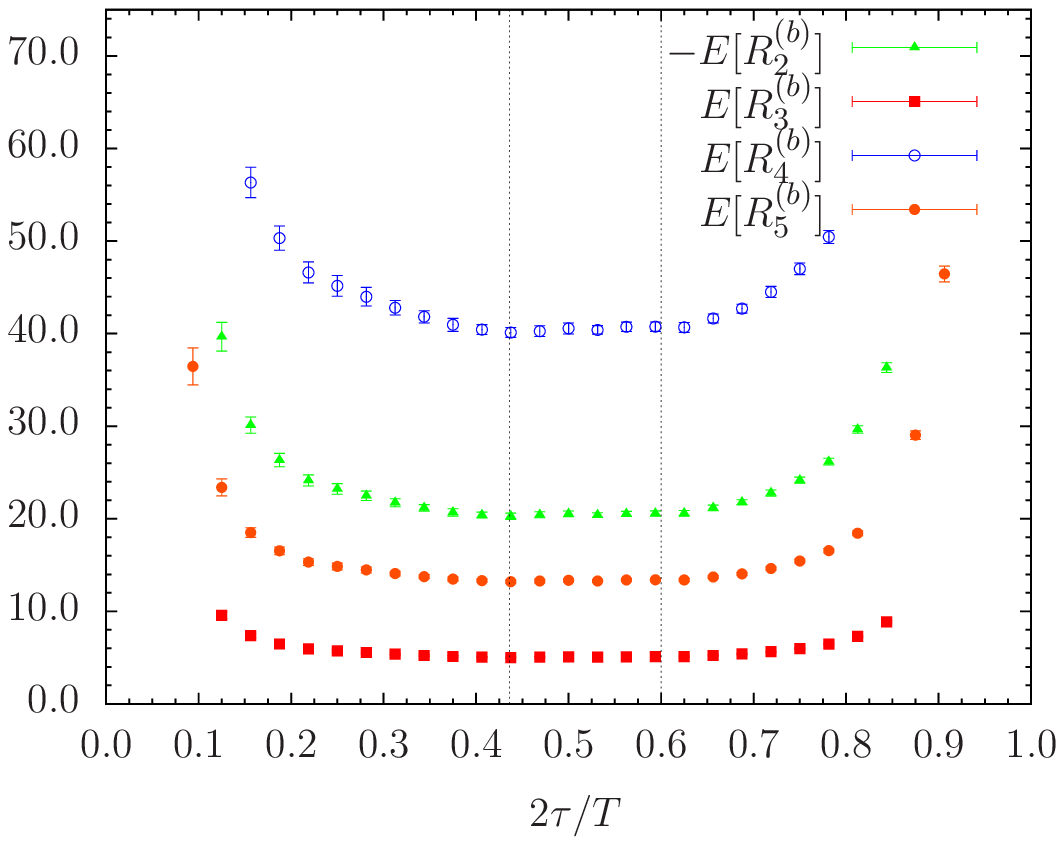}}
\hspace*{3.0cm} \subfigure[]{\includegraphics[scale=0.50,angle=-0]{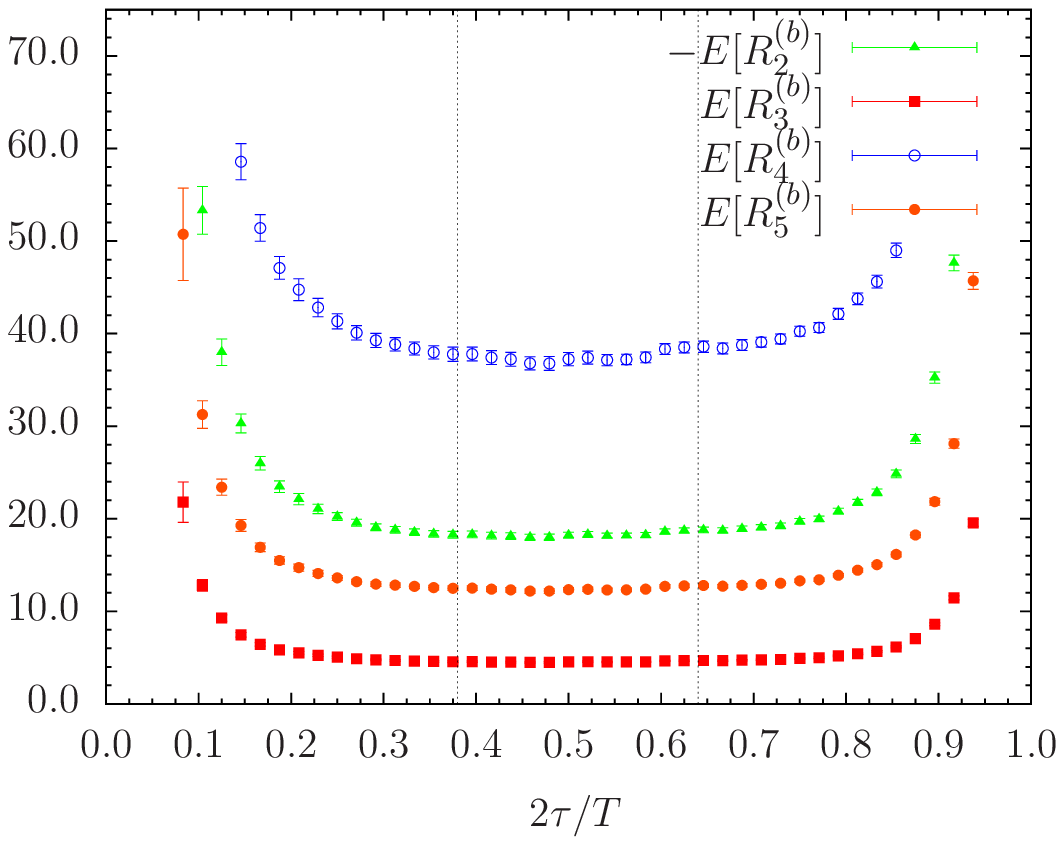}}
\vskip -0.5cm
\begin{center}
\caption{\sl Data and time-plateaux for $E[R_i^{(b)}](\tau)$ ($i=2, \dots, 5$) 
(see Eq.~(\ref{ROiratio}))  plotted vs. $2 \tau/T \equiv 2(x_0-y_0)/T$. 
In panel (a) we show data for $\beta = 3.80$ and $(a\mu_\ell, a\mu_{``s"}) = (0.0080, 0.0165)$  
on a $24^3 \times 48$ lattice; in panel (b)
for $\beta = 3.90$, $(a\mu_\ell, a\mu_{``s"}) = (0.0040, 0.0150)$ on a 
$24^3 \times 48$ lattice; in panel (c) for $\beta = 4.05$ and $(a\mu_\ell, a\mu_{``s"}) = (0.0030, 0.0120)$ 
on a $32^3 \times 64$ lattice; 
in panel (d) for $\beta = 4.20$ and $(a\mu_\ell, a\mu_{``s"}) = (0.0020, 0.0129)$ on a $48^3 \times 96$ lattice. 
Vertical dotted lines delimit the plateau region. 
} 
\label{fig:ROi}
\end{center}
\end{figure}

For illustration in Fig.~\ref{fig:RBi} we display some examples of the $B_i^{(b)}$  
plateau quality at $\beta=$3.8, 3.9, 4.05 and 4.20. 
Vertical dotted lines indicate the plateau region where the $K^0$- and the $\bar K^0$-state dominate the 
three-point correlators. Similar examples of the plateau quality for the case of the four-fermion operator ratios are 
illustrated in Fig.~\ref{fig:ROi}. Both Figs.~\ref{fig:RBi} and~\ref{fig:ROi} display very good signals.

\vskip 0.0cm

\vskip 0.0cm

\subsection{Computation at the physical point}
\label{sec:CBHAT}

Extracting physical quantities from lattice data requires performing  extrapolations and/or interpolations of 
renormalized lattice estimators to the physical point (continuum limit and ``physical" value of quark masses).

\subsubsection{RCs computation and combined continuum-chiral extrapolation}
\label{sec:Z4f}

We have computed the full matrix of the four-fermion operator RCs in a mass independent scheme. 
We carry out the non-perturbative calculation adopting the RI-MOM approach. The implementation of the RI-MOM setup 
has been presented in Refs.~\cite{Constantinou:2010qv} and~\cite{Constantinou:2010gr}. 
We should 
mention that in our RC estimators cutoff effects, though parametrically of O($a^2$), are numerically reduced 
owing to the subtraction of perturbatively evaluated   O($a^2g^2$) contributions. 
After that, two different, but by now standard~\cite{Constantinou:2010gr}, procedures are employed to deal with 
O($a^2p^2$) discretization effects. The first, called M1, consists in linearly extrapolating to zero the residual 
(after the perturbative subtraction) O$(a^2p^2$) terms. The second one (so-called $p^2$-window method, or M2 for short) 
leads to RC estimates obtained by averaging data over a fixed (in physical units) and very narrow momentum interval.

We carry out continuum and chiral extrapolations in a combined way. For all bag parameters, $B_i$, and ratios, 
$\tilde{R}_i$ (see Eq.~(\ref{REN_Riratio})), 
we have tried out a fit ansatz of the following general form
\begin{equation}
Y = \sum_{n=0}^{2} A_Y^{(n)}(r_0\hat{\mu}_{s})\, [r_0\,\hat{\mu}_{\ell}]^{n} + D_Y(r_0\hat{\mu}_{s})\, 
\LSB \dfrac{a}{r_0}\RSB^2 \, ,
\label{Yfit}
\end{equation} 
where we have made explicit the dependence of the fit parameters  $A_Y^{(n)}$ and $D_Y$ on the renormalized strange 
quark mass\footnote{We use the symbol ( $\hat{} $ ) to denote renormalized quark masses 
in the $\overline{\rm{MS}}$ scheme at  $2\, \rm{GeV}$.} in units of $r_0$ ($r_0\hat{\mu}_s$). 
We studied separately the cases of linear and polynomial ansatz. 
We have also considered NLO ChPT fit functions for $B_i$ based on the formulae given 
in~\cite{Becirevic:2004qd} in the case of SU(3). Those formulae transformed to NLO SU(2) ChPT read:
\begin{equation}
B_i = B_i^{\chi}(r_0\hat{\mu}_{s}) \left[ 1 + b_i(r_0\hat{\mu}_{s}) \mp 
\frac{2\hat B_0 \hat\mu_{\ell}}{2(4\pi f_0)^2}\log\frac{2\hat B_0\hat\mu_{\ell}}
{(4 \pi f_0)^2}  \right] + D_{Bi}^{'}(r_0\hat{\mu}_{s}) \LSB \dfrac{a}{r_0}\RSB^2 
\label{ChPTBi}
\end{equation}  
with $\hat B_0=2.84(11)$ GeV (renormalized in $\overline{\rm{MS}}$ at 2 GeV) 
and $f_0=121.0(1)$ MeV, as we used in \cite{Constantinou:2010qv}.
The sign before the logarithmic term is minus (-) for $i=1,2,3$ and plus (+) for $i=4,5$.
As for $\tilde{R}_i$ and $i=2,3$ the ChPT fit formula at NLO coincides with the linear fit ansatz, while for the 
cases $i=4,5$ we use
\begin{equation}
\tilde R_i = \tilde R_i^{\chi}(r_0\hat{\mu}_{s}) \left[ 1 + c_i(r_0\hat{\mu}_{s}) + 
\frac{2 \hat B_0 \hat\mu_{\ell}}{(4\pi f_0)^2}\log\frac{2\hat B_0\hat\mu_{\ell}}
{(4 \pi f_0)^2}  \right] + D_{Ri}^{'}(r_0\hat{\mu}_{s}) \LSB \dfrac{a}{r_0}\RSB^2 
\label{ChPTRi}
\end{equation}
The $(r_0/a)$ values are
\begin{equation}
\frac{r_0}{a}{\Big |}_\beta = \{4.54(7), \,\, 5.35(4), \,\, 6.71(4), \,\, 8.36(6)\} 
\end{equation}
at $\beta=\{3.80, \, 3.90, \, 4.05, \, 4.20\}$ respectively. \\
The $u/d$ and $s$ quark masses have been computed in Ref.~\cite{Blossier:2010cr}.  
Their values in the $\overline{\rm{MS}}$ scheme at 2~GeV are
\begin{equation}
\mu_{u/d}^{\overline{\rm{MS}}} (2\, \rm{GeV})\, = \, 3.6(2)\, ~\rm{MeV}, ~~~~~  
\mu_{s}^{\overline{\rm{MS}}} (2\, \rm{GeV})\, = \, 95(6)\, ~\rm{MeV} 
\label{uds_mass}
\end{equation} 

\begin{figure}[!ht]
\subfigure[]{\includegraphics[scale=0.55,angle=-0]{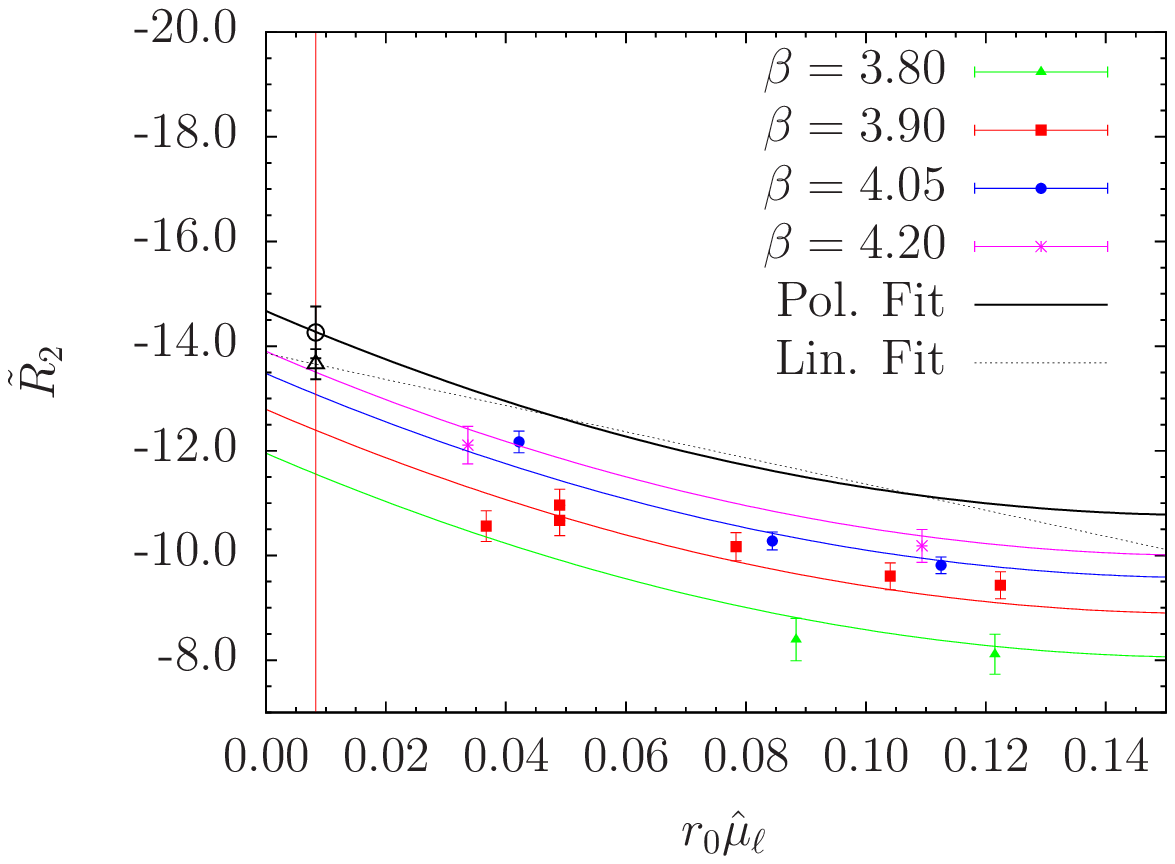}}
\subfigure[]{\includegraphics[scale=0.55,angle=-0]{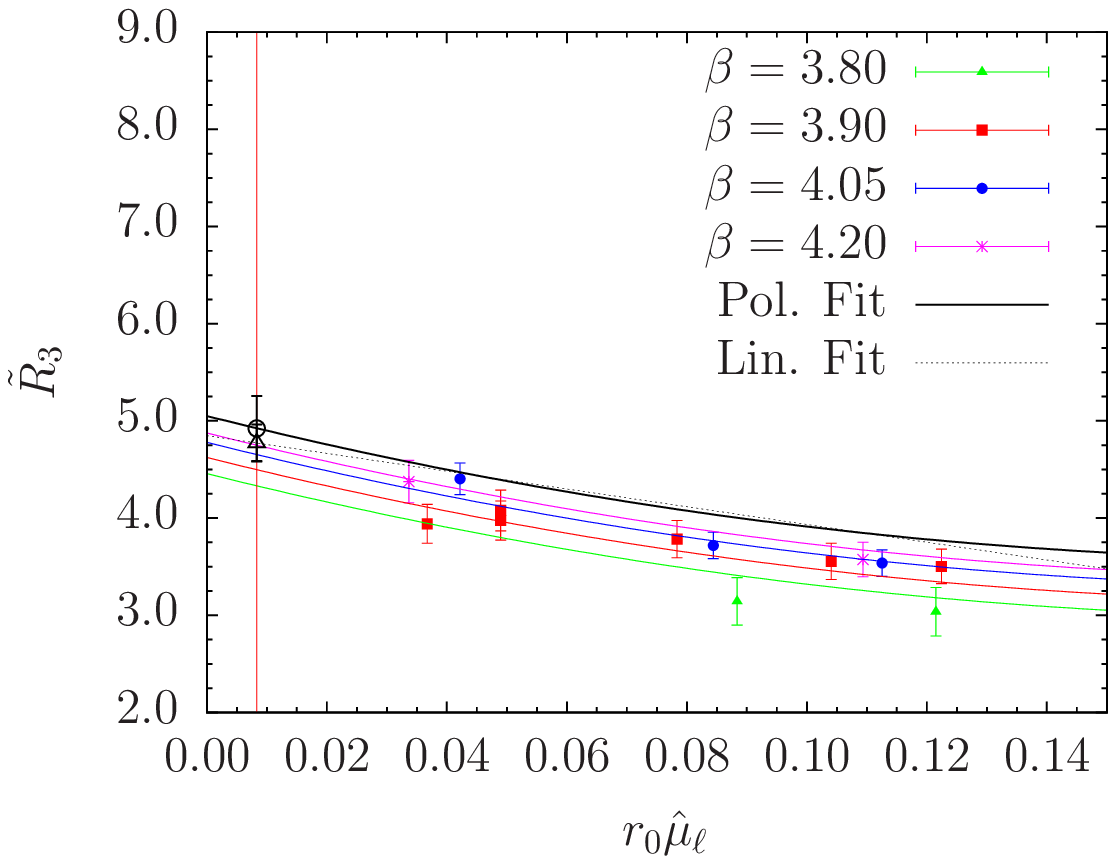}}
\hspace*{0.1cm} \subfigure[]{\includegraphics[scale=0.55,angle=-0]{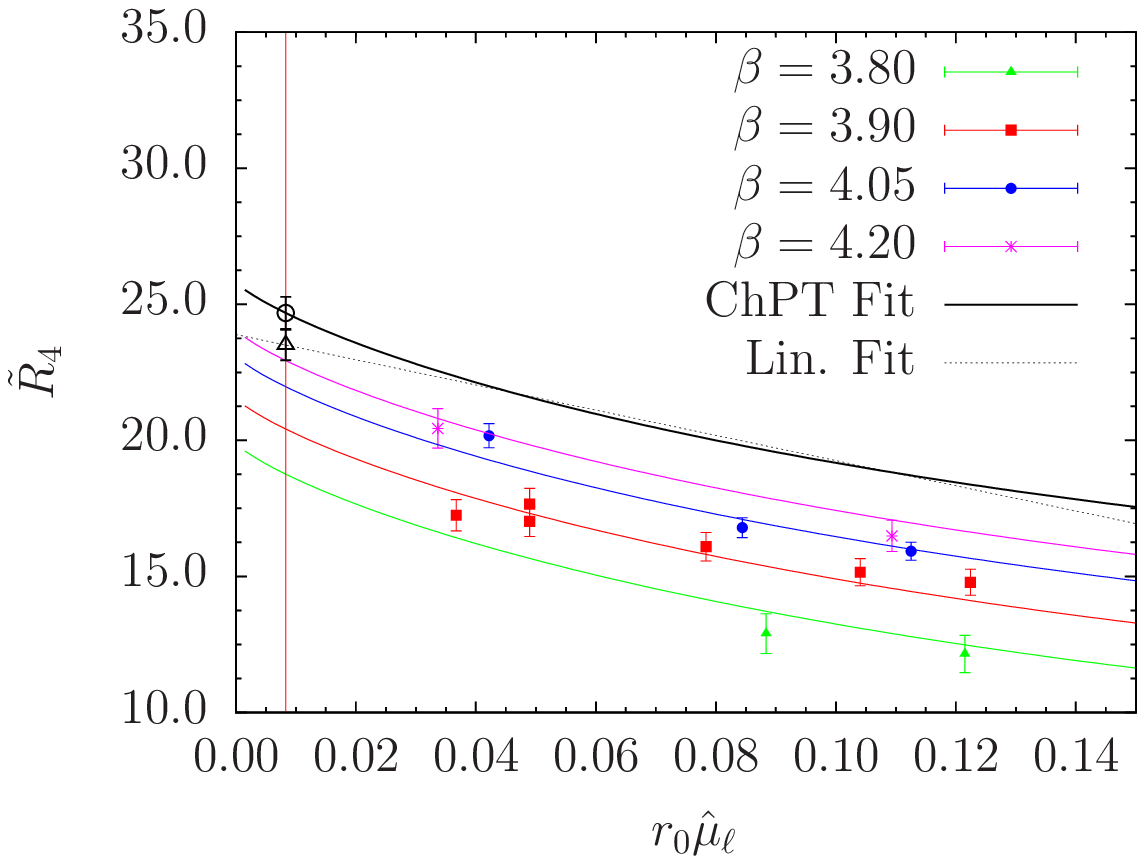}}
\hspace*{1.cm} \subfigure[]{\includegraphics[scale=0.55,angle=-0]{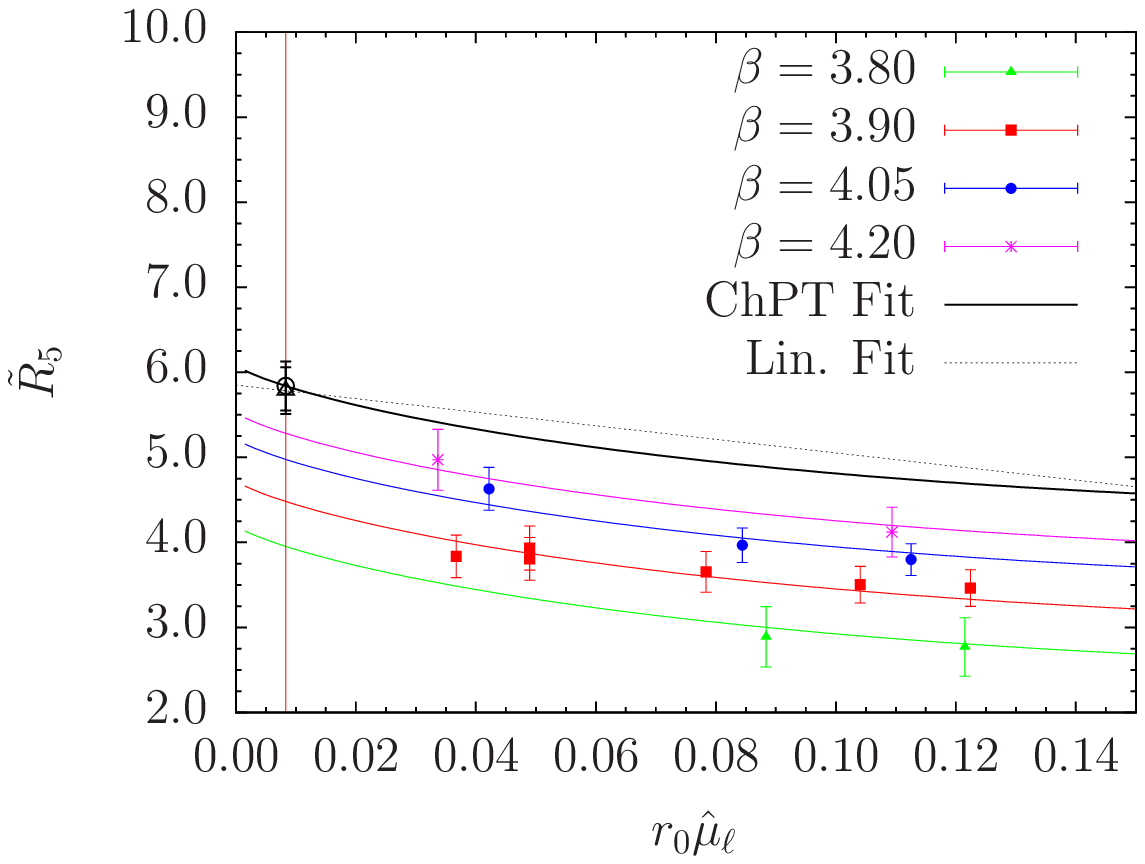}}
\vskip -0.5cm
\begin{center}
\caption{\sl Solid lines in panels (a) and (b) show the behaviour vs.~the renormalized light quark mass of the 
combined chiral and continuum fits (according to the polynomial formula~(\ref{Yfit}) with $n=2$) of the $\tilde{R}_i$ ratios, 
with $i=2$ and $i=3$  respectively, renormalized in the $\overline{\rm{MS}}$ scheme of 
Ref.~\cite{mu:4ferm-nlo} at 2~GeV with the M1-type RCs. The full black line is the continuum limit curve.
In panels (c) and (d), solid lines, instead, show the combined chiral and continuum fit described by NLO-ChPT, 
Eq.~(\ref{ChPTRi}) for $i=4$ and $i=5$, respectively. The full black line is the continuum limit curve. 
The dashed black line represents the continuum limit curve in 
the case of the linear fit ansatz. Black open circles and triangles stand for the results at the physical point 
corresponding to the  polynomial (panels (a) and (b)) and ChPT fit (panels (c) and (d)),   
and linear fit ansatz, respectively.  
}
\label{fig:Rivsmul}
\end{center}
\end{figure}

\begin{figure}[!ht]
\subfigure[]{\includegraphics[scale=0.55,angle=-0]{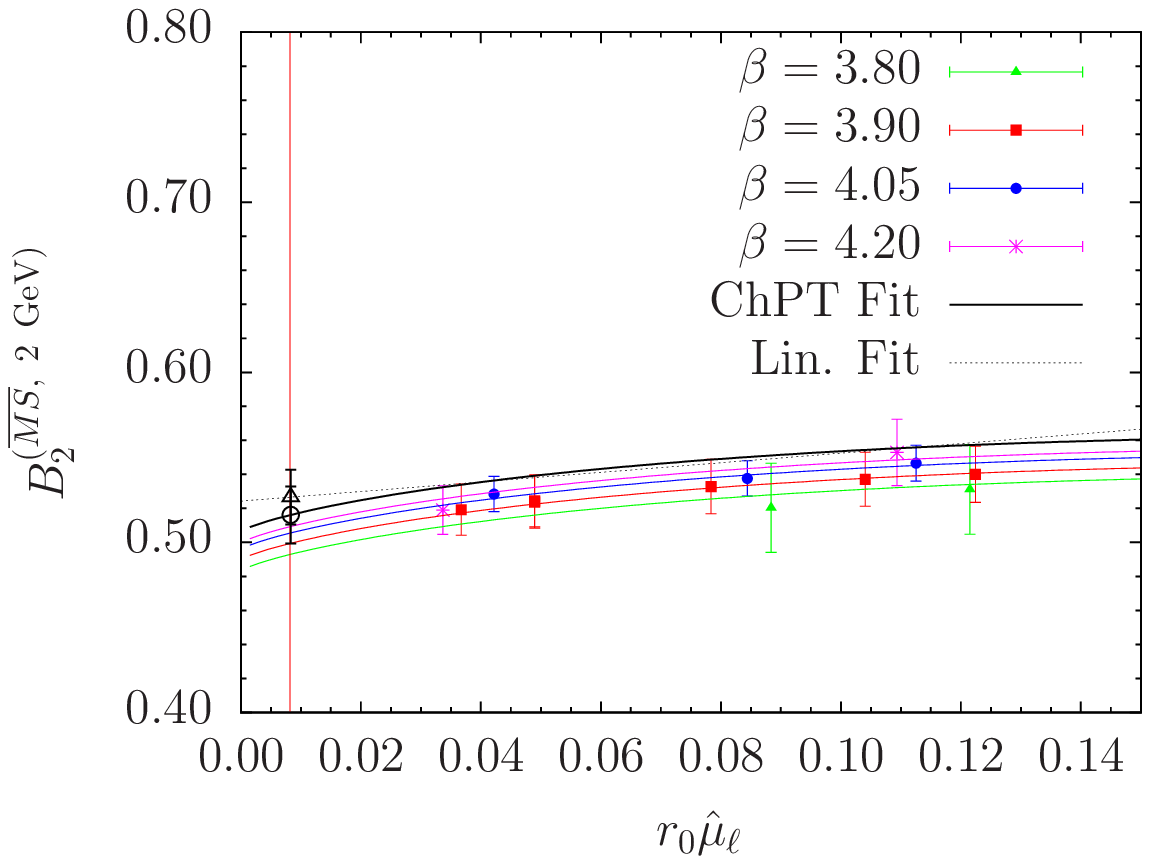}}
\subfigure[]{\includegraphics[scale=0.55,angle=-0]{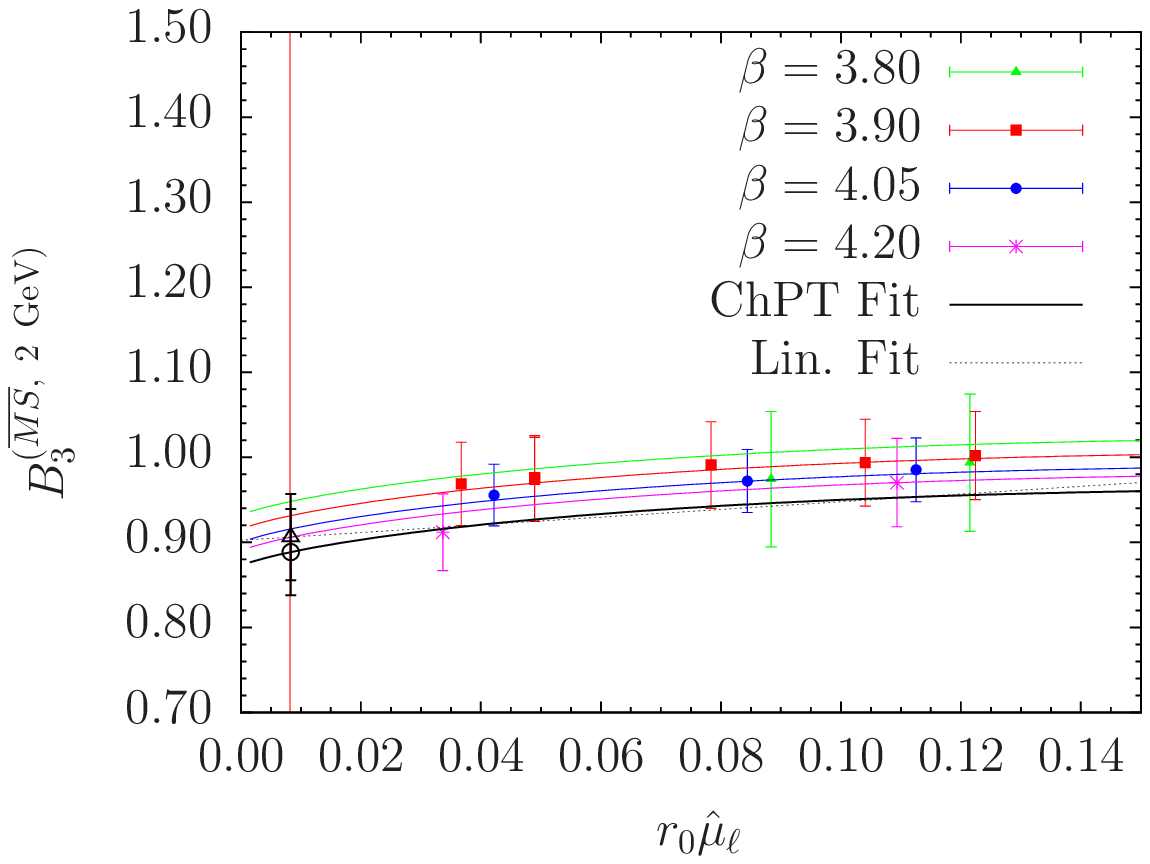}}
\hspace*{0.1cm} \subfigure[]{\includegraphics[scale=0.55,angle=-0]{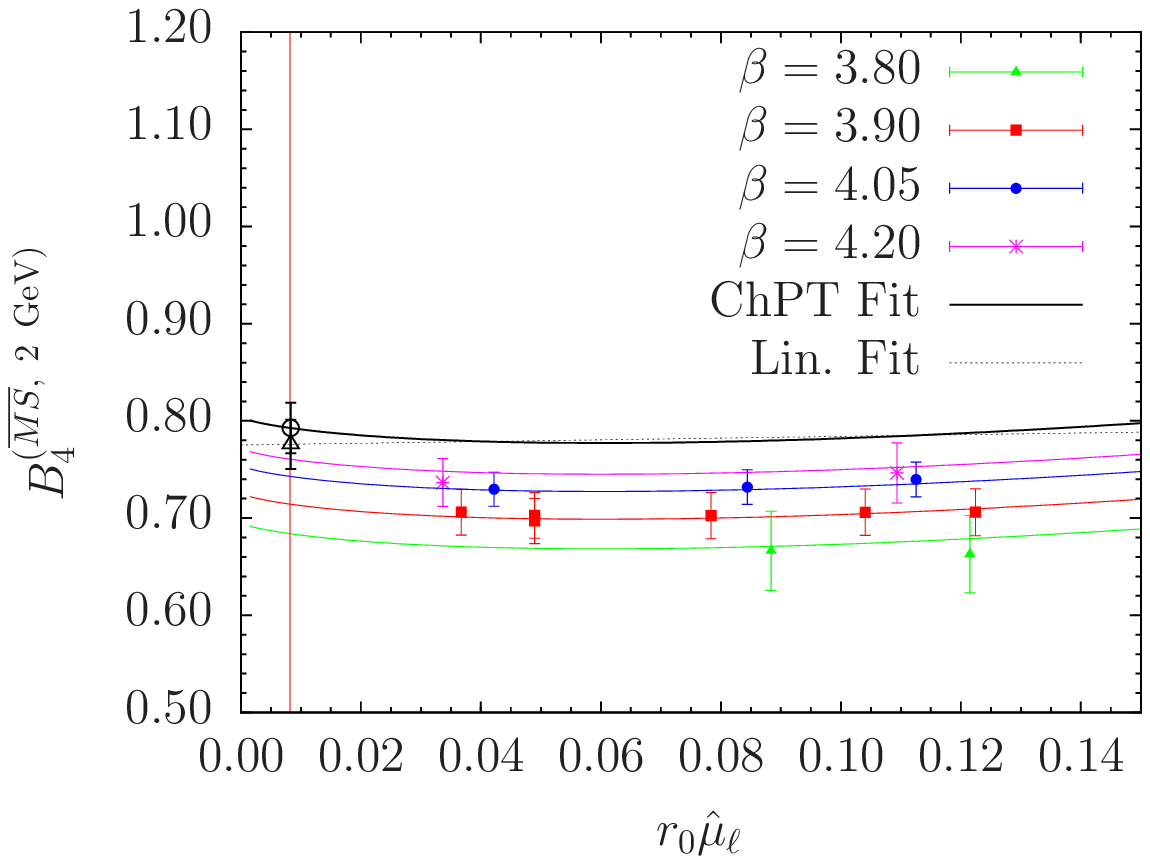}}
\hspace*{1.cm} \subfigure[]{\includegraphics[scale=0.55,angle=-0]{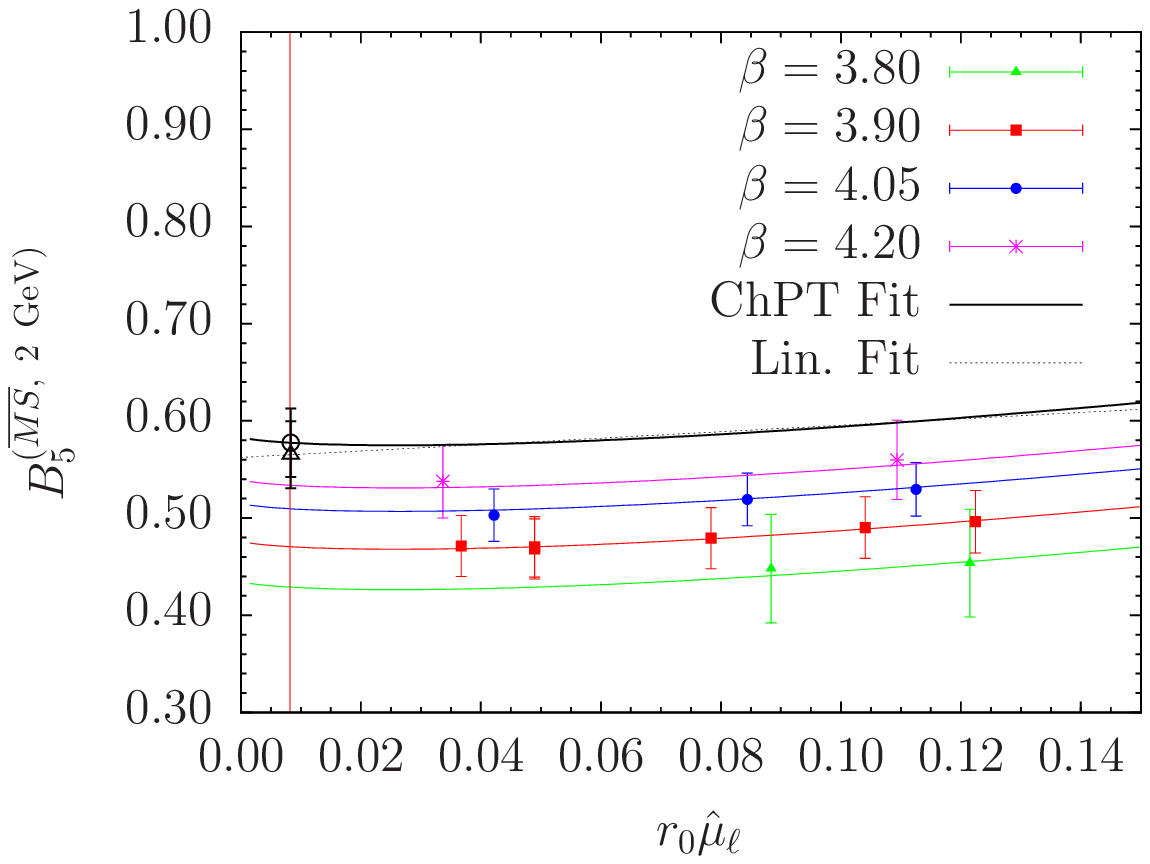}}
\vskip -0.5cm
\begin{center}
\caption{\sl Solid lines in panels (a) to (d) show the behaviour vs.\ the renormalized light quark mass of the
combined chiral and continuum fits (according to the ChPT fit formula ~(\ref{ChPTBi})) for the $B_i$ parameters
with $i=2,\ldots,5$ respectively, renormalized in the $\overline{\rm{MS}}$ scheme of Ref.~\cite{mu:4ferm-nlo} 
at 2~GeV with the M1-type RCs.
The full black line is the continuum limit curve~(\ref{Yfit}).
The dashed black line represents the continuum limit curve in
the case of the linear fit ansatz. Black open circles and triangles stand for the results at the physical point
corresponding to the  ChPT fit and linear fit ansatz, respectively. 
}
\label{fig:Bivsmul}
\end{center}
\end{figure} 

In the four panels of Fig.~\ref{fig:Rivsmul} we show the combined chiral and continuum 
fit (see Eq.~(\ref{Yfit}) and Eq.~(\ref{ChPTRi})) for the  ratios $\tilde{R}_i$  against the renormalized 
light quark mass for $i=2,\ldots,5$, 
respectively. The RCs used in these plots are the ones computed with the M1-method and are expressed 
in the $\overline{\rm{MS}}$ 
scheme of Ref.~\cite{mu:4ferm-nlo} at 2 GeV. Lattice data correspond to points taken at the pair of quark masses 
$(r_0\hat{\mu}_\ell, r_0\hat{\mu}_s)$.

In panels (a) and (b) (corresponding to cases with $i=2,3$ respectively) 
we display the curves that correspond to the polynomial fit function~(\ref{Yfit}) 
at the four $\beta$ values we are considering in this paper. The black solid line represents the continuum limit curve. 
The dashed black line represents the continuum limit curve that is obtained if a linear fit ansatz in 
$\hat{\mu}_{\ell}$ is used.
Black open circles and triangles stand for the results at the physical quark mass point from the polynomial 
and the linear fit ansatz, respectively. Recall that for $\tilde R_i$  with $i=2,3$ ChPT fit formula at NLO 
coincides with a linear fit ansatz. 
In panels (c) and (d) (corresponding to cases with $i=4,5$ respectively) we display the curves 
corresponding to the ChPT fit formula and the linear fit function. 
In this case black open circles and triangles stand for the results at the physical quark mass point from the ChPT fit  
and the linear fit ansatz, respectively.  

Similarly, in Fig.~\ref{fig:Bivsmul} we present the combined chiral and 
continuum fit for the $B_i$-parameters, again renormalized in the $\overline{\rm{MS}}$ scheme of Ref.~\cite{mu:4ferm-nlo} at 2~GeV.
In all four panels we display  the curves 
corresponding to the ChPT fit formula at the four $\beta$ values. The black solid line 
represents the continuum limit curve. 
The dashed black line represents the continuum limit curve that is obtained if a linear fit ansatz in 
$\hat{\mu}_{\ell}$ is used.

Note the nice agreement (within one standard deviation) of the two fit ans\"{a}tze  
for both the $\tilde{R}_i$ ratios and the $B_i$ parameters. 
    
In Tables~\ref{results_MSbar_intro} and~\ref{results_RIMOM_intro} of Section~\ref{sec:introII}   
we have gathered our final continuum results 
for $R_i$ and $B_i$ in the $\overline{\rm{MS}}$ of Ref.~\cite{mu:4ferm-nlo} and RI-MOM scheme at 2~GeV respectively. 
The final value of $B_i$ for $i=2, \ldots, 5$ has been computed by averaging the estimates obtained from the 
three kinds of fit ansatz discussed above, and using bootstrap error analysis. 
The half difference between the two more distant results has been taken as an estimate of the systematic error 
associated to the  extrapolation procedure. The total uncertainty is obtained by adding in quadrature 
the statistical and the systematic error. 
For $i=1$  we update the result 
for $B_K$ published in Ref.~\cite{Constantinou:2010qv}; note that the difference between the two 
results is about half standard deviation.

In Appendix~\ref{App_CL} we provide more detailed results obtained from the various fitting procedure 
we have investigated. We also show that 
the continuum extrapolated quantities that are eventually obtained by employing M1-type and M2-type RCs 
turn out to be perfectly consistent between each other within statistical errors.
Also in Appendix~\ref{App_CL},  see Tables~\ref{results_MSbar_3GeV} and~\ref{results_RIMOM_3GeV}, we quote 
our continuum results for $B_i$ and $R_i$
in the $\overline{\rm{MS}}$ and  the RI-MOM scheme respectively at 3~GeV.

As already stated, an alternative (indirect) way to compute the ratio of the kaon  
matrix elements of the renormalized operators 
$O_i$, $i=2,\ldots,5$ to that of $O_1$ is based on the formula (see eqs.~(\ref{B1}) and~(\ref{Bi}))
\begin{equation}
\dfrac{\langle \bar{K}^{0} |  O_i(\mu) | K^{0} \rangle }{\langle \bar{K}^{0} | O_1(\mu) | K^{0} \rangle } = 
\dfrac{\xi_i B_i(\mu)}{\xi_1 B_1(\mu)} \dfrac{m_K^2}{(\hat{\mu}_s(\mu) + \hat{\mu}_d(\mu))^2}\, .
\label{ratio_indirect} 
\end{equation}
This of course requires  knowledge of the $B_i$ parameters and the renormalized quark masses.

We find that the two evaluations (indirect and direct, based on Eq.~(\ref{ratio_indirect}) 
and Eq.~(\ref{REN_Riratio}), respectively) 
lead to compatible results within  errors. However, the indirect estimates suffer from much larger final 
uncertainties. This is due to several reasons. 
One is related to the quadratic dependence on the quark mass, which makes the relative error on the mass to give a 
significant contribution to 
the final error. Furthermore in the indirect method one has to consider extra uncertainties due to the error 
of the bilinear operators' RCs that are used to compute the $B$-parameters. 
A comparison of the direct and indirect results obtained for the ratios $R_i$ is provided in Appendix~\ref{App_CL}.

\newpage
\section{Conclusions}
\label{sec:concl}

 Accurate measurements of the $K^0-\bar K^0$ mixing amplitudes
can yield useful hints on New Physics 
if theory can provide comparatively accurate calculations of quantities parametrizing beyond the SM effects. 
This requires a precise, first principle, 
evaluation of the kaon matrix elements of the full basis of four-fermion operators 
entering the most general effective 
$\Delta S=2$ weak Hamiltonian. 

In this paper we have  presented the first unquenched lattice QCD determination in the continuum limit    
of  the matrix elements of the 
full $\Delta S=2$ four-fermion operator basis. We have used $N_f=2$ unquenched tm-LQCD gauge configurations produced by the
ETM Collaboration in combination  with maximally twisted valence quarks of the OS type. 

The mixed action setup proposed in Ref.~\cite{Frezzotti:2004wz} offers the possibility of 
obtaining automatically O($a$) 
improved results and an operator renormalization pattern identical to that of a chirally invariant 
regularization at the rather cheap price of mere O($a^2$) unitarity violations. 
Using data at four lattice spacings (with $a$ in the interval [0.05, 0.1]~fm) and a number of
 pseudoscalar masses 
(``pions") in 
the range [280, 500]~MeV, we are able to safely carry out the continuum and the light quark mass 
limit of the observables 
of interest. All results are non-perturbatively renormalized in the RI/MOM scheme. 

We get in this way the most accurate estimates to date of $\Delta S=2$ effective weak Hamiltonian matrix elements. 
The total error on the $R_i$ ratios is between $4\%$ and $6\%$ and on the bag parameters, $B_i$, 
between $3\%$ and $7\%$.

Tables~\ref{comparison_Bi_RIMOM} and~\ref{comparison_Ri_RIMOM} show a comparison between 
our results for $R_i$ and $B_i$ (in RI/MOM at 2~GeV) and the data at fixed lattice spacings coming 
from the two old 
quenched calculations of Refs.~\cite{Donini:1999nn} and~\cite{Babich:2006bh}~\footnote{In this 
comparison we do not 
include the (preliminary) quenched results at one value of the lattice spacing given in 
Ref.~\cite{Nakamura:2006eq}.}.    

For the $B$-parameters, one finds large differences between the central values of our results 
and those of Refs.~\cite{Donini:1999nn} and
\cite{Babich:2006bh}, which vary between 5\% and 25\% (though the errors are typically comparably large). 
With respect to Ref.~\cite{Donini:1999nn}, the differences are even larger when the results are compared 
in terms of the ratios $R_i$, presumably due to a combined effect, in this case, of having overestimated 
the values for both $B_1$ and the strange quark mass in the computation of~\cite{Donini:1999nn}. 
We emphasize that, with respect to the old quenched calculations, having performed in the present 
study simulations at four values of the lattice spacing and quite smaller values of the pion masses 
provides us with a much better control over the main sources of systematic uncertainties, besides the 
quenched approximation. 
Current experience suggests that the possible systematic errors related to 
the quenching of the strange and charm quarks, which still affect our calculation, are negligible 
within the present uncertainties. 
Forthcoming results from simulations in the continuum limit with $N_f=2+1$ and $N_f=2+1+1$ 
dynamical flavours will provide a check of this expectation. 
We should add that our continuum limit results for $R_i$ and 
$B_i$ ($i=2,\ldots,5$) are in the same ballpark with the numbers given at one lattice spacing 
in Ref.~\cite{Boyle:2012temp} 
where $N_f=2+1$ dynamical quarks are employed. 
 
As an interesting phenomenological application of the results obtained in this paper we have 
carried out a new UT analysis 
along the lines of the work of Ref.~\cite{Bona:2007vi}. Thanks to the improved accuracy of the 
present determination of the 
$\Delta S=2$ $B$-parameters, we could substantially strengthen the existing upper bounds on the
 Wilson coefficients of the 
operators of the non-standard sector of the effective weak Hamiltonian, and 
consequently increase the lower bound on the New Physics scale.

\begin{table}[!h]
\begin{center}
\begin{tabular}{|c|c||c|c||c|c|}
\hline 
\multicolumn{6}{|c|}{$B_{i}$ (RI-MOM at 2 GeV)}\tabularnewline
\hline 
\multicolumn{2}{|c||}{This work} & \multicolumn{2}{c||}{Ref.~\cite{Babich:2006bh}} & \multicolumn{2}{c|}{
Ref.~\cite{Donini:1999nn}}\tabularnewline
\hline
\hline 
\multicolumn{1}{|c}{} & \multicolumn{1}{c||}{{CL}} & $a=0.09$ fm & $a=0.13$ fm & $a=
0.07$ fm & $a=0.09$ fm\tabularnewline
\hline 
\multicolumn{6}{|c|}{}\tabularnewline
\hline 
1 & 0.52(2) & 0.56(5)      & 0.53(4)      & \,\,0.68(21) & \,\,0.70(15)\tabularnewline
\hline 
2 & 0.70(2) & 0.87(7)      & \,\,0.90(10) & 0.67(7)      & 0.72(9)\tabularnewline
\hline 
3 & 1.22(7) & \,\,1.41(12) & \,\,1.53(40) &\,\, 0.95(15) & \,\,1.21(10)\tabularnewline
\hline 
4 & 1.00(4) & 0.94(5)      & \,\,0.90(13) & 1.00(9)      & 1.15(5)\tabularnewline
\hline 
5 & 0.69(5) & 0.62(5)      & \,\,0.56(14) & \,\,0.66(11) & 0.88(6)\tabularnewline
\hline
\end{tabular}
\caption{Comparison between the unquenched results for $B_i$ obtained in the present work and 
the quenched values of Refs.~\cite{Babich:2006bh} and~\cite{Donini:1999nn}. 
Numbers are for renormalized quantities in the RI-MOM scheme at 2~GeV.}
\label{comparison_Bi_RIMOM}
\end{center}
\end{table}

\begin{table}[!h]
\begin{center}
\begin{tabular}{|c|c||c|c||c|c|}
\hline 
\multicolumn{6}{|c|}{$R_{i}$ (RI-MOM at 2 GeV)}\tabularnewline
\hline 
\multicolumn{1}{|c}{} & This work  & \multicolumn{2}{c||}{Ref.~\cite{Babich:2006bh}} & \multicolumn{2}{c|
}{Ref.~\cite{Donini:1999nn}}\tabularnewline
\hline 
\multicolumn{1}{|c}{} & \multicolumn{1}{c||}{{CL}} & $a=0.09$ fm & $a=0.13$ fm & $a=
0.07$ fm & $a=0.09$ fm\tabularnewline
\hline 
\multicolumn{6}{|c|}{}\tabularnewline
\hline 
1 & 1 & 1 & 1 & 1 & 1\tabularnewline
\hline 
2 & -12.9(4)  & -16.1(3.0) & -15.8(2.9) & -6.7(1.8)  & -6.6(1.1)\tabularnewline
\hline 
3 & ~~~4.5(2) & ~5.2(9)   & ~5.4(8)   & ~1.9(5)   & ~2.3(4)\tabularnewline
\hline 
4 & ~21.2(7)  & ~20.7(3.0) & ~18.8(2.8) & ~12.1(3.3) & ~12.6(2.1)\tabularnewline
\hline 
5 & ~~~4.7(3) & ~4.6(6)   & ~3.9(1.3)  & ~2.6(7)   & ~3.3(5)\tabularnewline
\hline
\end{tabular}
\caption{Same as in Table~\ref{comparison_Bi_RIMOM} for the $R_i$ ratios. 
}
\label{comparison_Ri_RIMOM}
\end{center}
\end{table}

\newpage

\vspace{0.4cm}
{\bf{Acknowledgments - }} We wish to thank all the other members 
of ETMC for their interest in this work and a most enjoyable and fruitful collaboration. 
Part of this work has been completed thanks to allocation of  CPU time on BlueGene/Q -Fermi based on the
agreement between INFN and CINECA and the specific initiative INFN-RM123.
N.C. and V.G. thank the MICINN and MINECO (Spain) for partial
support under Grants n.~FPA2008-03373 and FPA2011-23897, respectively, and the Generalitat Valenciana (Spain) for partial support
under Grant n.~GVPROMETEO2009-128. F.M. acknowledges the financial support from projects FPA2010-20807, 2009SGR502 and Consolider CPAN., and CSD2007-00042.
We acknowledge partial support from ERC Ideas Starting Grant n.~279972 ``NPFlavour'' and ERC
Ideas Advanced Grant n.~267985 ``DaMeSyFla". We also thank MIUR (Italy) for partial support under the contract PRIN08. M.P. acknowledges financial support by a 
Marie Curie European Reintegration Grant of the 7th European Community Framework
Programme under contract number PERG05-GA-2009-249309.  M.C. is associated to Dipartimento di Fisica,
Universit\`a di Roma Tre. L.S. is associated to Dipartimento di Fisica, Universit\`a
di Roma ``Sapienza".

\clearpage
\begin{appendices}
\vspace{1.2cm}
\section{Renormalization properties of $\Delta S=2$ four-fermion operators}
\label{APP_Op_Bas}

In this appendix we want to spell out the renormalization properties of the
four-fermion operators of interest for the description of $\bar K^0-K^0$ oscillations
in the mixed action (MA) lattice setup of Section~\ref{sec:tmQCD-gen}. 
We will do this by exploiting the results of Ref.~\cite{Donini:1999sf}. 
In  particular, we show that the operators in Eq.~(\ref{OMAPM_v2}), that we report here for the reader convenience,
\begin{eqnarray}
&&O^{MA}_{1[\pm]}= 2\big{\{}\big{(}[\bar q_1^\alpha\gamma_\mu q_2^\alpha][\bar q_3^\beta\gamma_\mu q_4^\beta]+[\bar q_1^\alpha\gamma_\mu \gamma_5 q_2^\alpha][\bar q_3^\beta\gamma_\mu \gamma_5 q_4^\beta]\big{)}\pm \big{(}2\leftrightarrow 4\big{)}\big{\}}\nonumber \\
&&O^{MA}_{2[\pm]}=2\big{\{}\big{(}[\bar q_1^\alpha q_2^\alpha][\bar q_3^\beta q_4^\beta]+[\bar q_1^\alpha\gamma_5 q_2^\alpha][\bar q_3^\beta\gamma_5 q_4^\beta]\big{)}\pm \big{(}2\leftrightarrow 4\big{)}\big{\}} \nonumber \\
&&O^{MA}_{3[\pm]}= 2\big{\{}\big{(}[\bar q_1^\alpha q_2^\beta][\bar q_3^\beta q_4^\alpha]+[\bar q_1^\alpha\gamma_5 q_2^\beta][\bar q_3^\beta\gamma_5 q_4^\alpha]\big{)}\pm \big{(}2\leftrightarrow 4\big{)}\big{\}}\nonumber \\
&&O^{MA}_{4[\pm]}= 2\big{\{}\big{(}[\bar q_1^\alpha q_2^\alpha][\bar q_3^\beta q_4^\beta]-[\bar q_1^\alpha\gamma_5 q_2^\alpha][\bar q_3^\beta\gamma_5 q_4^\beta]\big{)}\pm \big{(}2\leftrightarrow 4\big{)}\big{\}}\nonumber \\
&&O^{MA}_{5[\pm]}=2\big{\{}\big{(}[\bar q_1^\alpha q_2^\beta][\bar q_3^\beta q_4^\alpha]-[\bar q_1^\alpha\gamma_5 q_2^\beta][\bar q_3^\beta\gamma_5 q_4^\alpha]\big{)}\pm \big{(}2\leftrightarrow 4\big{)}\big{\}}\, ,
\label{OMAPM}
\end{eqnarray}
exhibit the same renormalization pattern as the corresponding continuum operators.
We recall that the Wilson $r$-parameters of valence quarks  in Eq.~(\ref{OMAPM}) are taken as
specified in Eq.~(\ref{RPAR}).
The normalization we have chosen in the definitions~(\ref{OMAPM}) is such that, when the 
operators $O^{MA}_{i[+]}$ are taken between the pseudoscalar operators 
$P^{12}=\bar q_1\gamma_5q_2$ and $P^{43}=\bar q_4\gamma_5 q_3$, one gets the same Wick contraction 
multiplicities one would obtain in QCD upon evaluating the kaon matrix elements 
of the operators~(\ref{def_Oi}). 
Naturally, apart from the issue of renormalization, the physical matrix elements 
will be obtained (in the continuum limit) by finally setting in our MA 
setup $\mu_1=\mu_3=\mu_s$ and $\mu_2=\mu_4=\mu_\ell$, with $\mu_s$ (resp.\ $\mu_\ell$) corresponding to the bare
strange (resp.\ degenerate up-down) quark mass.

A key result of Ref.~\cite{Donini:1999sf} (stated there in the quark basis that is
most natural for untwisted Wilson fermions)
is that the combinations that enjoy simple renormalization properties are not the operators 
listed in Eqs.~(\ref{OMAPM}), but those that are obtained after performing a Fierz transformation 
on the operators $O^{MA}_{3[\pm]}$ and $O^{MA}_{5[\pm]}$. This transformation has the effect of 
rewriting $O^{MA}_{3[\pm]}$ and $O^{MA}_{5[\pm]}$ in terms of operators where both spin and 
color indices are contracted within the same pair of quarks. With the definitions
\begin{eqnarray}
&&Q^{MA}_{1[\pm]}= 2\big{\{}\big{(}[\bar q_1\gamma_\mu q_2][\bar q_3\gamma_\mu q_4]+[\bar q_1\gamma_\mu \gamma_5 q_2][\bar q_3\gamma_\mu \gamma_5 q_4]\big{)}\pm \big{(}2\leftrightarrow 4\big{)}\big{\}}\nonumber \\
&&Q^{MA}_{2[\pm]}=2\big{\{}\big{(}[\bar q_1\gamma_\mu q_2][\bar q_3\gamma_\mu q_4]-[\bar q_1\gamma_\mu \gamma_5 q_2][\bar q_3\gamma_\mu \gamma_5 q_4]\big{)}\pm \big{(}2\leftrightarrow 4\big{)}\big{\}}\nonumber \\
&&Q^{MA}_{3[\pm]}= 2\big{\{}\big{(}[\bar q_1 q_2][\bar q_3 q_4]-[\bar q_1\gamma_5 q_2][\bar q_3\gamma_5 q_4]\big{)}\pm \big{(}2\leftrightarrow 4\big{)}\big{\}}\nonumber \\
&&Q^{MA}_{4[\pm]}= 2\big{\{}\big{(}[\bar q_1 q_2][\bar q_3 q_4]+[\bar q_1\gamma_5 q_2][\bar q_3\gamma_5 q_4]\big{)}\pm \big{(}2\leftrightarrow 4\big{)}\big{\}}\nonumber \\
&&Q^{MA}_{5[\pm]}=2\big{\{}\big{(}[\bar q_1\sigma_{\mu\nu}q_2][\bar q_3\sigma_{\mu\nu}q_4]\big{)}\pm \big{(}2\leftrightarrow 4\big{)}\big{\}}\,\,\, (\rm{for}\, \mu > \nu) ,
\label{QMAPM}
\end{eqnarray}
where $\sigma_{\mu\nu} = [\gamma_\mu,\gamma_\nu]/2$, one gets
\begin{equation}
O^{MA}_{i[\pm]}=\Lambda_{ij}^{[\pm]}Q^{MA}_{j[\pm]}\, ,\qquad\quad\Lambda^{[\pm]} = 
\left(\begin{array}{ccccc}
1 & 0 & 0 & 0 & 0 \\
0 & 0 & 0 & 1 & 0 \\
0 & 0 & 0 & \mp 1/2 & \pm 1/2 \\
0 & 0 & 1 & 0 & 0 \\
0 & \mp 1/2 & 0 & 0 & 0 \\
\end{array} \right)
\label{REL}
\end{equation}
In Eq.~(\ref{QMAPM}) we have omitted color indices as they are always contracted within each square parenthesis. 

In order to make direct contact with the formulae of Ref.~\cite{Donini:1999sf} we must  pass 
from the $q_f$-basis, in which the valence quark action~(\ref{OSvalact}) was written 
and where the Wilson term is (maximally) twisted, to the $\chi_f$-basis, where the Wilson 
term takes its standard form. This is achieved by the chiral transformation
\begin{equation}
q_f \longrightarrow \chi_f = e^{-i\pi r_f \gamma_5/4} q_f, ~~~~ 
\bar{q}_f \longrightarrow \bar{\chi}_f = \bar{q}_f e^{-i\pi r_f \gamma_5/4},
\label{chi_rotation}
\end{equation} 
under which 
\begin{eqnarray}
\hspace{-.8cm}&&S^{\rm OS}_{val} \to  \tilde S^{\rm OS}_{val} =\label{OSvalactC}\\
\hspace{-.8cm}&& = a^4 \sum_{x,f} \bar \chi_f (x) \Big \{ \frac{1}{2} \sum_\mu \gamma_\mu(\nabla_\mu 
+ \nabla^\ast_\mu ) + \big [ M_{\rm cr} - 
\frac{a}{2} \sum_\mu \nabla^\ast_\mu \nabla_\mu \big ] + i \gamma_5 r_f \mu_f \Big \} \chi_f(x)\, , \nonumber 
\end{eqnarray}
and  assuming $r_4 = \pm 1$ (as used in the present work),
\begin{eqnarray}
\hspace{-.8cm} r_4 Q^{MA}_{1[\pm]} &\to& \tilde{Q}^{MA}_{1[\pm]}=2i\big{\{}\big{(}[\bar \chi_1\gamma_\mu \chi_2][\bar \chi_3\gamma_\mu \gamma_5 \chi_4]+[\bar \chi_1\gamma_\mu \gamma_5 \chi_2][\bar \chi_3\gamma_\mu  \chi_4]\big{)}\pm \big{(}2\leftrightarrow 4\big{)}\big{\}}\nonumber \\
\hspace{-.8cm} r_4 Q^{MA}_{2[\pm]} &\to& \tilde{Q}^{MA}_{2[\mp]}=2i\big{\{}\big{(}[\bar \chi_1\gamma_\mu \chi_2][\bar \chi_3\gamma_\mu \gamma_5 \chi_4]-[\bar \chi_1\gamma_\mu \gamma_5 \chi_2][\bar \chi_3\gamma_\mu \chi_4]\big{)}\mp \big{(}2\leftrightarrow 4\big{)}\big{\}}\nonumber \\
\hspace{-.8cm} -r_4 Q^{MA}_{3[\pm]} &\to& \tilde{Q}^{MA}_{3[\mp]}=2i\big{\{}\big{(}[\bar \chi_1 \gamma_5 \chi_2][\bar \chi_3 \chi_4]-[\bar \chi_1 \chi_2][\bar \chi_3\gamma_5 \chi_4]\big{)}\mp \big{(}2\leftrightarrow 4\big{)}\big{\}}\nonumber \\
\hspace{-.8cm} -r_4 Q^{MA}_{4[\pm]} &\to& \tilde{Q}^{MA}_{4[\pm]}= 2i\big{\{}\big{(}[\bar \chi_1 \gamma_5 \chi_2][\bar \chi_3 \chi_4]+[\bar \chi_1 \chi_2][\bar \chi_3\gamma_5 \chi_4]\big{)}\pm \big{(}2\leftrightarrow 4\big{)}\big{\}}\nonumber \\
\hspace{-.8cm} -r_4 Q^{MA}_{5[\pm]} &\to& \tilde{Q}^{MA}_{5[\pm]}=2i\big{\{}\big{(}[\bar \chi_1\sigma_{\mu\nu}\chi_2][\bar \chi_3\sigma_{\mu\nu}\gamma_5\chi_4]\big{)}\pm \big{(}2\leftrightarrow 4\big{)}\big{\}}\, .
\label{QQT}
\end{eqnarray}
According to Ref.~\cite{Donini:1999sf}, for the renormalized operators $\tilde Q^{MA}_{i}$ one gets
\begin{equation}
\left( \begin{array}{c}
 \tilde{Q}^{MA}_{1[\pm]} \\
 \tilde{Q}^{MA}_{2[\pm]} \\
 \tilde{Q}^{MA}_{3[\pm]} \\
 \tilde{Q}^{MA}_{4[\pm]}\\
 \tilde{Q}^{MA}_{5[\pm]}
 \end{array} \right)^{\rm ren} = 
\left( \begin{array}{ccccc}
{\cal Z}_{11} & 0 & 0 & 0 & 0 \\
0 & {\cal Z}_{22} & {\cal Z}_{23} & 0 & 0 \\
0 & {\cal Z}_{32} & {\cal Z}_{33} & 0 & 0 \\
0 & 0 & 0 & {\cal Z}_{44} & {\cal Z}_{45} \\
0 & 0 & 0 & {\cal Z}_{54} & {\cal Z}_{55} \\
\end{array} \right)^{[\pm]}
\left( \begin{array}{c}
 \tilde{Q}^{MA}_{1[\pm]} \\
 \tilde{Q}^{MA}_{2[\pm]} \\
 \tilde{Q}^{MA}_{3[\pm]} \\
 \tilde{Q}^{MA}_{4[\pm]}\\
 \tilde{Q}^{MA}_{5[\pm]} \end{array} \right)^{(b)}
\label{renorm_pattern}
\end{equation}
Since at $\mu_f=0$ the fermion action $\tilde S^{\rm OS}_{val}$ is indistinguishable from a standard 
massless Wilson fermion action, in any mass independent renormalization scheme, the operators 
$Q^{MA}_{i}$ in the l.h.s. of Eqs.~(\ref{QQT}) 
enjoy the same renormalization properties of the corresponding operators $\tilde Q_i^{MA}$
into which they are trasformed under~(\ref{chi_rotation}). The $\tilde Q_i^{MA}$ operators have, up to lattice artefacts,
the same RCs of the corresponding operators in the standard Wilson's formulation of lattice QCD.
This result could also have been proved using the somewhat more elaborated approach 
of Ref.~\cite{Frezzotti:2004wz}. 

From Eqs.~(\ref{QQT})--(\ref{renorm_pattern}) and recalling Eq.~(\ref{REL}), 
one finally arrives for 
the operators $O^{MA}_{i[+]}$ of interest to us in this paper 
at the renormalization formulae
\begin{eqnarray}
&&O^{MA}_{i[+]}\Big{|}^{\rm ren}= Z_{ij}O^{MA}_{j[+]}\Big{|}^{(b)}\, ,\label{RENO}\\
&&Z=\Lambda^{[+]} Z_Q (\Lambda^{[+]})^{-1}\, ,\label{RENZ}
\end{eqnarray} 
\begin{equation}
Z_Q = 
\left( \begin{array}{ccccc}
{\cal Z}_{11}^{[+]} & 0 & 0 & 0 & 0 \\
0 & {\cal Z}_{22}^{[-]} & -{\cal Z}_{23}^{[-]} & 0 & 0 \\
0 & -{\cal Z}_{32}^{[-]} & {\cal Z}_{33}^{[-]} & 0 & 0 \\
0 & 0 & 0 & {\cal Z}_{44}^{[+]} & {\cal Z}_{45}^{[+]} \\
0 & 0 & 0 & {\cal Z}_{54}^{[+]} & {\cal Z}_{55}^{[+]} \\
\end{array} \right)
\label{ZQ}
\end{equation}
\vskip 0.15cm
From the renormalizability of (correlation functions evaluated in) the MA
lattice setup of Section~\ref{sec:tmQCD-gen} and the exact conservation of
the individual valence flavours it immediately follows that the operator
renormalization pattern of eqs.~(\ref{RENO})--(\ref{ZQ}) is independent (up
to cutoff effects, as usual) from the values of sea and valence quark masses.
It is hence possible to determine the relevant renormalization constants in
any mass-indipendent renormalization scheme by extrapolating to the chiral
limit suitable renormalization constant estimators evaluated at non-vanishing 
quark masses. Following this strategy we computed non-perturbatively in the RI-MOM
scheme the renormalization matrix $Z_Q$, see Eq.~(\ref{ZQ}), as detailed 
in Appendix~\ref{App_RIMOM} and summarized in Appendix~\ref{App_RCS}.

At this point the matrix elements 
$\langle P^{43}| O^{MA}_{i[+]}|^{\rm ren} | P^{12}\rangle|_{\rm ren}$, 
built using Eqs.~(\ref{OMAPM}), (\ref{RENO}) and~(\ref{ZQ}) and evaluated in our MA setup 
with $\mu_1=\mu_3=\mu_s$ (strange quark mass) and $\mu_2=\mu_4=\mu_\ell$ (up-down quark mass), 
tend in the limit $a\to 0$ to the matrix elements $\langle \bar K^0| O_{i}|K^0\rangle$ of the 
operators~(\ref{def_Oi}) in QCD with mere O($a^2$) discretization errors~\cite{Frezzotti:2004wz}.

\clearpage
\vspace{1.2cm}
\section{RI/MOM computation of renormalization constants of four-fermion operators}
\label{App_RIMOM}

In order 
to convert our lattice results for the bag parameters $B_{i}$ to their physical continuum
counterparts, in the same renormalization scheme and at the same scale as the corresponding 
perturbative Wilson coefficients used in the phenomenological analysis, we need the
renormalization constants (RCs)  of the operators $O^{MA}_{i[+]}$, $i=1,2, \dots , 5$,
see Eq.~(\ref{OMAPM_v2}), or equivalently Eq.~(\ref{OMAPM}). As discussed in Section~4, 
in our mixed action (MA) setup for lattice
correlation functions these operators represent the analogs 
of the parity-even parts of the $\Delta S=2$ 
four-fermion operators~(\ref{def_Oi}) that are relevant in the formal continuum theory.

In this Appendix, we give details on the non-perturbative computation of the RCs performed using the 
RI'-MOM scheme~(\cite{Franco:1998bm},~\cite{Constantinou:2010gr}).

As explained in detail in Ref.~\cite{Donini:1999sf}, instead of using 
the operator basis $O_{i [+]}^{MA}$ of Eq.~(\ref{OMAPM}), it is more convenient 
to employ the Fierz transformed operators $Q_{i [+]}^{MA}$, $i=1,2, \dots , 5$,
defined in Eq.~(\ref{QMAPM}). 
To lighten our notation, in the following we will drop the superscript
and the sign subscript, denote these operators simply by $Q_{i}$ and
assemble them in the array $\mathbf{Q}$. The generic renormalization
pattern of the bare operators $\mathbf{Q}^{(b)}$ is of the form  
\be
\mathbf{Q}^{\rm ren}\, =\, \mathbf{Z}\, \left[\, \mathbf{I}\, +\, \pmb{\Delta}\, \right]\, \mathbf{Q}^{(b)}
\label{Qrenpatt}
\ee
where  the scale-dependent renormalization matrix $\mathbf{Z}$ is block-diagonal, 
with a continuum-like block structure (the same as for e.g.\ the matrix in 
Eq.~(\ref{renorm_pattern})), while $\pmb{\Delta}$ is a sparse off-diagonal 
and scale-independent matrix  of the form 
\be
\pmb{\Delta}\, =\, \left[\, \begin{array}{ccccc}
0&\Delta_{12}&\Delta_{13}&\Delta_{14}&\Delta_{15}\\
\Delta_{21}&0&0&\Delta_{24}&\Delta_{25}\\
\Delta_{31}&0&0&\Delta_{34}&\Delta_{35}\\
\Delta_{41}&\Delta_{42}&\Delta_{43}&0&0\\
\Delta_{51}&\Delta_{52}&\Delta_{53}&0&0\\
\end{array} \right]  \; .
\ee
However, as shown in Appendix A, 
using the MA lattice setup of Section 4, the ``wrong chirality mixing" 
terms $\Delta_{i j}$, are reduced to mere O($a^2$) effects, the renormalization
matrix $\mathbf{Z}$ coincides with the matrix $Z_Q$ of Eq.~(\ref{ZQ}) and we 
recover a continuum-like renormalization pattern. This is a very important advantage of 
our approach, which we will implement in practice using the following strategy:
compute the quark propagators in the $q_{f}$-basis (also called physical basis
of tmQCD at maximal twist, in which the critical Wilson term is twisted, see 
Eq.~(\ref{OSvalact})), impose RI-MOM renormalization conditions on the operators $Q_{i}$
and extract the renormalization matrix ($\mathbf{Z}$) and, for check purposes, 
the mixing matrix ($\pmb{\Delta}$).  

\subsection{Procedure for extracting the RCs}

To determine the matrices $\mathbf{Z}$ and $\pmb{\Delta}$ in 
Eq.~(\ref{Qrenpatt}) we proceed as follows. We start           
by computing the lattice quark propagator
\be
S_{q_{f}}(p)\, =\, a^4\, \sum_{x}\, e^{-i p x}\; \langle\, q_{f}(x)\, \bar{q}_{f}(0)\, \rangle
\ee
and the four-point Green functions with an insertion of the operator $Q_{i}$, namely 
\begin{eqnarray}
\label{Geqdef}
\hspace{-.7cm}&&\lefteqn{G_{i}(p,\, p,\, p,\, p)^{a\, b\, c\, d}_{\alpha\, \beta\, \gamma\, \delta} \,=} \\
\hspace{-.7cm}&&\phantom{=}a^{16} \!\! \!\sum_{x_{1},x_{2},x_{3},x_{4}} \!e^{-i p (x_{1}-x_{2}+x_{3}-x_{4})} \langle \,
\left[q_{1}(x_{1})\right]^{a}_{\alpha}
\left[\bar{q}_{2}(x_{2})\right]^{b}_{\beta}\, Q_{i}(0)\, 
\left[q_{3}(x_{3})\right]^{c}_{\gamma}
\left[\bar{q}_{4}(x_{4})\right]^{d}_{\delta} \,\rangle\, .\nonumber
\end{eqnarray}
The lower(upper) case Greek (Latin) symbols denote uncontracted spin (color) indices. The corresponding amputated Green functions are given by
\begin{eqnarray}
\label{Lambdaeqdef}
\hspace{-.7cm}&&\lefteqn{\Lambda_{i}(p,p,p,p)^{abcd}_{\alpha\beta\gamma\delta}\, =}\\
\hspace{-.7cm}&&\phantom{=}\left[S_{q_{1}}(p)^{-1}\right]^{a a'}_{\alpha \alpha'} \left[S_{q_{3}}(p)^{-1}\right]^{c c'}_{\gamma \gamma'}
 G_{i}(p,p,p,p)^{a'b'c'd'}_{\alpha'\beta'\gamma'\delta'}\left[S_{q_{2}}(p)^{-1}\right]^{b' b}_{\beta' \beta} \left[S_{q_{4}}(p)^{-1} \right]^{d' d}_{\delta' \delta}\, .\nonumber
\end{eqnarray}
For the sake of clarity, we will use matrix notation, denoting the matrices by boldface symbols and omitting color and spin indices.
The amputated Green functions will be collected in the $1\times 5$ row vector 
\be
\pmb{\Lambda}(p)\, =\,\left(\, \Lambda_{1},\, \Lambda_{2},\, \Lambda_{3},\, \Lambda_{4},\, \Lambda_{5}\, \right)(p,p,p,p)\, .
\ee 
Setting
\be
\label{rensett}
\pmb{\hat{\Lambda}}(ap, a\mu)\, =\, Z_{q}^{-2}(ap)\, \pmb{\Lambda}(a p)\, \left[\, \mathbf{I}\, 
+\, \pmb{\Delta}^{T}\, \right]\, \mathbf{Z}(a\mu)^{\rm T} \, , 
\ee
the renormalization matrix $\mathbf{Z}(a\mu)$ is determined by solving the 
renormalization conditions~\cite{Donini:1999sf}, namely
\be
\mathbf{P}\, \pmb{\hat{\Lambda}}(p)\mid_{p^2\, =\, \mu^2}\, =\, \mathbf{I} \, .
\label{rencond}
\ee
In Eq.~(\ref{rencond}), $Z_{q}$ is the quark field RC and $\pmb{\Delta}$ is, as we said before, the mixing matrix. We have also introduced the $5 \times 1$ 
column vector of spin projectors (see Eq.~(37) of Ref.~\cite{Donini:1999sf} for the explicit form of these projectors)
\be
\mathbf{P}^{\rm T}\, =\,\left(\, P_{1},\, P_{2},\, P_{3},\, P_{4},\, P_{5}\, \right)
\ee
which act on the amputed Green functions by $(i,j\, =\, 1\, \cdots,\, 5)$,
\[
P_{i}\, \Lambda_{j}\, \equiv\, {\rm Tr}\; P_{i}\, \Lambda_{j}(a p)
\]
where the trace is taken over spin and colour, and obey the orthogonality relations 
\be
{\rm Tr}\; P_{i}\, \Lambda_{j}^{(0)}(a p)\, =\, \delta_{i j}\, ,
\ee
with $\Lambda^{(0)}_{j}$ the tree level amputated Green function of the operator $Q_{j}$.
It is convenient to express $\pmb{\Lambda}$ in terms of a ``dynamics" matrix ${\mathbf D}$, defined by 
\be
\pmb{\Lambda}(a p)\, =\, \pmb{\Lambda}^{(0)}(a p)\; \mathbf{D}(a p) \,.
\ee
This matrix equation can be solved for $\mathbf{D}$ using the spin projectors $\mathbf{P}$, getting
\be
\label{dynaP}
\mathbf{D}(p)\, =\, \mathbf{P}\; \pmb{\Lambda}(p) \, . 
\ee
Combining Eqs.~(\ref{rensett}), (\ref{rencond}) and~(\ref{dynaP}), we see that, once 
the dynamics matrix is known, we can                 
determine both the renormalization and the mixing 
matrices from the relation 
\be
Z_{q}^{-2}\; \mathbf{D}\, \left[\, \mathbf{I}\, +\, \pmb{\Delta}^{\rm T}\, \right]\, \mathbf{Z}^{\rm T}\, =\, \mathbf{I}\, \rightarrow\,
\mathbf{Z}\, \left[\, \mathbf{I}\, +\, \pmb{\Delta}\, \right]\, =\, Z_{q}^{2}\, \left(\, \mathbf{D}^{\rm T}\, \right)^{-1} \, .
\ee
This matrix equation can be solved for $\mathbf{Z}$ and $\pmb{\Delta}$ by exploiting the block diagonal structure of the $\mathbf{Z}$ matrix. In fact, it is easy to 
see that the three diagonal blocks of the renormalization matrix ($i,j=1$; $i,j=2,3$ and $i,j=4,5$) are given by
\be
\label{ZfromD}
Z_{ij} \, =\, Z_{q}^{2}\, \left(\, \mathbf{D}^{\rm T}\, \right)^{-1}_{ij} \qquad (i,j=1) \,\,\,(i,j=2,3)\,\,\, (i,j=4,5)
\ee
whereas the mixing coefficients are easily obtained from the equations,
\begin{eqnarray}
\label{DeltafromD}
Z_{11}\, \Delta_{1i} &=& Z_{q}^{2}\, \left(\, \mathbf{D}^{\rm T}\, \right)^{-1}_{1i}\qquad\qquad\;\; i=2,\cdots,5\\[0.25cm]
\left(\!\begin{array}{cc} Z_{i i}&Z_{i\, i+1}\\[0.2cm] Z_{i+1\, i}&Z_{i+1\, i+1}\end{array} \!\right)
\left( \!\begin{array}{c} \Delta_{i j}\\[0.2cm] \Delta_{i+1\, j}\end{array}  \!\right) &=& Z_{q}^{2}\, \left( \!\begin{array}{c} \left(\, \mathbf{D}^{\rm T}\,
\right)^{-1}_{ij}\\[0.2cm] \left(\,
\mathbf{D}^{\rm T}\, \right)^{-1}_{i+1\, i}\end{array} \! \right)
\,\,\,\left\{\begin{array}{cc}i=2&j=1,4,5\\[0.2cm]i=4&j=1,2,3\\ \end{array}\right\} \, .\nonumber
\end{eqnarray}

We can now summarize our procedure to determine the renormalization matrix of the parity-even part of the four-fermion operators of the SUSY basis of 
Eq.~(\ref{def_Oi}). 

\begin{description}
\item[Step 1] The Green functions~(\ref{Geqdef}) and ~(\ref{Lambdaeqdef}), and from them the dynamics matrix $\mathbf{D}$, are evaluated in the Landau gauge for a 
sequence of sea, $\mu_{sea}$, and valence, $\mu_{val}$, quark mass values at each of the four lattice spacings we consider here. The bare parameters and the 
statistics of this computation are detailed in Table 2 of Ref.~\cite{Constantinou:2010gr}. 
One can also find there (see Eqs.~(3.6) and~(3.7)) the set of discrete lattice momenta, 
$p_{\nu}$ ($p_{1,2,3}\, =\, (2 \pi/L)\, n_{1,2,3}$, 
$p_{4}\, =\, (2 \pi/T)\, (n_{4}\, +\, 1/2)$),
that we include in the present calculation. To minimize the 
contributions of Lorentz non-invariant discretization artifacts, we take into 
consideration only momenta satisfying the constraint
\be
\label{cut28}
\sum_{\rho}\, \tilde{p}_{\rho}^4\, <\, 0.28\, \left(\, \sum_{\nu}\, \tilde{p}_{\nu}^{2}\, \right)^2,\;\;\;\;\;\;\;\;\;a\, \tilde{p}_{\nu}\, \equiv\, \sin(a p_{\nu}) \, .
\ee
In the following, we shall often use the short-hand $\tilde{p}^2\, =\, \sum_{\nu}\, \tilde{p}_{\nu}^2$.

\item[Step 2] For each $\beta$ and each choice of the scale $\tilde{p}^2$, 
the renormalization relation~(\ref{ZfromD}) is enforced at all values of $\mu_{sea}$ and 
$\mu_{val}$ given in Table 2 of Ref.~\cite{Constantinou:2010gr}. By doing so, we obtain at nonzero quark masses the estimators $Z_{ij}^{\rm RI'}(\tilde{p}^2; 
a^2 \tilde{p}^2; \mu_{val}; \mu_{sea})$, which are then extrapolated 
to $\mu_{val}\, =\, 0$ (see Section~\ref{sec:ValenceLimits}) and 
$\mu_{sea}\, =\, 0$ (see Section~\ref{sec:SeaChiralLimits}).

\item[Step 3] Improved estimates of $Z_{ij}^{\rm RI'}(\tilde{p}^2; a^2 \tilde{p}^2; 0; 0)$, are obtained by subtracting the perturbatively leading cutoff effects (see
Section~\ref{sec:RemoveOa2}). 

\item[Step 4] Using the NLO continuum QCD evolution of the renormalization matrix $\mathbf{Z}$ calculated in Refs.~\cite{rm1:4ferm-nlo,mu:4ferm-nlo}, the first argument 
of $Z_{ij}^{\rm RI'}(\tilde{p}^2; a^2 \tilde{p}^2; 0; 0)$ is brought to a reference scale $\mu_{0}^2$. 
In this step, we assume that the scales
$\tilde{p}^2$ and $\mu_{0}^2$ are large enough to make NLO perturbation theory accurate. 
This is the same level of accuracy achieved in the determination of the Wilson coefficients.

\item[Step 5] The residual $a^2 \tilde{p}^2$ dependence in 
$Z_{ij}^{\rm RI'}(\mu_{0}^2; a^2 \tilde{p}^2; 0; 0)$ is
attributed to lattice artifacts, which we treat according to
either the M1 or M2 methods, introduced in Ref.~\cite{Constantinou:2010gr} 
(see Section~\ref{sec:FinalEstimates}).

\item[Step 6] In order to reduce the statistical error, the lattice RC estimators are averaged over two equivalent patterns of Wilson parameters 
$(r_{1}, r_{2}, r_{3}, r_{4})$, namely $(1,1,1,-1)$ and $(-1,-1,-1,1)$, as well as over different lattice momenta corresponding to the same $\tilde{p}^2$. 
We have checked that performing these averages before or after taking the chiral limit leads to consistent results. 
\end{description}

\subsection{Valence chiral limit}
\label{sec:ValenceLimits}
In view of the relation~(\ref{ZfromD}) and since the extraction of $Z_q$ 
(see Ref.~\cite{Constantinou:2010gr}) poses no particular problems,
our discussion will be mainly focused here on the quark mass dependence of the
dynamics matrix $\mathbf{D}$.                              
At fixed values of $\beta$, $a^2 \tilde{p}^2$ and $a \mu_{sea}$, we fit the dynamics matrix elements $D_{i j}$ to the ansatz
\be
D_{i j}(p; \mu_{val}; \mu_{sea})\, =\, A(\tilde{p}^2; \mu_{sea})\, +\, B(\tilde{p}^2; \mu_{sea})\, \mu_{val}\, +\, C(\tilde{p}^2; \mu_{sea})/\mu_{val}\, .
\label{valfitRC}
\ee
Here we have introduced a term with a pole in $\mu_{val} \sim m_{PS}^2$ to cope 
with the expected Goldstone boson (GB) pole contribution to the elements of the 
$\bf D$ matrix. The existence of such a GB-pole term can be understood as follows. 
At asymptotically large $p^2$, non-perturbative effects giving contributions 
potentially divergent in the chiral limit to the Green functions~(\ref{Geqdef})
do vanish and the latter 
turn out to be polynomial in the quark mass parameters~\cite{renorm_mom:paper1}. 
At finite values of $p^2$, however, the contributions to (the spectral decomposition
of) these Green functions from one-GB intermediate state with momentum $q$
and mass $m_{PS}$, give rise to terms proportional to $(q^2+m^2_{PS})^{-1}$ and
suppressed by some power of $1/p^2$. If several one-GB intermediate states
contribute to the spectral representation of the Green functions~(\ref{Geqdef})
several terms, each behaving as $(q^2+m^2_{PS})^{-1}$ and
suppressed by some power of $1/p^2$, will show up. These results follow
straightforwardly from the ``polology' study of the Green functions (see
e.g.\ the discussion in the book~\cite{Weinberg:1996}) or from the well known
Lehmann-Symanzik-Zimmermann (LSZ) reduction formalism. 
Now, since in the Green functions~(\ref{Geqdef}) the four-fermion operator
is inserted at zero four-momentum transfer ($q=0$), one expects, from the time
orderings where two quark fields can create from the vacuum a pseudoscalar 
(i.e.\ GB) one-particle state, a contribution proportional to $1/m^2_{PS}$,  
suppressed by some power of $1/p^2$. Similarly, from those time orderings where
two quark fields create a GB-state and two further quark fields destroy another
GB-state, contributions do arise that behave as $(1/m^2_{PS})^2$ and are twice 
more strongly suppressed at large $p^2$.
 
In conclusion, by exploiting (along the lines of Appendix~A of 
Ref.~\cite{renorm_mom:paper1}) the large-$p^2$ behaviour of
the matrix element~\footnote{Here $q_f'(p)$ ($\bar{q}_f(p)$) 
denotes the four-dimensional Fourier transform of
the quark field $q_f'(x)$ ($\bar{q}_f(x)$), while
$| P^{f'f} \rangle$ is the pseudoscalar meson state with valence
quarks of flavour $f$ and $f'$.}
$\langle 0 | \bar{q}_f(p) q_{f'}(-p) | P^{f'f} \rangle$
and taking also into consideration the four factors of $S_q(p)^{-1}$
that stem from the relations~(\ref{Lambdaeqdef}),   
one finds that the dynamics matrix $D_{ij}$ contains 
GB-pole contributions of the following kinds: 
\begin{eqnarray}
\hspace{-0.85cm}&&D^{{\rm Single}\; 12}_{ij} \, \sim \, P_i
S_{q_1}^{-1}(p) S_{q_2}^{-1}(-p) \langle 0 | \bar{q}_{2} q_1 | P^{12} \rangle\; 
\frac{\displaystyle 1}{\displaystyle p^4 \left(M^{12}\right)^2}\;
\langle P^{12} | Q_{j}(0) | q_{3}(p) \bar{q}_{4}(-p) \rangle
\nonumber\\  
\hspace{-0.85cm}&&D^{{\rm Single}\; 34}_{ij} \, \sim \, P_i
\langle q_{1}(p) \bar{q}_{2}(-p) | Q_{j}(0) | P^{34} \rangle\; 
\frac{\displaystyle 1}{\displaystyle \left(M^{34}\right)^2 p^4 }\; 
\langle P^{34} | \bar{q}_{4} q_{3} | 0 \rangle S_{q_3}^{-1}(p) S_{q_4}^{-1}(-p)
\nonumber\\
\hspace{-0.85cm}&&D^{{\rm Double}\, 12\, 34}_{ij} \, \sim \, P_i   
S_{q_1}^{-1}(p) S_{q_2}^{-1}(-p) \langle  0 | \bar{q}_{2} q_{1} | P^{12} \rangle\, 
\frac{\displaystyle \langle P^{12} | Q_{j}(0) | P^{34} \rangle}{\displaystyle 
p^4 \left(M^{12} \right)^2\,
\left(M^{34}\right)^2 p^4 } \times 
\nonumber\\
\hspace{-0.85cm}&& \phantom{D^{{\rm Double}\, 12\, 34}_{ij} \, \sim \, } \times
\langle P^{34} | \bar{q}_{4} q_{3} | 0\rangle S_{q_3}^{-1}(p) S_{q_4}^{-1}(-p) \, .
\label{POLOLOGY}
\end{eqnarray}

We recall that the kinematics of our Green functions corresponds to an
exceptional momentum configuration where $p_1=p_3=-p_2=-p_4=p$, and thus $q = 0$.
As in the chiral limit $S_{q}^{-1}(p) \sim \gamma_\mu p_\mu$, the
result~(\ref{POLOLOGY}) implies that single and double GB-pole  
terms are suppressed by $1/p^2$ and $1/(p^2)^2$ factors, respectively.
A second observation is that thanks to the choice $r_4=-r_3$ in our
MA setup the lattice axial current 
$\bar{q}_{4}\gamma_{\mu} \gamma_{5}q_{3}$ is conserved (only broken 
by soft mass terms) and hence   
the matrix elements of the operator $Q_{1}$ in $D^{{\rm Single} \; 34}_{i1}$ 
and $D^{{\rm Double}\, 12\, 34}_{i1}$ vanish as 
$\left(M^{34}\right)^2 \sim \left(\mu_3 +\mu_ 4\right)$ 
in the limit $\mu_{3,4}\rightarrow 0$. This implies that 
no double pole occurs in $D_{i1}$. 
At non-vanishing lattice spacing there exists, however, an O($a^2$) 
single pole contribution in $D^{{\rm Single}\; 12}_{i1}$ because, owing
to $r_2=r_1$, the lattice axial current 
$\bar{q}_{2}\gamma_{\mu} \gamma_{5}q_{1}$ is broken by 
discretization effects (see Ref.~\cite{FrezzoRoss1} and 
Appendix~A of Ref.~\cite{Constantinou:2010qv}). 

For the case $j \neq 1$, when no similar GB-pole simplifications can occur,
double GB-pole terms strongly suppressed (like $1/(p^2)^2$) at large $p^2$
are to be expected in $D_{ij}(p)$. 
However, precisely owing to this strong suppression in practice, within our statistical
errors and in the ranges of quark masses and $p^2$ we explore (see Table 2 of 
Ref.~\cite{Constantinou:2010gr} and sect.~\ref{sec:FinalEstimates}), 
we hardly see in our lattice data any effects that can reliably be ascribed
to double GB-pole contributions. On the contrary, we do find clear numerical
evidence for single GB-pole contributions, which indeed at high $p^2$
are only suppressed as $1/p^2$. We thus decided to ignore
double GB-pole terms in our valence mass chiral extrapolations.
 
\begin{figure}[H]
\mbox{}\vskip -3cm
\subfigure{\includegraphics[scale=0.19,angle=-0]{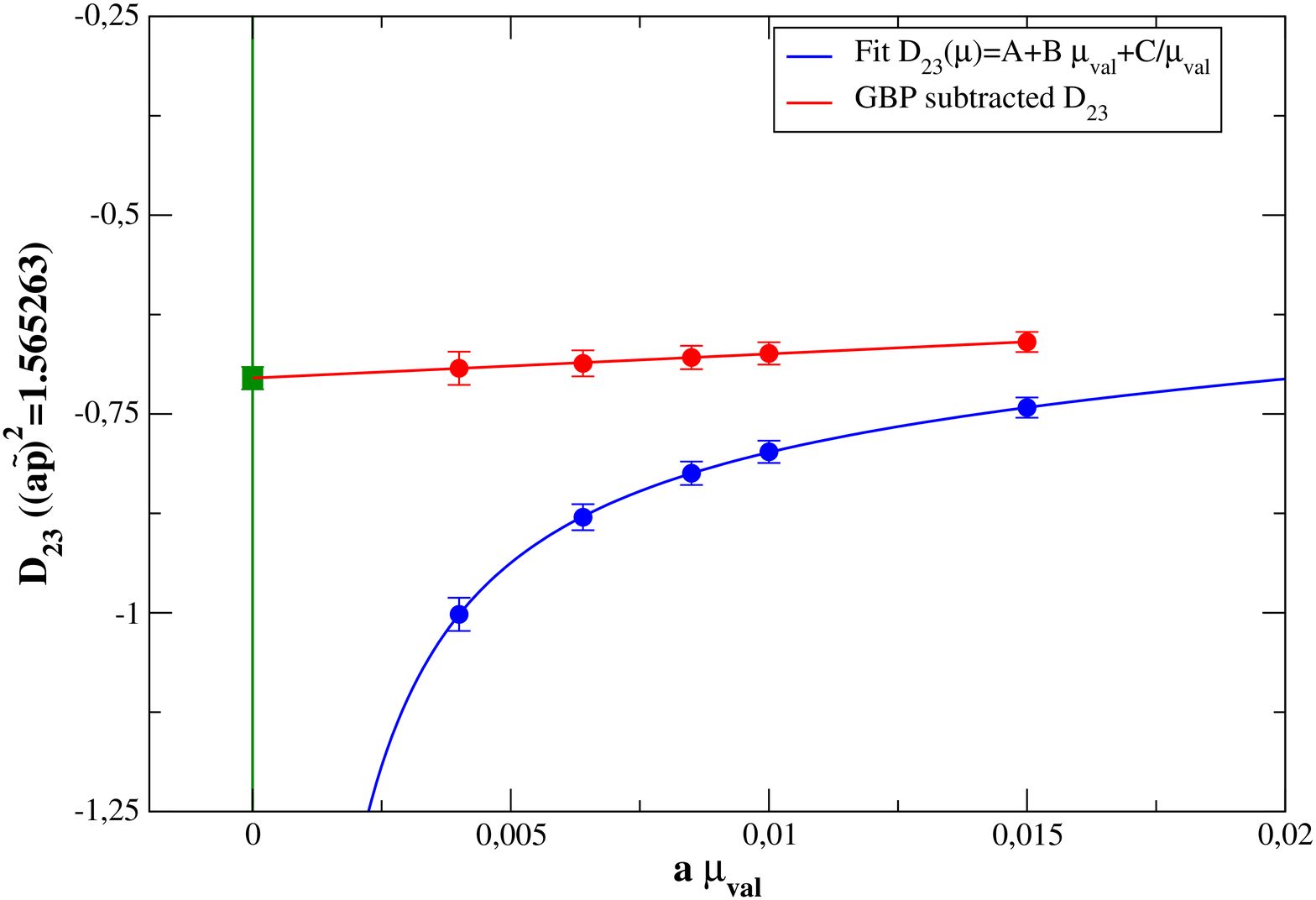}}\hspace*{0.15cm}
\subfigure{\includegraphics[scale=0.19,angle=-0]{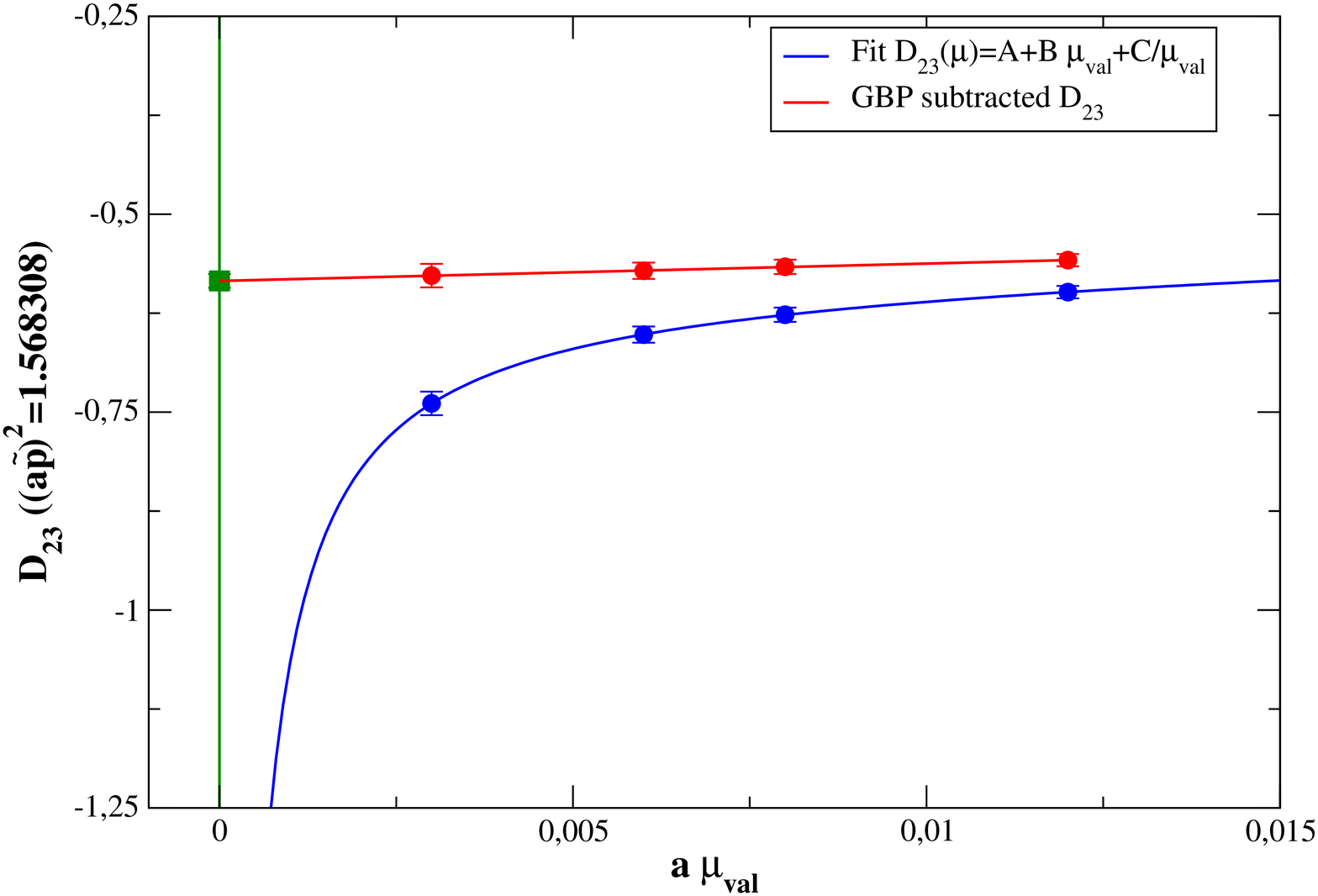}}\\[0.75cm]
\subfigure{\includegraphics[scale=0.19,angle=-0]{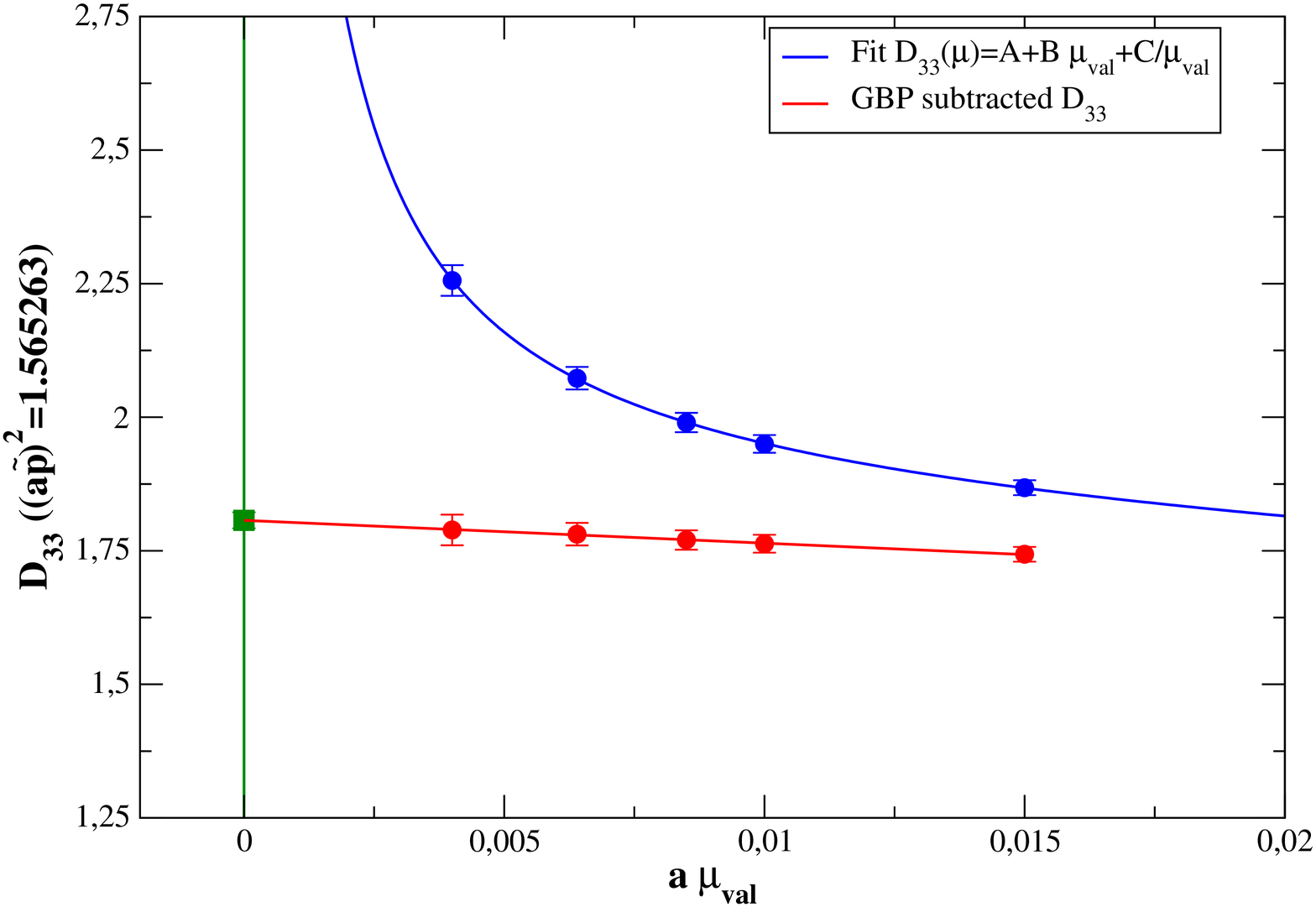}}\hspace*{0.15cm}
\subfigure{\includegraphics[scale=0.19,angle=-0]{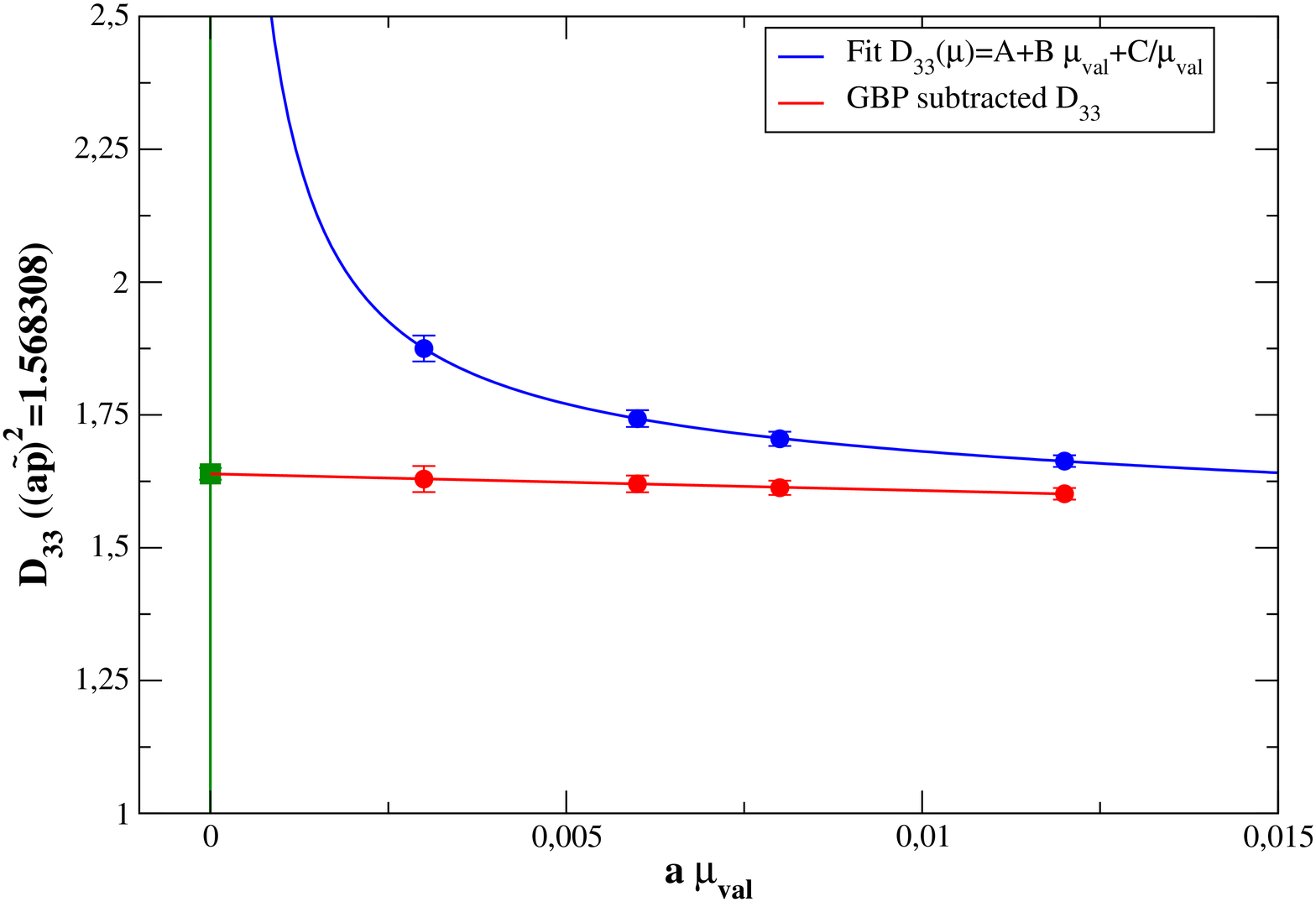}}\\[0.75cm]
\subfigure{\includegraphics[scale=0.19,angle=-0]{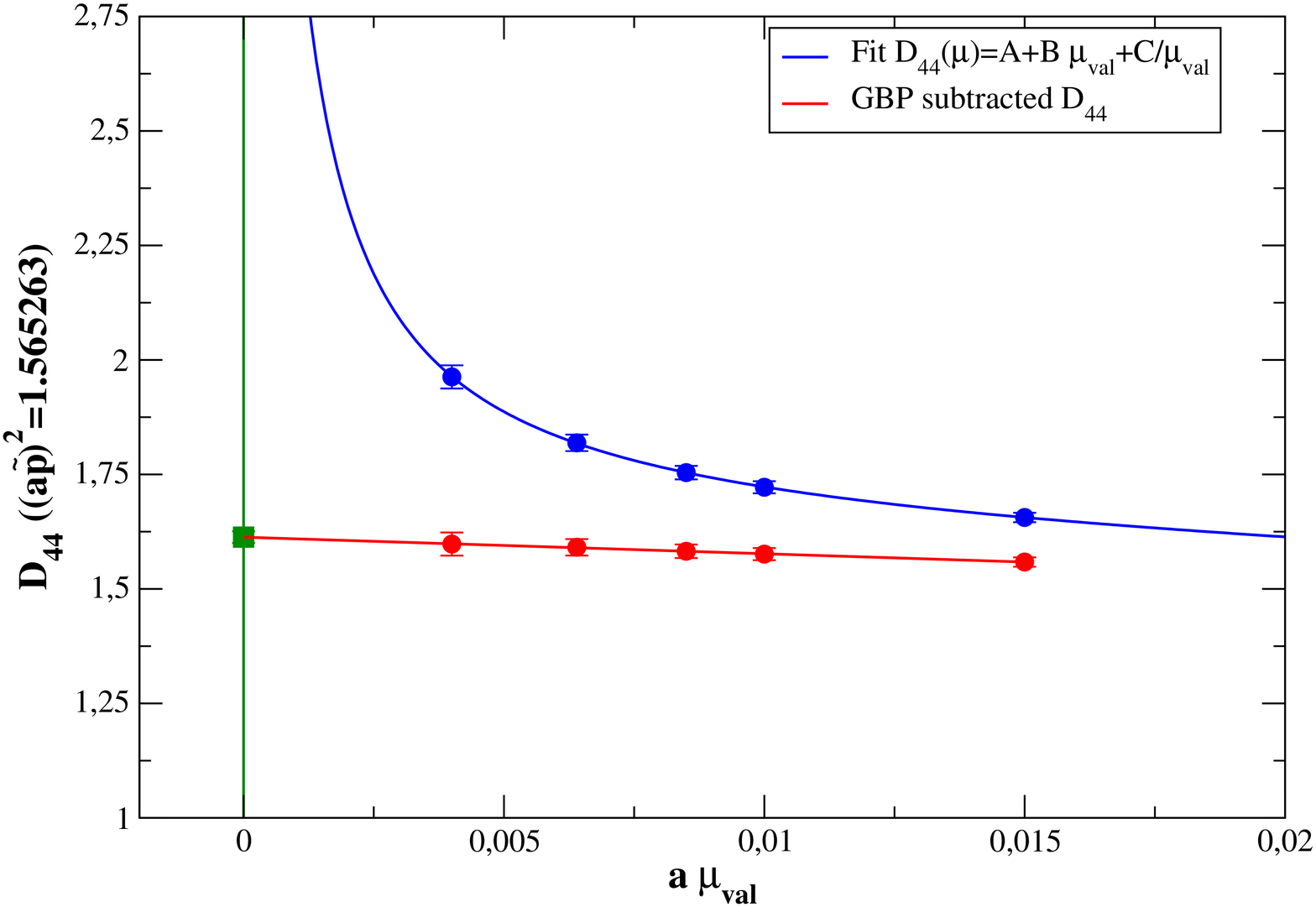}}\hspace*{0.15cm}
\subfigure{\includegraphics[scale=0.19,angle=-0]{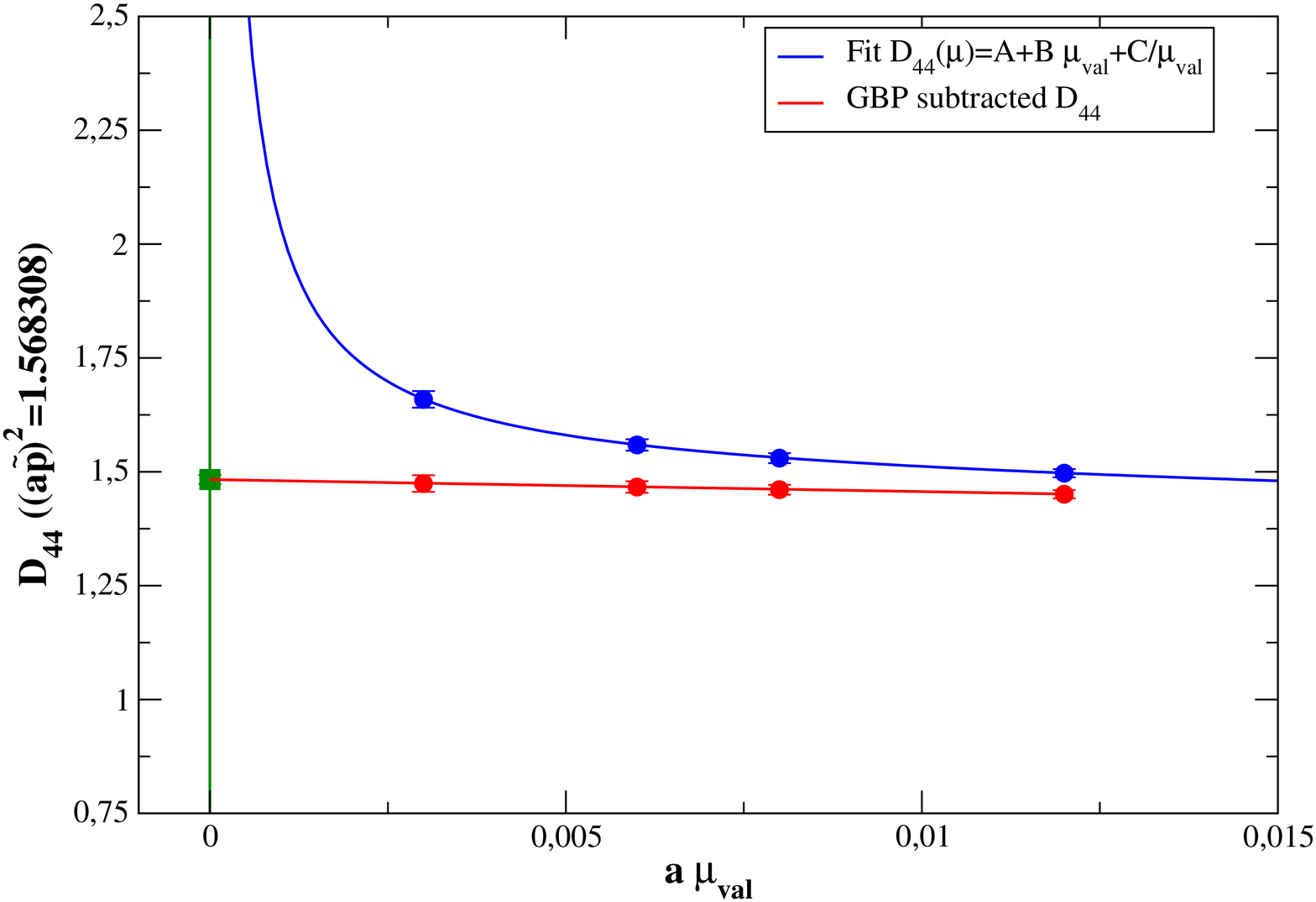}}\\[0.75cm]
\subfigure{\includegraphics[scale=0.19,angle=-0]{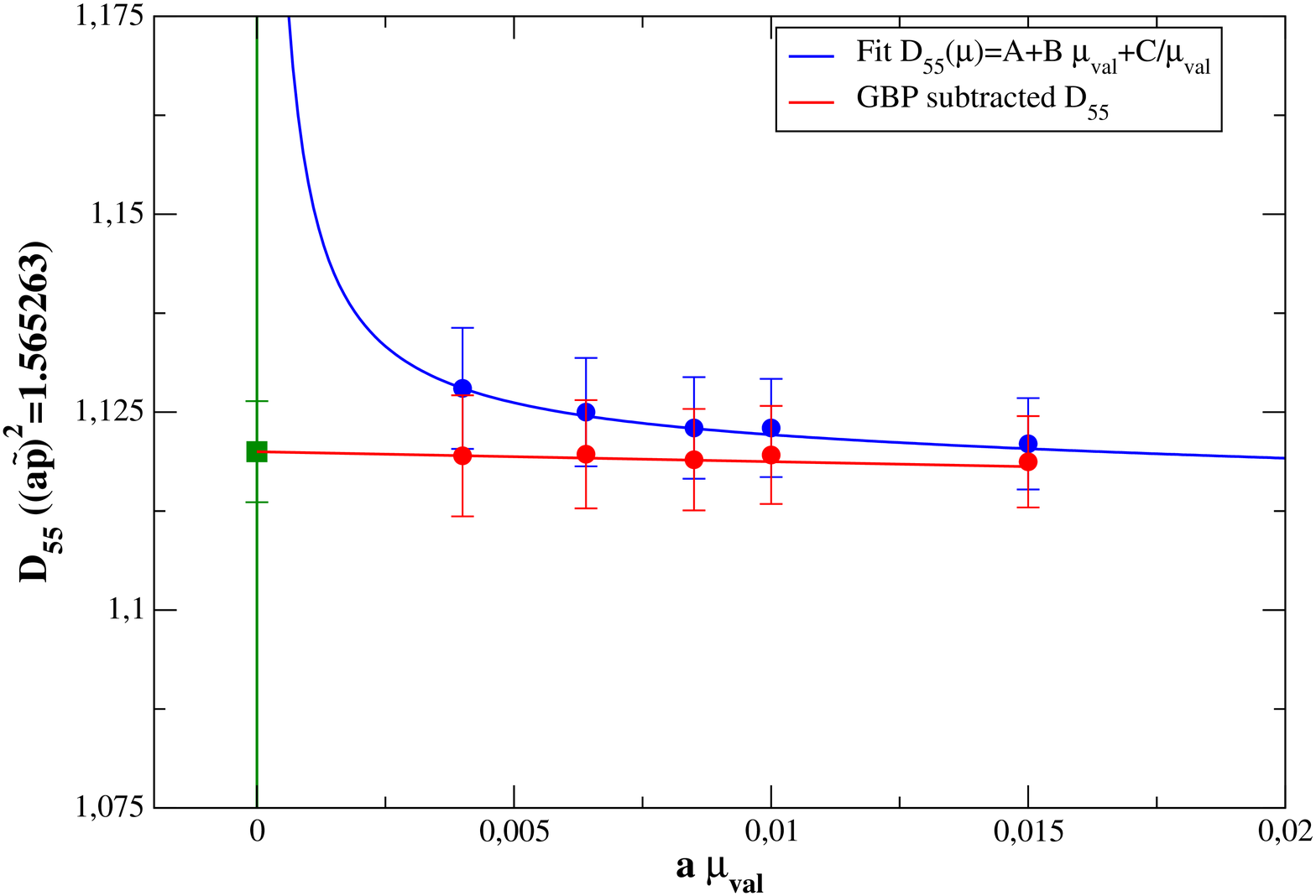}}\hspace*{0.15cm}
\subfigure{\includegraphics[scale=0.19,angle=-0]{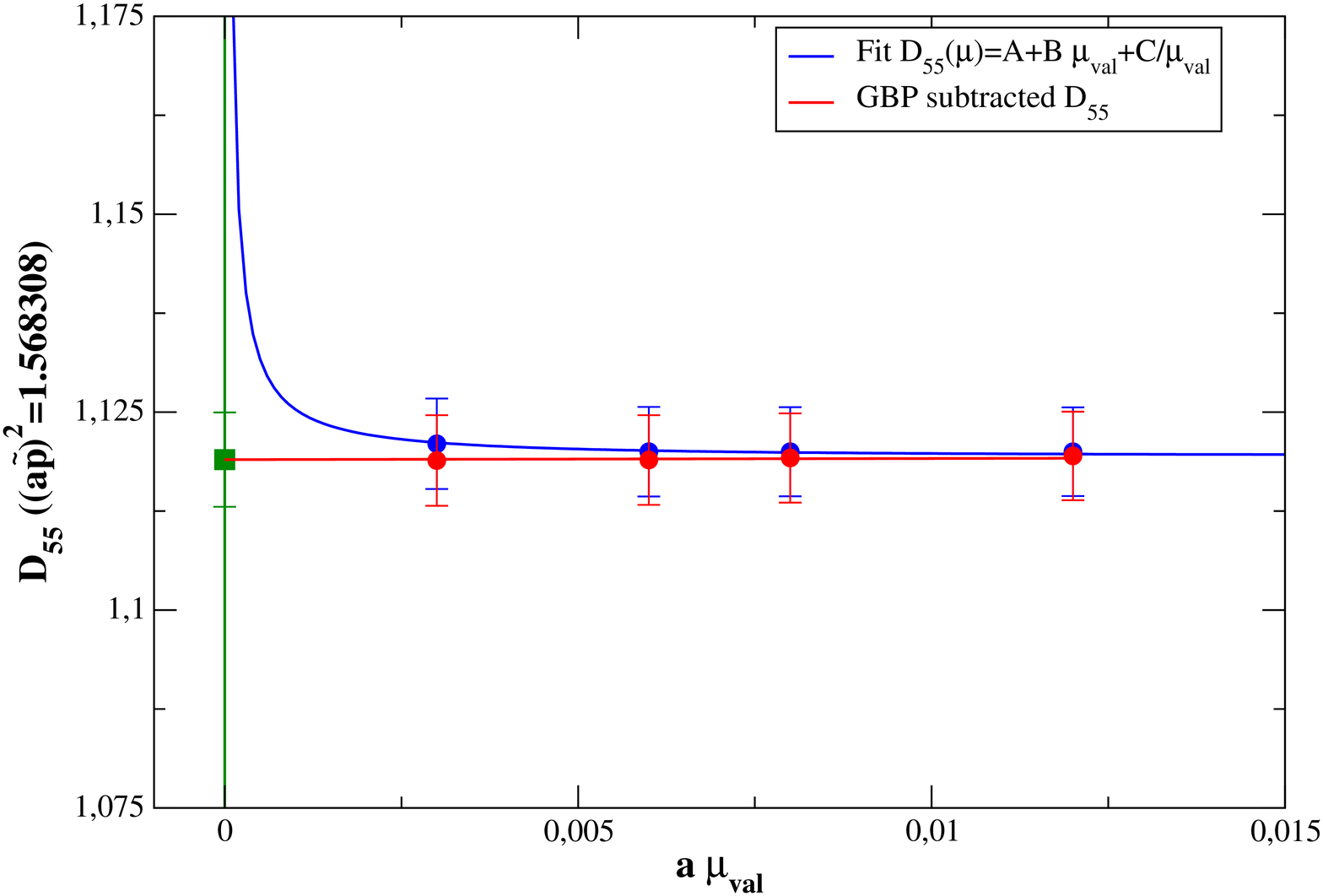}}
\vskip -1.0cm
\begin{center}
\caption{\sl GB-pole subtraction and valence chiral limit of $D_{2 3}$, $D_{3 3}$, $D_{4 4}$ and $D_{5 5}$ plotted
versus $a \mu_{val}$, for $\beta\, =\, 3.9$, $a \mu_{sea}\, =\, 0.0040$ and $(a \tilde{p})^2\, \approx\, 1.565$ (left column) and
$\beta\, =\, 4.05$, $a \mu_{sea}\, =\, 0.0030$ and $(a \tilde{p})^2\, \approx\, 1.568$ (right column).
}
\label{fig:GBb39405}
\end{center}
\end{figure}

This choice is also justified a posteriori by the results of the valence
chiral fits based on the ansatz~(\ref{valfitRC}). A subset of these
results is illustrated in Fig.~\ref{fig:GBb39405}. There
we display typical examples of the effect of GB-pole subtractions in the 
matrix elements of the dynamics matrix at two values of $\beta$. 
As can be seen, after the subtraction, a smooth dependence upon $\mu_{val}$ (or equivalently on $M_{12}^2$) is observed. Combining the valence chiral limit lattice 
estimator of $D_{i j}$ and $Z_{q}$, we are able to get reliable estimates of 
the intermediate quantities $Z^{\rm lat}_{i j}(\tilde{p}^2;\, a^2 \tilde{p}^2; 0, 
a \mu_{sea})$.

\subsection{Sea chiral limit}
\label{sec:SeaChiralLimits}

\begin{figure}[!ht]
\begin{center}
\mbox{}\vskip -3cm
\subfigure[]{\includegraphics[scale=0.20,angle=0]{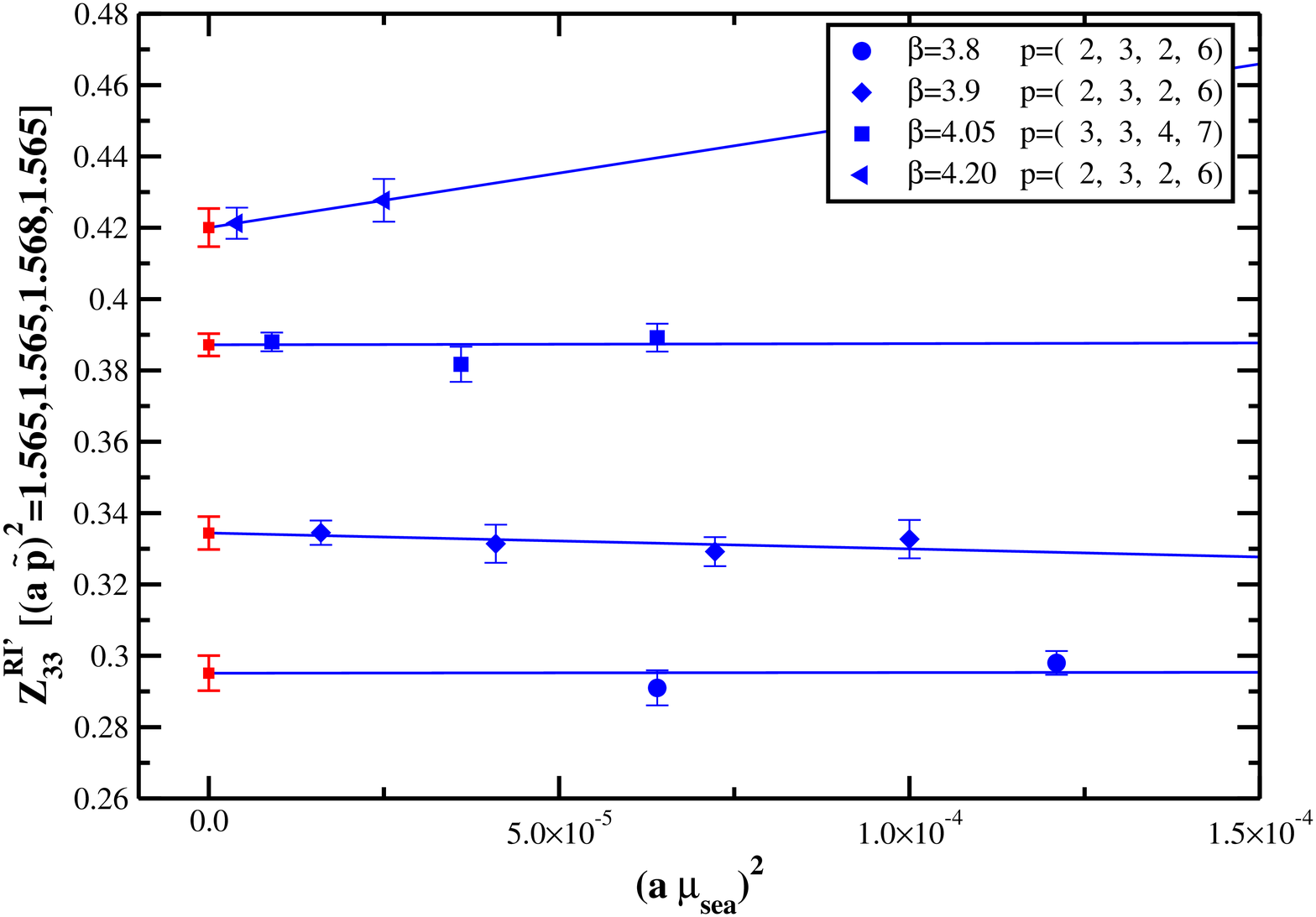}}\hspace*{0.15cm}
\subfigure[]{\includegraphics[scale=0.20,angle=0]{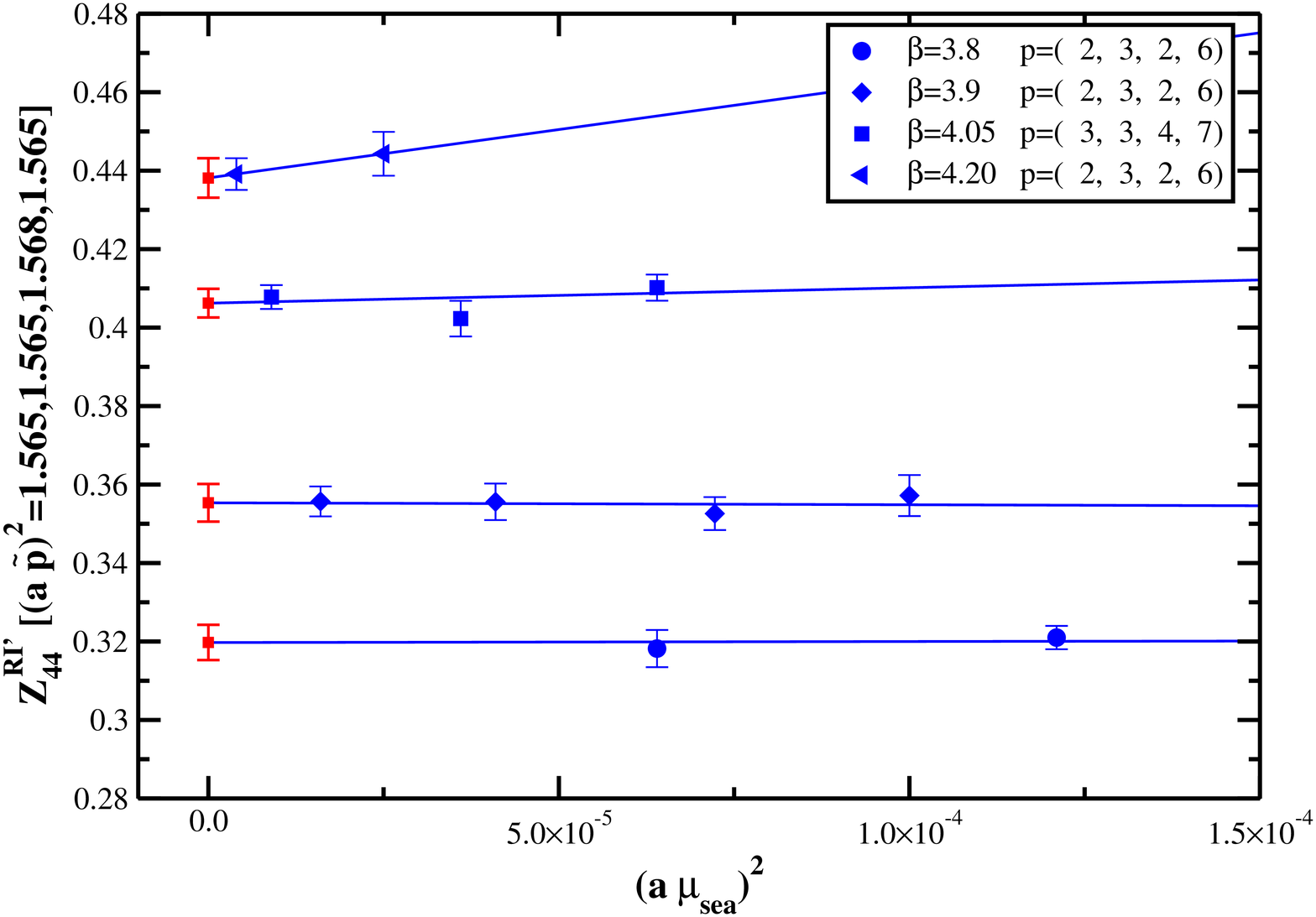}}
\end{center}
\vskip -1.0cm
\begin{center}
\caption{\sl The quantities $Z^{\rm RI'}_{33}(\tilde{p}^2; a^2 \tilde{p}^2; 0, a \mu_{sea})$ 
and $Z^{\rm RI'}_{44}(\tilde{p}^2; a^2 \tilde{p}^2; 0, a \mu_{sea})$, 
taken at the valence chiral limit, as functions of $a^2 \mu_{sea}^2$, 
for a typical lattice momentum choice (see inset) giving $a^2 \tilde{p}^2 \sim 1.56$, for 
four $\beta$ values ($\beta=$3.80, 3.90, 4.05 and 4.20).}
\label{fig:SeaChiral}
\end{center}
\end{figure}

At fixed $\beta$ and $a^2 \tilde{p}^2$, we fit 
$Z^{\rm lat}_{i j}(\tilde{p}^2; a^2 \tilde{p}^2; 0, a\mu_{sea})$ 
data to a first order polynomial in $a^2 \mu_{sea}^2$. This choice 
is dictated by the expectation that effects of spontaneous chiral
symmetry, which may induce a dependence on $|\mu_{sea}|$, are
strongly suppressed, and in practice immaterial
within errors, in quantities like our RC-estimators that are evaluated
at momentum scales $\tilde{p}^2 \gg \Lambda_{QCD}^2$. In fact we find
that the dependence on the sea quark mass is hardly visible within our statistical error bars, as shown in 
Fig.~\ref{fig:SeaChiral}. Moreover, we have 
checked that repeating the whole analysis using a constant fit leads to similar RC results, 
though affected by smaller errors and often, but not always, yielding 
acceptable $\chi^2$'s. Hence, we conservatively decided to perform the sea chiral extrapolation 
using a linear fit in $a^2 \mu_{sea}^2$. We construct in this way the 
RC estimators $Z^{\rm RI'}_{i j}(\tilde{p}^2; a^2 \tilde{p}^2; 0, 0)$.

\subsection{Removal of O($a^2 g^2$) cutoff effects}
\label{sec:RemoveOa2}

We will obtain improved chiral limit RC estimators, $Z^{\rm RI'-impr}_{i j}(\tilde{p}^2; a^2 \tilde{p}^2)$, by removing from our
$Z^{\rm RI'}_{i j}(\tilde{p}^2; a^2 \tilde{p}^2; 0, 0)$ lattice data perturbative discretization errors. 
This can be done up to $O(a^2 g^2)$ exploiting the one-loop 
perturbative results obtained~\cite{Constantinou:2009tr,
Constantinou:2010zs} in the massless lattice theory for the
quark propagator form factor $\Sigma_{1}$, related to the quark-field RC by 
$Z_{q}(p) = \Sigma_{1}(p)$, and the dynamics matrix elements, {\em v.i.z.}  
\begin{eqnarray}
\hspace{-1.5cm}&&\left[\, Z_{q}(p)\, \right]^{LPT}\! =1 + \frac{g^2}{16 \pi^2} a^2 \!\left[ \tilde{p}^2 \left( c_{q}^{(1)} + c_{q}^{(2)} \log(a^2 \tilde{p}^{2})\right) + c_{q}^{(3)} \frac{\sum_{\rho} 
\tilde{p}_{\rho}^4}{\tilde{p}^2} \right]\!+\!{\rm O}(a^4 g^2, g^4)\nonumber\\ 
\hspace{-1.5cm}&&\left[\, D_{i j}(p)\, \right]^{LPT} =  1\, +\, \frac{g^2}{16 \pi^2}\, \left[\, b_{i j}^{(1)}\, +\,b_{i j}^{(2)}\, \log(a^2 \tilde{p}^{2})\, \right]\nonumber\\
\hspace{-1.5cm}&&\phantom{=}+\, \frac{g^2}{16 \pi^2}\, a^2\, \Big{[}\, \tilde{p}^2\, \left(\, c_{i j}^{(1)}\, +\, c_{i j}^{(2)}\, \log(a^2 \tilde{p}^2)\,\right)
+\, 
c_{i j}^{(3)}\, \frac{\sum_{\rho}\, \tilde{p}_{\rho}^4}{\tilde{p}^2}\,
\Big{]}\, +\, {\rm O}(a^4 g^2, g^4) \label{DIJ}
\end{eqnarray}
The values of the coefficients $c_{q}^{(k)}$, $k = 1, 2, 3$ can be found in Eq.~(34) of Ref.~\cite{Constantinou:2010gr}, 
while the values of the coefficients 
$b_{i j}^{(k)}$ and $c_{i j}^{(k)}$ are available in 
Refs.~\cite{Constantinou:2009tr,Constantinou:2010zs}. 

\begin{figure}[!ht]
\begin{center}
\mbox{}\vskip -3cm
\subfigure[]{\includegraphics[scale=0.19,angle=-0]{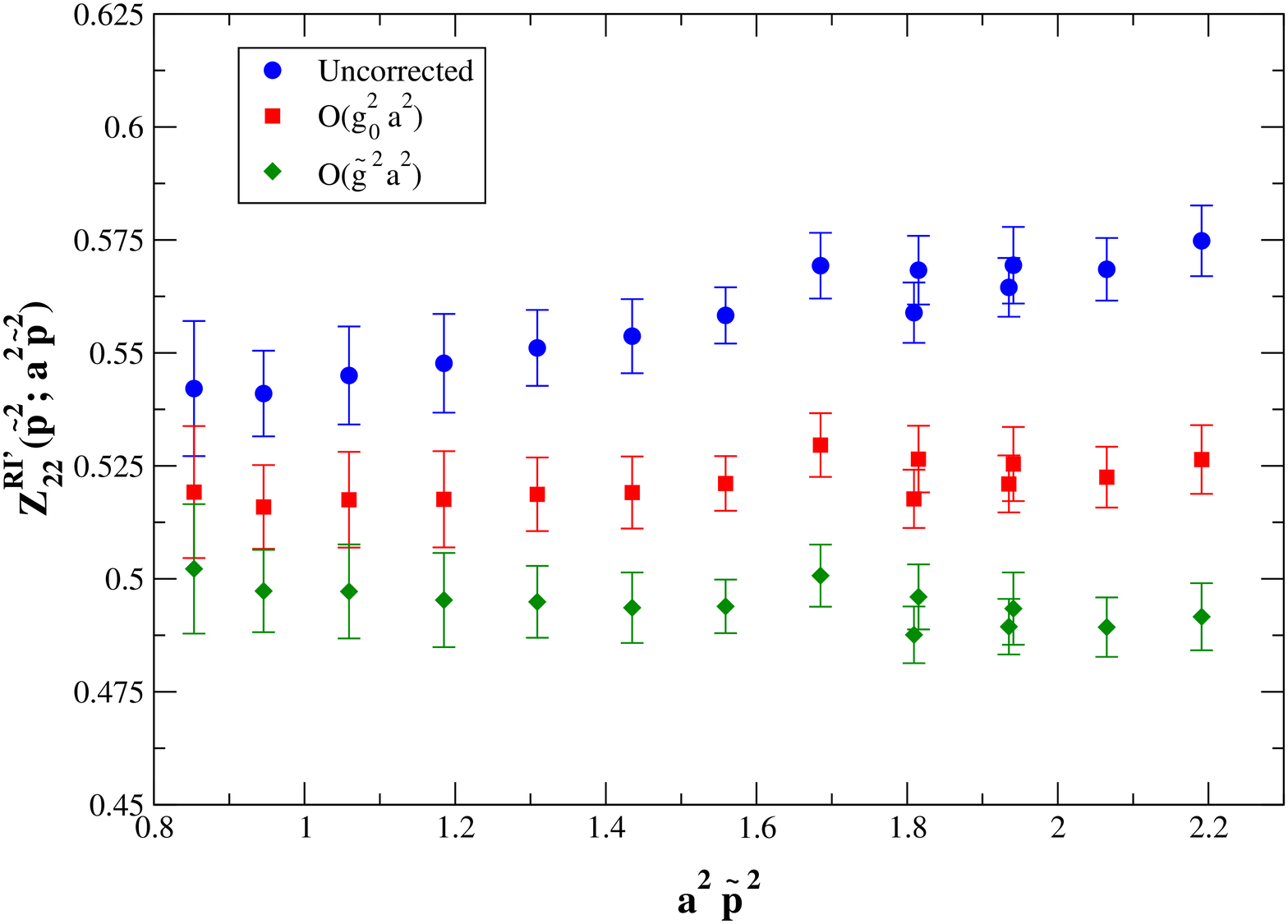}}\hspace*{0.15cm}
\subfigure[]{\includegraphics[scale=0.19,angle=-0]{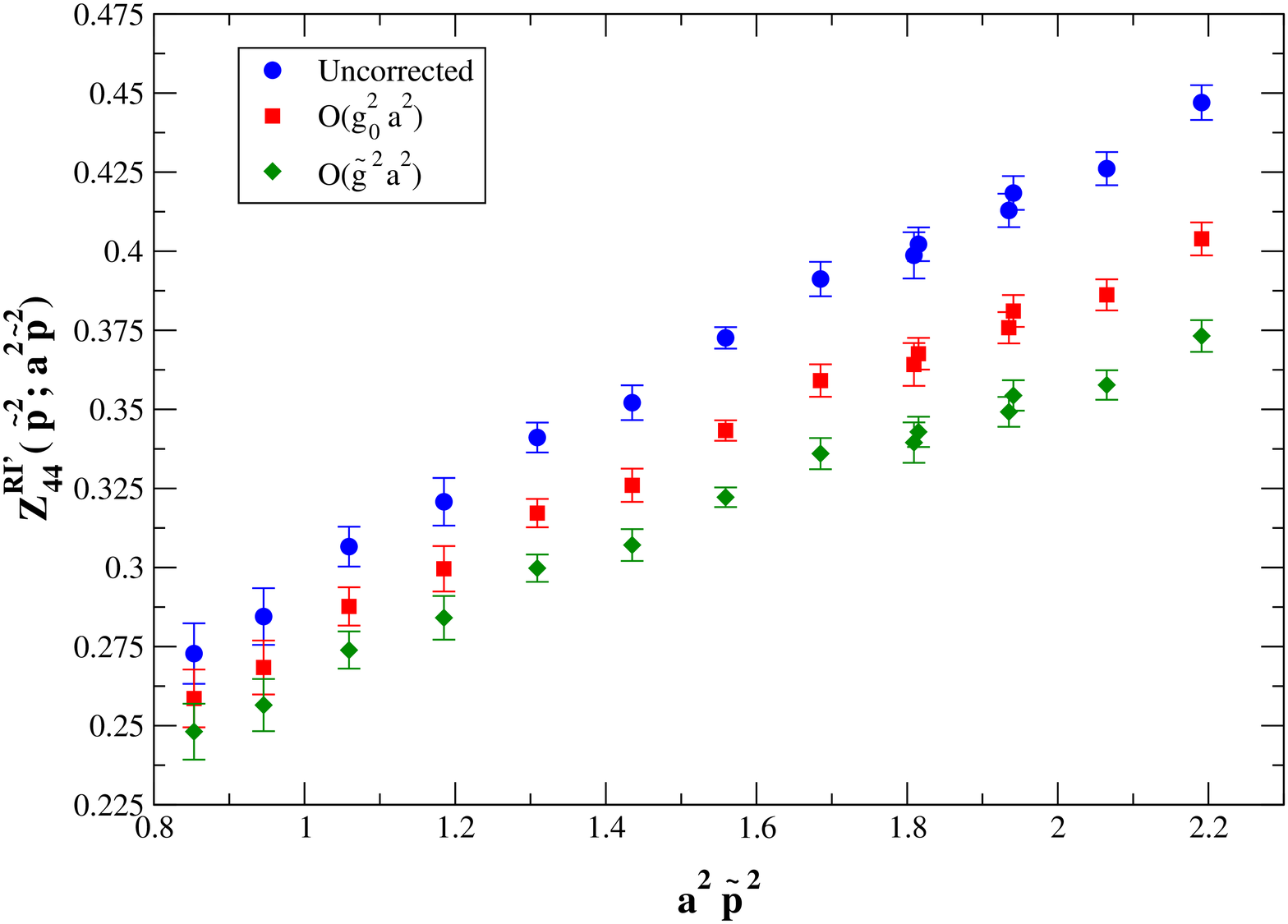}}
\end{center}
\vskip -1.0cm
\begin{center}
\caption{\sl The effect of subtracting from $Z^{\rm RI'}_{22}(\mu_0^2=a(\beta)^{-2}; a^2 \tilde{p}^2; 0, 0)$ 
and $Z^{\rm RI'}_{44}(\mu_0^2=a(\beta)^{-2}; a^2 \tilde{p}^2; 0, 0)$ at 
$\beta=3.8$ (blue dots) the O($a^2 g^2$) correction, setting either $g^2=g_{0}^2$ (red squares) 
or $g^2 = \tilde{g}^2$ (green diamonds).
}
\label{fig:oa2g2}
\end{center}
\end{figure}

In the numerical evaluation of the perturbative corrections, we take the coupling constant $g^2$ as the simple boosted coupling 
$\tilde{g}^2\, \equiv\, g_{0}^2/\langle P \rangle$. For the average plaquette $\langle P \rangle$ 
we employ the non-perturbative values 
$\left[0.5689, 0.5825, 0.6014, 0.6200 \right]$ 
corresponding to $\beta = \left[ 3.8, 3.9, 4.05, 4.20 \right]$, respectively. The important impact of the perturbative corrections in removing 
the unwanted $a^2 \tilde{p}^2$ dependence is illustrated, 
for the case of $\beta=3.8$, in Fig.~\ref{fig:oa2g2}. In this figure, the uncorrected values of 
$Z^{\rm RI'}_{i j}(\mu_0^2=a(\beta)^{-2}; a^2 \tilde{p}^2; 0, 0)$ 
are compared with the values of $Z^{\rm RI'-impr}_{i j}(\mu_0^2=a(\beta)^{-2}; a^2 \tilde{p}^2)$ obtained setting 
either $g^2\, =\, g_{0}^{2}\, =\, 6/\beta$ or (as we did in the end) $g^2\, =\, \tilde{g}^2$.

\subsection{Absence of wrong chirality mixings}
\label{sec:AbsenceWrongMixings}

\begin{figure}[!ht]
\begin{center}
\mbox{}\vskip -2.25cm
\includegraphics[scale=0.4,angle=-0]{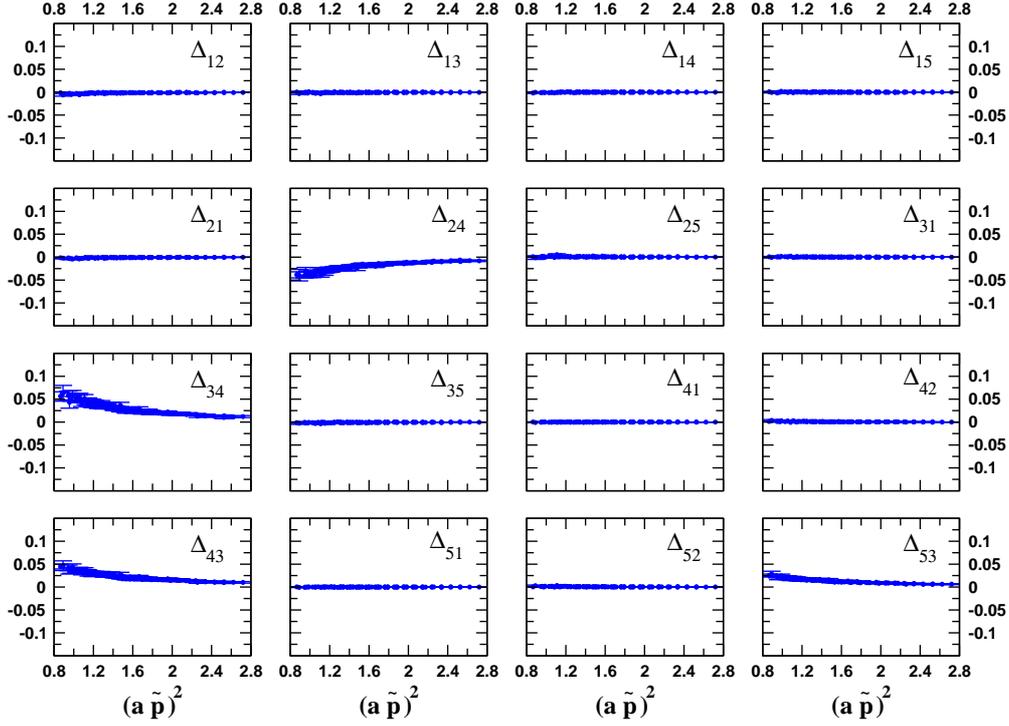}
\end{center}
\vskip -1.0cm
\begin{center}
\caption{\sl The behaviour of the mixing coefficients $\Delta_{i j}$, as a function of $a^2 \tilde{p}^2$ for $\beta=4.05$.}
\label{fig:Delta405}
\end{center}
\end{figure}

In Fig.~\ref{fig:Delta405}, one can clearly see that for all the operators of interest
the mixing coefficients $\Delta_{ij}$ are very small (in fact vanishing 
within errors in the range of 
$\tilde{p}^2$ that we eventually use for extracting RCs). We also find that this
is systematically more and more so as $\beta$ increases, well
in line with our expectation that in our lattice setup wrong chirality mixing 
effects are reduced to mere O($a^2$) artifacts. For these reasons the effects
of $\pmb{\Delta}$ have been neglected in our final RC analysis, where we have
assumed a fully continuum-like relation between renormalized and bare operators.
In addition we checked that repeating the whole analysis with the tiny effects of
$\pmb{\Delta}$ on the relation~(\ref{Qrenpatt}) properly taken into account
leads to no significant changes in the values of RCs.

\subsection{Final RC estimates from M1 and M2 method}
\label{sec:FinalEstimates}

Having extrapolated the (improved) RCs estimators to the valence and sea chiral limit 
at each value of the momenta, we evolve $Z^{\rm RI'-impr}_{i j}(\tilde{p}^{2};\, a^2
\tilde{p}^2)$ from the scale $\tilde{p}^2$ to a common scale $\mu_{0}^2$ by using the
known matrix formula for the NLO running of the operators $Q_{i}$~\cite{rm1:4ferm-nlo,mu:4ferm-nlo}, obtaining $Z^{\rm RI'-impr}_{i j}(\mu_{0}^{2};\, a^2 \tilde{p}^2)$.
This step is necessary in order to disentangle the O($a^2 \tilde{p}^2$) cutoff effects 
from the genuine continuum $p^2$ dependence. Notice also that the actual value of $\mu_{0}$ has no impact on the RGI results of the RC's. As is customary, 
 we take $\mu_{0}\, =\, a^{-1}(\beta)$ for each $\beta$, with 
$a^{-1}(3.8, 3.9, 4.05, 4.20)\,=\, [\, 2.0,\, 2.3,\, 3.0,\,  3.7\,]$ GeV. 

Of course we still allow for a residual dependence on $a^2\tilde{p}^2$. In order 
to deal with these cutoff effects, following Ref.~\cite{Constantinou:2010gr}, 
we use two methods.
Method M1 consists in fitting $Z^{\rm RI'-impr}_{i j}(\mu_{0}^{2};\, a^2 \tilde{p}^2)$ 
to the linear ansatz
\be
\label{MethodM1}
Z^{\rm RI'-impr}_{i j}(\mu_{0}^{2};\, a^2 \tilde{p}^2)\, =\, Z^{\rm RI'-impr}_{i j}(\mu_{0}^{2})\, +\, \lambda_{i j}\; ( a \tilde{p})^2
\ee
in the large momentum region, $1.0 \le a^2 \tilde{p}^2 \le 2.2$. 
As expected, the slopes $\lambda_{i j}$  depend smoothly on $\beta$.

According to the ansatz~(\ref{MethodM1}), with $\lambda_{ij}=\lambda_{ij}^{(0)}+\lambda_{ij}^{(1)}\tilde{g}^2$ 
($\tilde{g}^2$ is the boosted gauge coupling as in section B.4) a  linear extrapolation to $a^2 \tilde{p}^2 = 0$ 
 was performed simultaneously at all $\beta$'s.  The extrapolated values, 
$Z^{\rm RI'-impr}_{i j}(\mu_{0}^{2})$, are finally used to evaluate via the NLO running matrix formula of 
Ref.~\cite{mu:4ferm-nlo}, the quantities 
$Z_{i j}^{\overline{\rm MS}}(M1)$ and $Z_{i j}^{\rm RGI}(M1)$. Therefore, the $\overline{\rm MS}$ 
scheme we use here is the one defined by Buras {\it et al.} in 
Ref.~\cite{mu:4ferm-nlo}. This definition of the $\overline{\rm MS}$ scheme, which has become standard, 
differs from the one of Ref.~\cite{rm1:4ferm-nlo} proposed by Ciuchini {\it et al.} 
in the treatment of the four-fermion evanescent operators appearing in the calculation of the two-loop anomalous dimensions.

\begin{figure}[!hb]
\mbox{}\vskip -3cm
\subfigure{\includegraphics[scale=0.52,angle=-0]{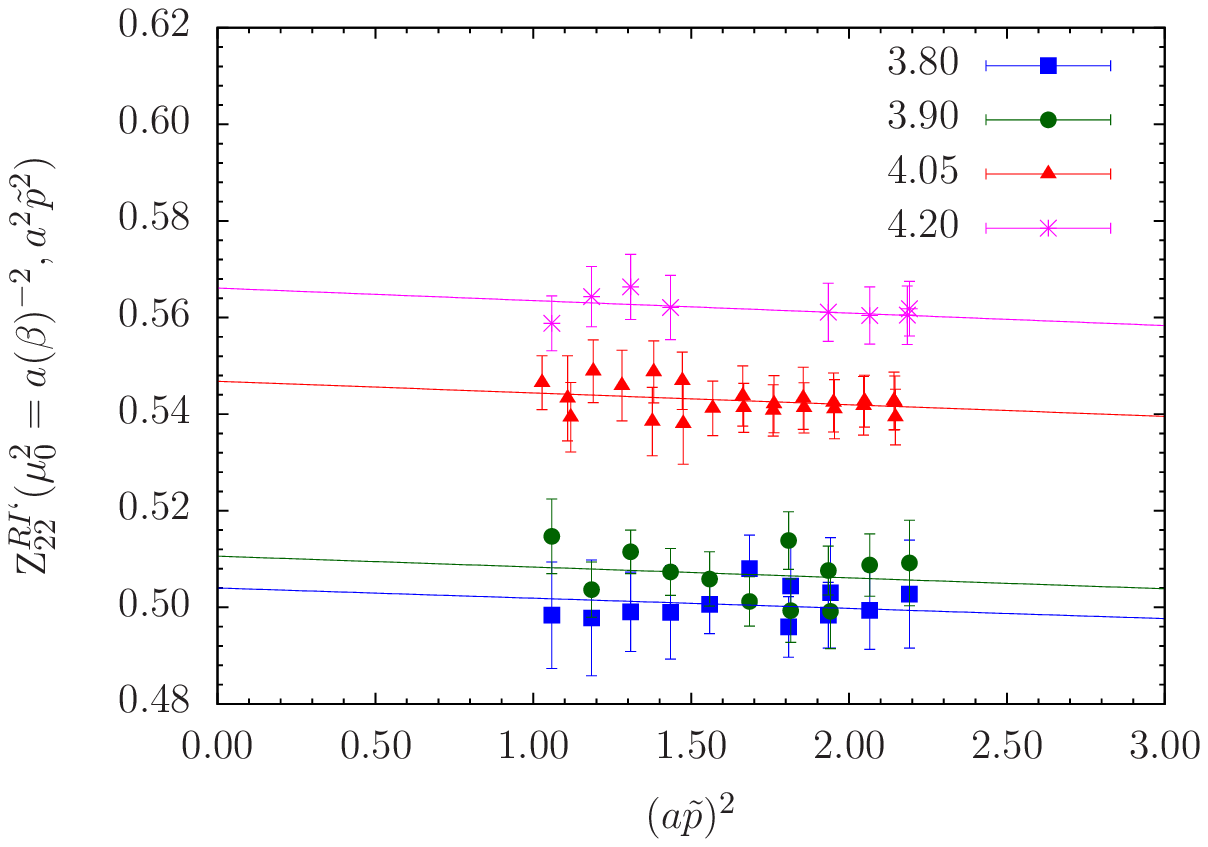}}\hspace*{0.15cm}
\subfigure{\includegraphics[scale=0.52,angle=-0]{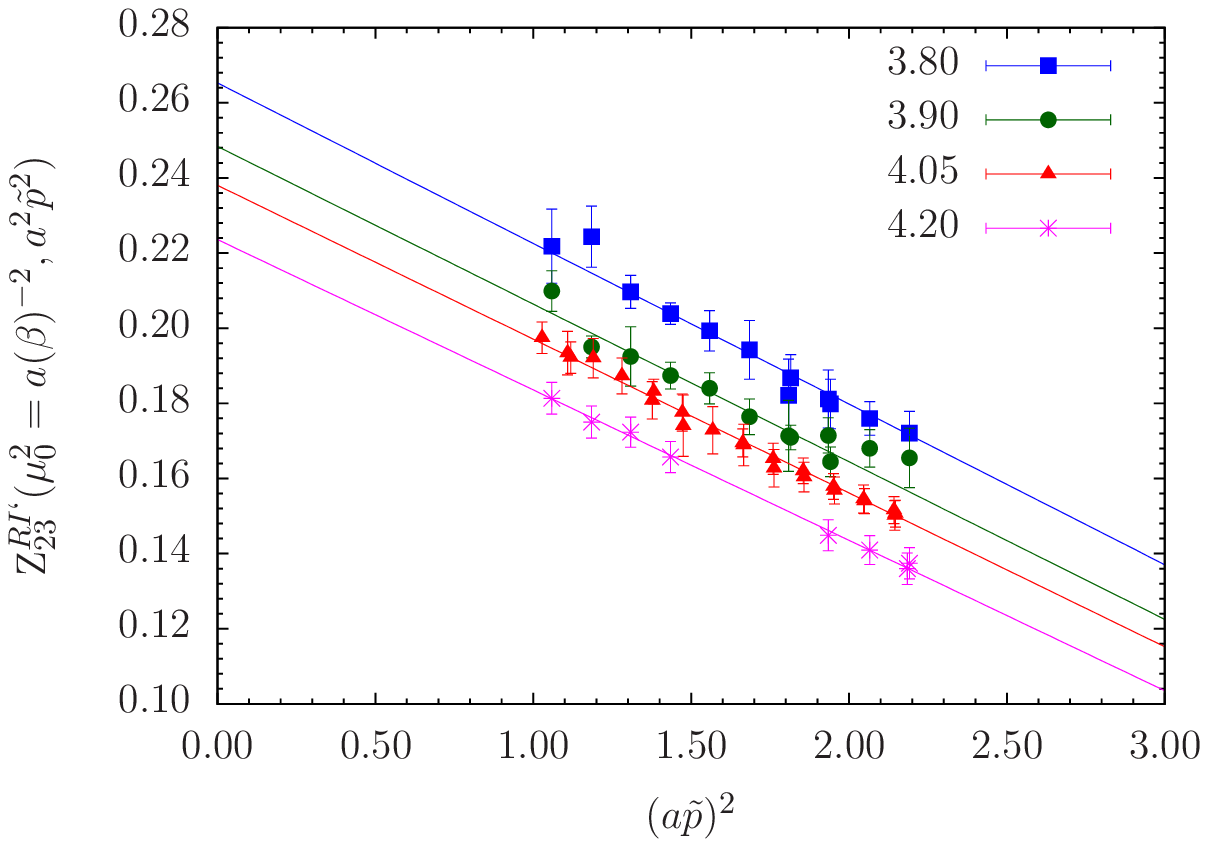}}\\[0.75cm]
\subfigure{\includegraphics[scale=0.52,angle=-0]{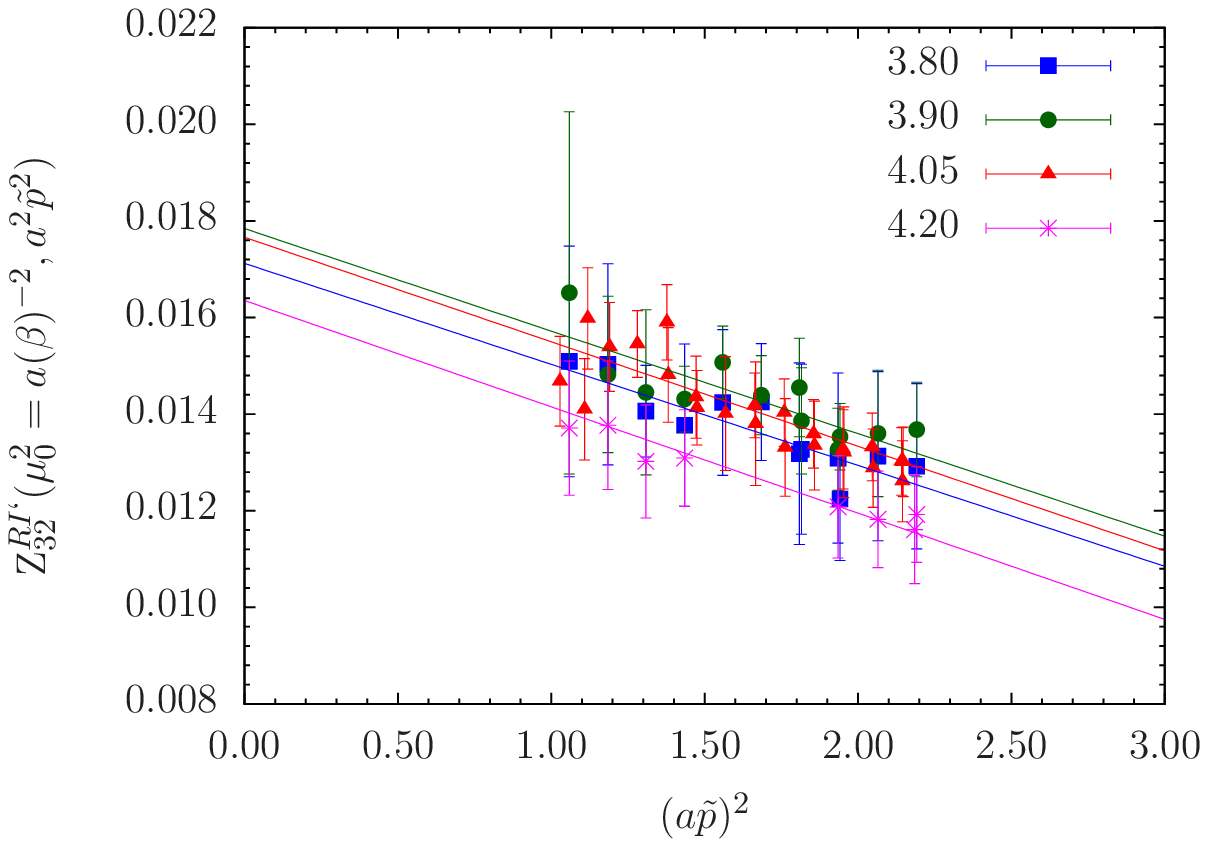}}\hspace*{0.15cm}
\subfigure{\includegraphics[scale=0.52,angle=-0]{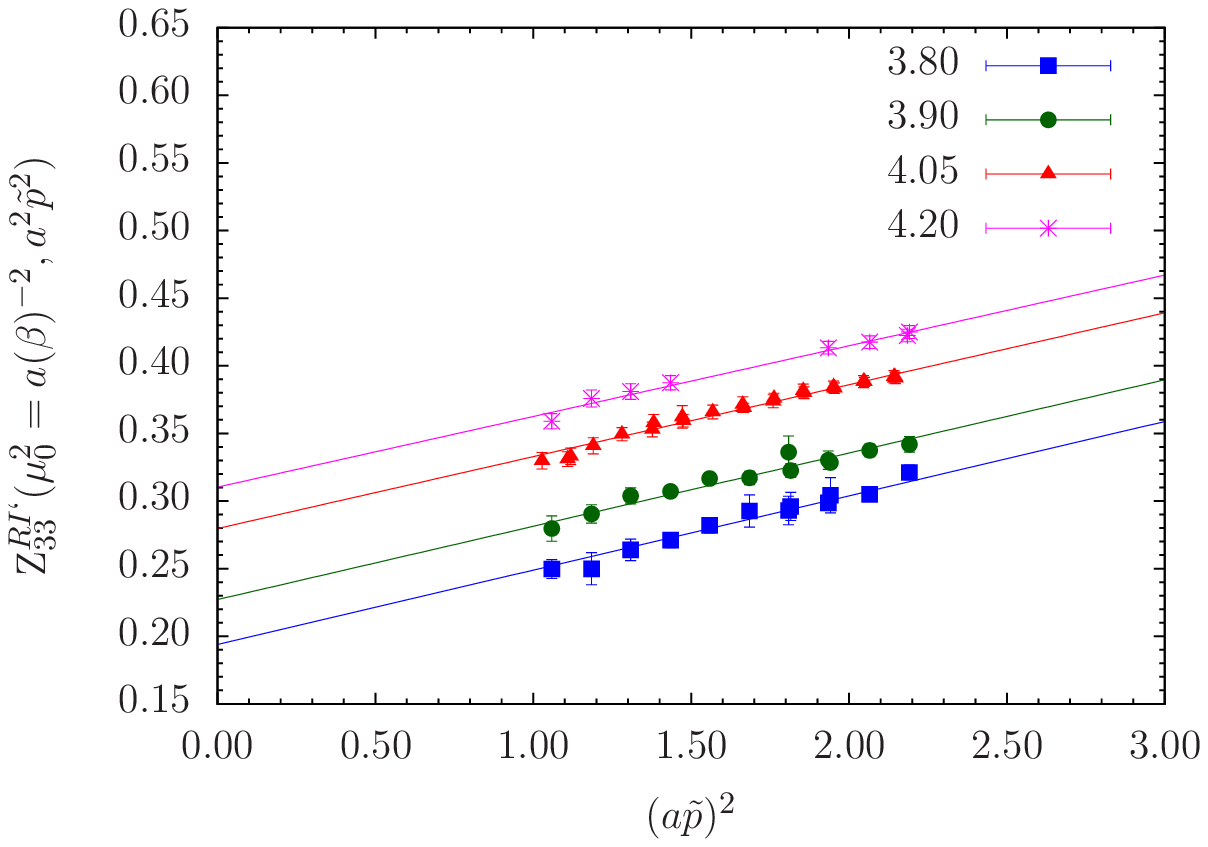}}\\[0.75cm]
\subfigure{\includegraphics[scale=0.52,angle=-0]{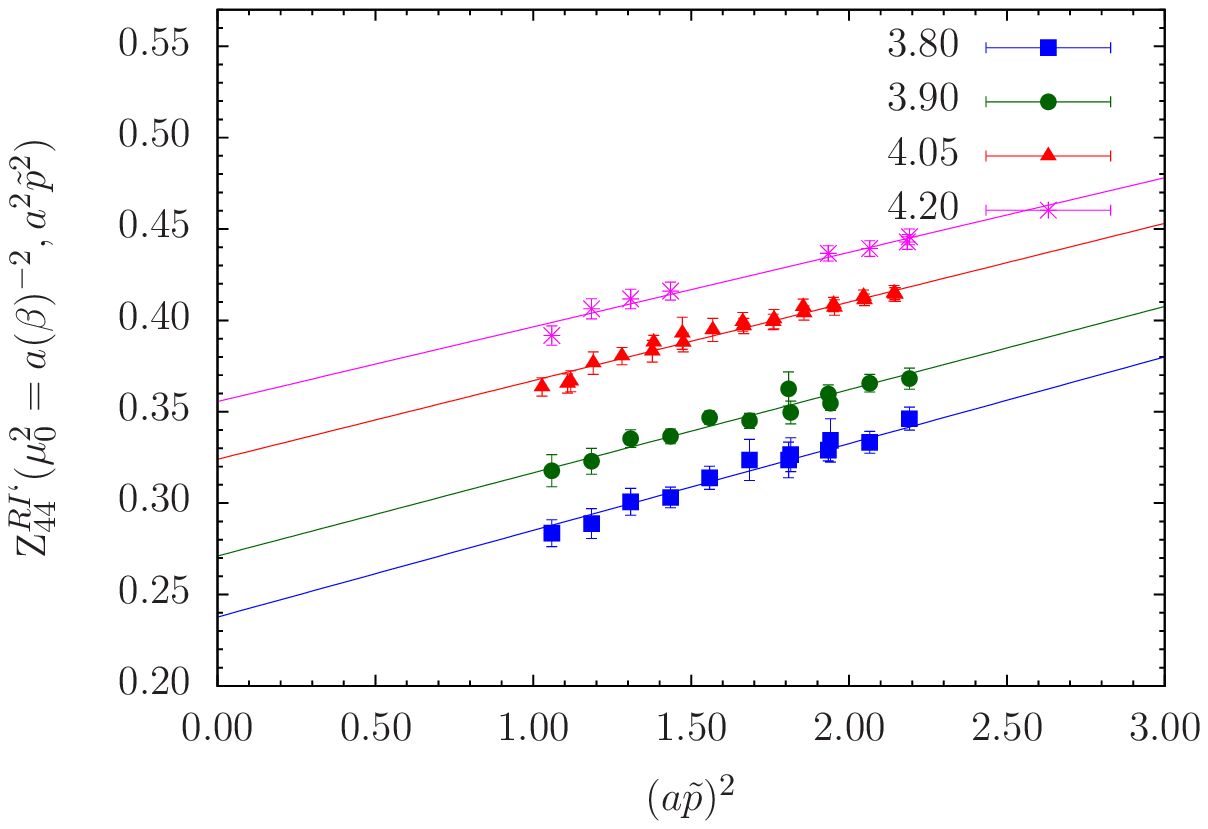}}\hspace*{0.15cm}
\subfigure{\includegraphics[scale=0.52,angle=-0]{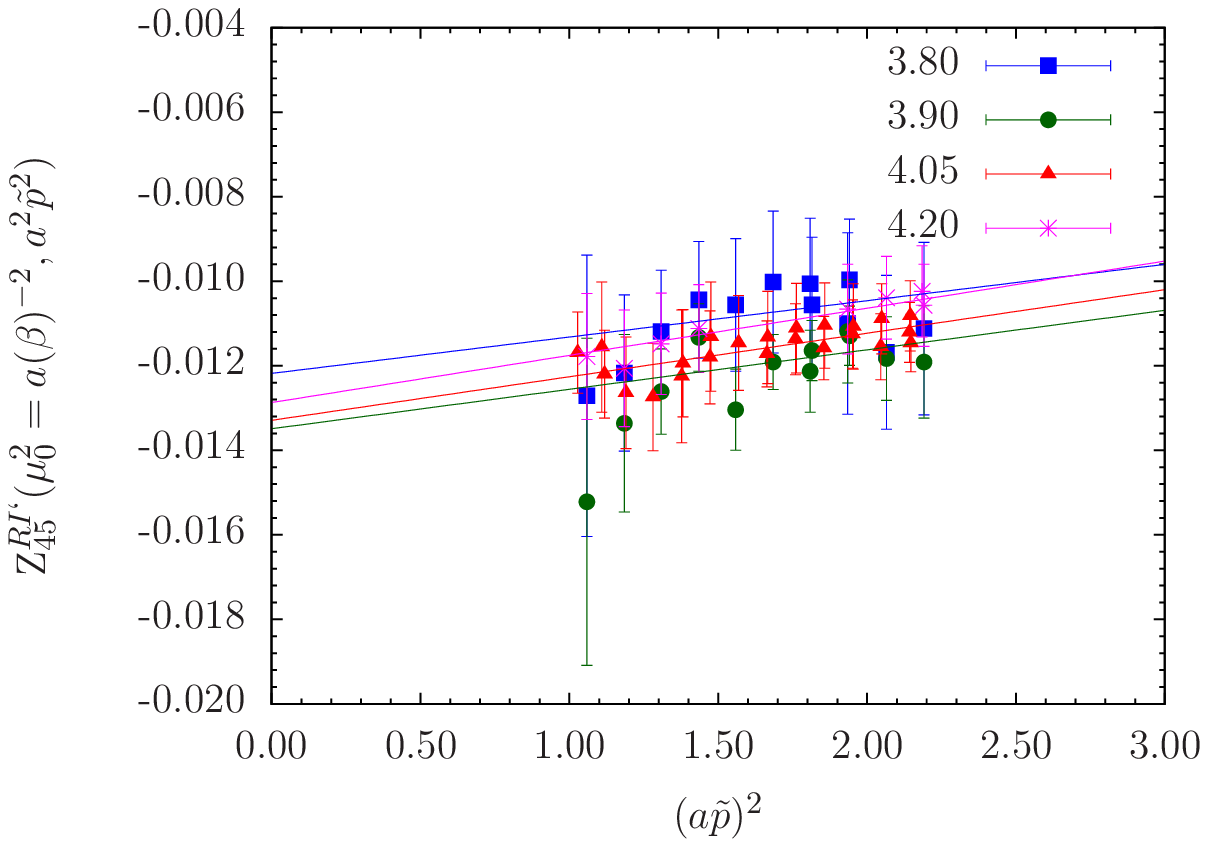}}\\[0.75cm]
\subfigure{\includegraphics[scale=0.52,angle=-0]{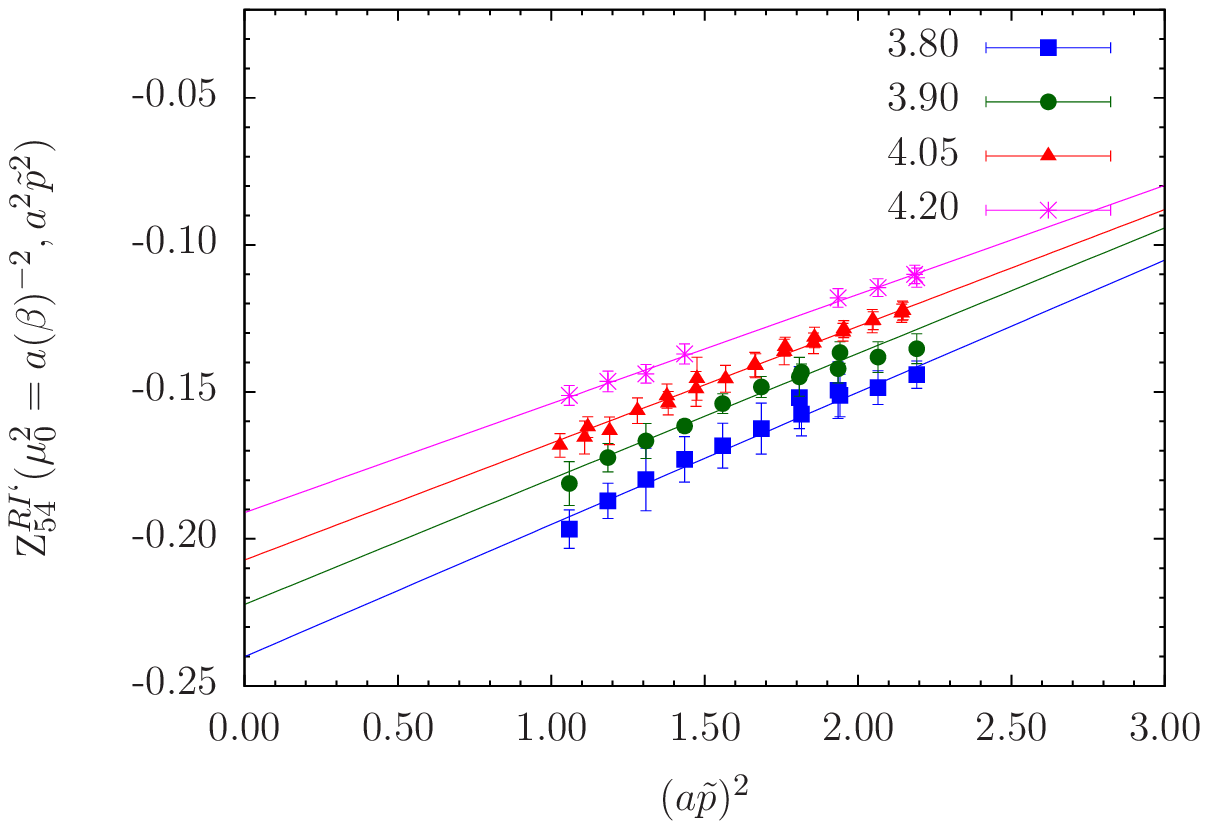}}\hspace*{0.15cm}
\subfigure{\includegraphics[scale=0.52,angle=-0]{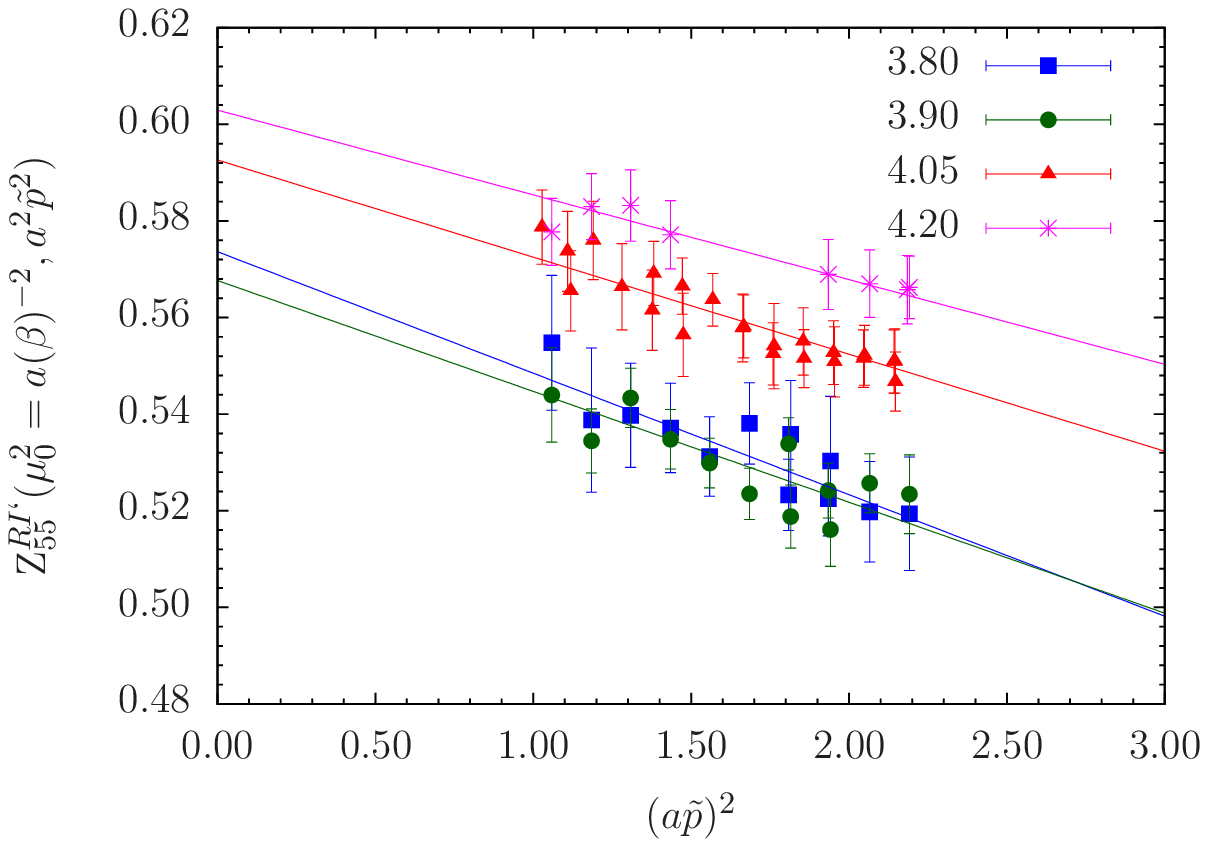}}
\vskip -1.0cm
\begin{center}
\caption{\sl $Z^{\rm RI'-impr}_{ij}(\mu_{0}^2=a(\beta)^{-2}; a^{2}\, \tilde{p}^2)$ 
for $\{ij\}=\{22, 23, 32, 33, 44, 45, 54, 55 \}$ as functions 
of $a^{2}\, \tilde{p}^2$ for the four $\beta$ values considered in our study. The 
straight lines represent the simultaneous linear fit to the  lattice data in 
the interval $1.0 \le a^2 \tilde{p}^2 \le 2.2$ at the four $\beta$'s.}
\label{fig:MethodM1}
\end{center}
\end{figure}

In Fig.~\ref{fig:MethodM1} the simultaneous best linear fits in $a^2 \tilde{p}^2$ 
at our four $\beta$'s of $Z_{ij}$ are shown. We recall that both in the analysis 
and in the figures of this Appendix, only data points corresponding to 
the momenta $\tilde{p}$ satisfying the constraint~(\ref{cut28}) are used and shown. 

The idea of the M2 method is instead to separately average at each $\beta$ the values 
of $Z^{\rm RI'-impr}_{i j}(\mu_{0}^{2};\, a^2 \tilde{p}^2)$ over 
 a narrow interval of momenta (ideally just one point), which has 
to be kept fixed in physical units for all $\beta$'s. We have chosen this
interval to be $\tilde{p}^2\in \left[\, 8.0,\, 9.5\, \right]$~GeV$^{2}$.
In this way, at the price of giving up the reduction of cutoff effects
implied by the M1 method, no assumptions are introduced in the RC analysis
about the detailed form of lattice artifacts and/or the adequacy of NLO
anomalous dimensions to describe the RC-evolution at scales below 
$\tilde{p}^2 \sim 9$~GeV$^{2}$. 

The $a^2$-scaling of renormalized quantities (in this work operator matrix 
elements) constructed using RCs determined with the M2 method will of course
be in general different from the one of their M1 method counterparts, but 
the continuum limit results for these quantities, if attainable from both methods 
with controlled errors, should be consistent with each other (see e.g.\ Appendix~E).

\clearpage
\vspace{1.2cm}

\section{  Renormalization Constant results}
\label{App_RCS}

In Tables~\ref{ZPZS_M1} and \ref{ZPZS_M2} we collect values of $Z_P$ and $Z_S$ calculated in the  RI-MOM scheme. 
Results are obtained with methods M1 and M2~\cite{Constantinou:2010gr}  at each value of the gauge 
coupling in $\overline {\rm{MS}}$ and RI-MOM at 2 GeV. We have used the three-loop conversion formula from RI-MOM 
to $\overline{\rm{MS}}$~\cite{Chetyrkin:1999pq}.

\begin{table}[!h]
\begin{center}
\begin{tabular}{|c|l|l|l|l|}
\hline 
RC(M1) & $\beta=3.80$ & $\beta=3.90$ & $\beta=4.05$ & $\beta=4.20$\tabularnewline
\hline
\hline 
\multicolumn{1}{|c}{} & \multicolumn{4}{c|}{$\overline {\rm{MS}}$ at 2 GeV }\tabularnewline
\hline 
$Z_{P}$ & 0.413(12) & 0.437(7)  & 0.477(6) & 0.498(5)\tabularnewline
\hline 
$Z_{S}$ & 0.728(16) & 0.712(10) & 0.702(5) & 0.694(8)\tabularnewline
\hline
\hline 
\multicolumn{5}{|c|}{\hspace*{1cm} RI-MOM at 2 GeV}\tabularnewline
\hline 
$Z_{P}$ & 0.339(9)  & 0.359(6) & 0.391(4) & 0.409(4)\tabularnewline
\hline 
$Z_{S}$ & 0.598(13) & 0.585(9) & 0.576(4) & 0.570(7)\tabularnewline
\hline
\end{tabular}
\caption{$Z_P$ and $Z_S$ results, using the M1 method  at  $\beta=3.80,\, 3.90$, 4.05 and 4.20 
in  $\overline{\rm{MS}}$ and RI-MOM at 2 GeV.}
\label{ZPZS_M1}
\end{center}
\end{table}

\begin{table}[!h]
\begin{center}
\begin{tabular}{|c|l|l|l|l|}
\hline 
RC(M2) & $\beta=3.80$ & $\beta=3.90$ & $\beta=4.05$ & $\beta=4.20$\tabularnewline
\hline
\hline 
\multicolumn{1}{|c}{} & \multicolumn{4}{c|}{$\overline {\rm{MS}}$ at 2 GeV }\tabularnewline
\hline 
$Z_{P}$ & 0.532(5) & 0.518(6) & 0.520(4) & 0.503(5)\tabularnewline
\hline 
$Z_{S}$ & 0.813(7) & 0.776(6) & 0.735(4) & 0.708(10)\tabularnewline
\hline
\hline 
\multicolumn{5}{|c|}{\hspace*{1cm} RI-MOM at 2 GeV}\tabularnewline
\hline 
$Z_{P}$ & 0.437(4) & 0.426(5) & 0.427(4) & 0.413(4)\tabularnewline
\hline 
$Z_{S}$ & 0.668(6) & 0.637(5) & 0.603(4) & 0.582(8)\tabularnewline
\hline
\end{tabular}
\caption{$Z_P$ and $Z_S$ results, using the M2 method, at $\beta=$3.80, 3.90,  4.05 and 4.20 
in  $\overline{\rm{MS}}$ and RI-MOM at 2 GeV.}
\label{ZPZS_M2}
\end{center}
\end{table}

The  RC matrices of the four-fermion operators $Z_Q$, (c.f. Eq.~(\ref{ZQ})),  
 are listed below. 
We present results obtained from both  M1 and M2 methods, in  
$\overline {\rm{MS}}$ and RI-MOM at 2 GeV.
\clearpage

\noindent {\bf  ($\mathbf{\overline{MS}}$, 2 GeV)}: 

$$
Z_Q(\beta=3.80; M1)=
\left( \begin{array}{lllcc}
0.425(15) & 0 & 0 & 0 & 0 \\
0 & 0.492(13) & 0.238(7) & 0 & 0 \\
0 & 0.022(02) & 0.227(10) & 0 & 0 \\
0 & 0 & 0 & ~~~0.257(9) & -0.006(2) \\
0 & 0 & 0 & -0.246(8) & \,\,\,\,0.600(14) \\
\end{array} \right)
$$

$$
Z_Q(\beta=3.90; M1)=
\left( \begin{array}{lllcc}
0.441(8) & 0 & 0 & 0 & 0 \\
0 & 0.503(9) & 0.231(4) & 0 & 0 \\
0 & 0.023(1) & 0.250(6) & 0 & 0 \\
0 & 0 & 0 & ~~~0.282(6) & -0.006(2) \\
0 & 0 & 0 & -0.244(5) & \,\,\,\,0.617(11) \\
\end{array} \right)
$$

$$
Z_Q(\beta=4.05; M1)=
\left( \begin{array}{lllcc}
0.491(5) & 0 & 0 & 0 & 0 \\
0 & 0.546(6) & 0.240(5) & 0 & 0 \\
0 & 0.023(1) & 0.281(5) & 0 & 0 \\
0 & 0 & 0 & ~~~0.319(4) & -0.004(1) \\
0 & 0 & 0 & -0.258(5) & \,\,\,0.692(8) \\
\end{array} \right)
$$

$$
Z_Q(\beta=4.20; M1)=
\left( \begin{array}{lllcc}
 0.523(10)& 0 & 0 & 0 & 0 \\
 0& 0.571(9) & 0.243(6) & 0 & 0 \\
0 & 0.021(1) & 0.292(8) & 0 & 0 \\
0 & 0 & 0 & ~~~0.336(7) & -0.002(2) \\
0 & 0 & 0 & -0.265(6) & \,\,\,\,0.727(10) \\
\end{array} \right)
$$

\clearpage
\noindent {\bf  ($\mathbf{\overline{MS}}$, 2 GeV)}:

$$
Z_Q(\beta=3.80; M2)=
\left( \begin{array}{lllcc}
0.440(9) & 0 & 0 & 0 & 0 \\
0 & 0.489(9) & 0.136(5) & 0 & 0 \\
0 & 0.017(2) & 0.362(5) & 0 & 0 \\
0 & 0 & 0 & ~~~0.367(5) &  -0.005(2) \\
0 & 0 & 0 & -0.148(4) & \,\,\,0.547(9) \\
\end{array} \right)
$$

$$
Z_Q(\beta=3.90; M2)=
\left( \begin{array}{lllcc}
0.447(5) & 0 & 0 & 0 & 0 \\
0 & 0.496(5) & 0.158(4) & 0 & 0 \\
0 & 0.019(1) & 0.348(5) & 0 & 0 \\
0 & 0 & 0 & ~~~0.361(4) & -0.005(1) \\
0 & 0 & 0 & -0.169(3) & \,\,\,0.576(5) \\
\end{array} \right)
$$

$$
Z_Q(\beta=4.05; M2)=
\left( \begin{array}{lllcc}
0.495(5) & 0 & 0 & 0 & 0 \\
0 & 0.543(6) & 0.197(4) & 0 & 0 \\
0 & 0.020(1) & 0.333(6) & 0 & 0 \\
0 & 0 & 0 & ~~~0.360(3) & -0.003(1) \\
0 & 0 & 0 & -0.214(4) & \,\,\,0.663(8) \\
\end{array} \right)
$$

$$
Z_Q(\beta=4.20; M2)=
\left( \begin{array}{lllcc}
 0.531(6)& 0 & 0 & 0 & 0 \\
 0& 0.579(6) & 0.239(3) & 0 & 0 \\
0 & 0.019(1) & 0.299(5) & 0 & 0 \\
0 & 0 & 0 & ~~~0.337(5) & -0.000(1) \\
0 & 0 & 0 & -0.259(4) & \,\,\,0.733(8) \\
\end{array} \right)
$$

\clearpage
\noindent {\bf (RI-MOM, 2 GeV)}: 

$$
Z_Q(\beta=3.80; M1)=
\left( \begin{array}{lllcc}
0.419(15) & 0 & 0 & 0 & 0 \\
0 & 0.504(13) & 0.265(7) & 0 & 0 \\
0 & 0.017(2) & 0.195(9) & 0 & 0 \\
0 & 0 & 0 & ~~~0.238(8) & -0.013(2) \\
0 & 0 & 0 & -0.240(8) & \,\,\,\,0.574(14) \\
\end{array} \right)
$$

$$
Z_Q(\beta=3.90; M1)=
\left( \begin{array}{lllcc}
0.434(8) & 0 & 0 & 0 & 0 \\
0 & 0.515(8) & 0.260(5) & 0 & 0 \\
0 & 0.018(1) & 0.215(5) & 0 & 0 \\
0 & 0 & 0 & ~~~0.261(6) & -0.013(1) \\
0 & 0 & 0 & -0.239(5) & \,\,\,\,0.589(11) \\
\end{array} \right)
$$

$$
Z_Q(\beta=4.05; M1)=
\left( \begin{array}{lllcc}
0.483(5) & 0 & 0 & 0 & 0 \\
0 & 0.559(6) & 0.273(5) & 0 & 0 \\
0 & 0.017(1) & 0.242(4) & 0 & 0 \\
0 & 0 & 0 & ~~~0.295(3) & -0.012(1) \\
0 & 0 & 0 & -0.253(5) & \,\,\,\,0.659(8) \\
\end{array} \right)
$$

$$
Z_Q(\beta=4.20; M1)=
\left( \begin{array}{lllcc}
 0.515(10)& 0 & 0 & 0 & 0 \\
 0& 0.585(9) & 0.278(7) & 0 & 0 \\
0 & 0.016(1) & 0.251(7) & 0 & 0 \\
0 & 0 & 0 & ~~~0.311(7) & -0.011(1) \\
0 & 0 & 0 & -0.260(6) & \,\,\,\,0.694(10) \\
\end{array} \right)
$$

\clearpage
\noindent {\bf (RI-MOM, 2 GeV)}: 

$$
Z_Q(\beta=3.80; M2)=
\left( \begin{array}{lllcc}
0.433(8) & 0 & 0 & 0 & 0 \\
0 & 0.501(9) & 0.175(5) & 0 & 0 \\
0 & 0.013(2) & 0.311(5) & 0 & 0 \\
0 & 0 & 0 & ~~~0.338(4) &  -0.011(2) \\
0 & 0 & 0 & -0.149(4) & \,\,\,0.522(9) \\
\end{array} \right)
$$

$$
Z_Q(\beta=3.90; M2)=
\left( \begin{array}{lllcc}
0.441(4) & 0 & 0 & 0 & 0 \\
0 & 0.508(5) & 0.196(4) & 0 & 0 \\
0 & 0.015(1) & 0.299(4) & 0 & 0 \\
0 & 0 & 0 & ~~~0.332(3) & -0.012(1) \\
0 & 0 & 0 & -0.169(3) & \,\,\,0.550(5) \\
\end{array} \right)
$$

$$
Z_Q(\beta=4.05; M2)=
\left( \begin{array}{lllcc}
0.487(5) & 0 & 0 & 0 & 0 \\
0 & 0.556(6) & 0.234(4) & 0 & 0 \\
0 & 0.015(1) & 0.287(5) & 0 & 0 \\
0 & 0 & 0 & ~~~0.331(3) & -0.011(1) \\
0 & 0 & 0 & -0.212(4) & \,\,\,0.632(8) \\
\end{array} \right)
$$

$$
Z_Q(\beta=4.20; M2)=
\left( \begin{array}{lllcc}
 0.523(6)& 0 & 0 & 0 & 0 \\
 0& 0.593(6) & 0.274(4) & 0 & 0 \\
0 & 0.014(1) & 0.257(4) & 0 & 0 \\
0 & 0 & 0 & ~~~0.308(4) & -0.009(1) \\
0 & 0 & 0 & -0.254(4) & \,\,\,0.700(7) \\
\end{array} \right)
$$

\clearpage
\vspace{1.2cm}
\section{Lattice data on masses and matrix elements}
\label{APP_RESULTS}
In the following Tables we gather our bare results at all  values of $\beta$ and combinations of 
quark masses for (i) pseudoscalar meson masses and pseudoscalar meson decay constants in lattice units 
(Tables~\ref{MpsFps_b380}, \ref{MpsFps_b390}, \ref{MpsFps_b405} and \ref{MpsFps_b420}); (ii) the ratio of the (bare) four-fermion operators  
 $R_i^{(b)} = \langle \bar{K}^0 | O_i | K^0 \rangle / \langle \bar{K}^0 | O_1 | K^0 \rangle$ ($i=2, \ldots, 5$) 
(see Tables~\ref{Rbare_b380}, \ref{Rbare_b390}, \ref{Rbare_b405} and \ref{Rbare_b420});  (iii) 
the quantities $\xi_i\, B_i^{(0)}$ ($i=2, \ldots, 5$) (see Tables~\ref{Bbare_b380}, \ref{Bbare_b390},
\ref{Bbare_b405} and \ref{Bbare_b420}).\vskip 1cm

\begin{table}[!h]
\begin{center}
\begin{tabular}{|c|c|l|l|l|l|}
\hline 
\multicolumn{6}{|c|}{\textbf{$\beta=3.80$ $(24^{3}\times48)a^{4}$}}\tabularnewline
\hline
\hline 
$a\mu_{\ell}=a\mu_{sea}$ & $a\mu_{``s"}$ & $aM^{34}$ & $aM^{12}$ & $aF^{34}$ & $aF^{12}$\tabularnewline
\hline
\hline 
 & 0.0165 & 0.2558(8)  & 0.3393(25)  & 0.0894(4) & 0.0883(15)\tabularnewline
\cline{2-6} 
0.0080 & 0.0200 & 0.2731(7)  & 0.3532(23)  & 0.0913(4) & 0.0895(15)\tabularnewline
\cline{2-6} 
 & 0.0250 & 0.2961(7) & 0.3718(21) & 0.0936(4) &  0.0909(15)\tabularnewline
\hline
\hline 
 & 0.0165 & 0.2712(4)  & 0.3508(16)  & 0.0924(3) & 0.0900(16)\tabularnewline
\cline{2-6} 
0.0110 & 0.0200 & 0.2877(4)  & 0.3644(14)  & 0.0942(3) & 0.0910(16)\tabularnewline
\cline{2-6} 
 & 0.0250 & 0.3098(4)  & 0.3828(12)  & 0.0966(3)  & 0.0924(16)\tabularnewline
\hline
\end{tabular}
\caption{ Pseudoscalar masses and decay constants at $\beta=3.80$.}
\label{MpsFps_b380}
\end{center}
\end{table}

\begin{table}[!h]
\begin{center}
\begin{tabular}{|c|c|l|l|l|l|}
\hline 
\multicolumn{6}{|c|}{\textbf{$\beta=3.90$ $(24^{3}\times48)a^{4}$}}\tabularnewline
\hline
\hline 
$a\mu_{\ell}=a\mu_{sea}$ & $a\mu_{``s"}$ & $aM^{34}$ & $aM^{12}$ & $aF^{34}$ & $aF^{12}$\tabularnewline
\hline
\hline 
 & 0.0150 & 0.2060(5)  & 0.2639(11)  & 0.0724(3)  & 0.0705(9)\tabularnewline
\cline{2-6} 
0.0040 & 0.0220 & 0.2401(5)  & 0.2915(11)  & 0.0757(3)  & 0.0725(9)\tabularnewline
\cline{2-6} 
 & 0.0270 & 0.2619(5)  & 0.3096(11)  & 0.0777(3) &  0.0759(9)\tabularnewline
\hline
\hline 
 & 0.0150 & 0.2179(8)  & 0.2762(16)  & 0.0755(5) &  0.0736(10)\tabularnewline
\cline{2-6} 
0.0064 & 0.0220 & 0.2506(7)  & 0.3028(15)  & 0.0785(5)  & 0.0759(9)\tabularnewline
\cline{2-6} 
 & 0.0270 & 0.2717(7) & 0.3204(14)  & 0.0805(4)  & 0.0774(9)\tabularnewline
\hline
\hline 
 & 0.0150 & 0.2283(7)  & 0.2849(17) & 0.0773(3)  & 0.0755(9)\tabularnewline
\cline{2-6} 
0.0085 & 0.0220 & 0.2598(7)  & 0.3109(15)  & 0.0804(3)  & 0.0779(8)\tabularnewline
\cline{2-6} 
 & 0.0270 & 0.2803(7)  & 0.3281(15)  & 0.0823(3) &  0.0794(9)\tabularnewline
\hline
\hline 
 & 0.0150 & 0.2351(7)  & 0.2892(14)  & 0.0784(4)  & 0.0761(9)\tabularnewline
\cline{2-6} 
0.0100 & 0.0220 & 0.2659(7)  & 0.3154(12)  & 0.0815(4)  & 0.0787(8)\tabularnewline
\cline{2-6} 
 & 0.0270 & 0.2860(6)  & 0.3328(12)  & 0.0834(4)  & 0.0802(9)\tabularnewline
\hline 
\multicolumn{6}{|c|}{\textbf{$\beta=3.90$ $(32^{3}\times64)a^{4}$}}\tabularnewline
\hline 
 & 0.0150 & 0.1982(4)  & 0.2558(13)  & 0.0720(3) &  0.0701(7)\tabularnewline
\cline{2-6} 
0.0030 & 0.0220 & 0.2329(4)  & 0.2838(11)  & 0.0750(3) &  0.0720(8)\tabularnewline
\cline{2-6} 
 & 0.0270 & 0.2550(4)  & 0.3021(11)  & 0.0770(3)  & 0.0745(8)\tabularnewline
\hline
\hline 
 & 0.0150 & 0.2041(4)  & 0.2644(15)  & 0.0727(3)  & 0.0702(11)\tabularnewline
\cline{2-6} 
0.0040 & 0.0220 & 0.2381(4)  & 0.2917(17)  & 0.0758(3)  & 0.0722(10)\tabularnewline
\cline{2-6} 
 & 0.0270 & 0.2599(4)  & 0.3096(15)  & 0.0777(3)  & 0.0753(9)\tabularnewline
\hline
\end{tabular}
\caption{Pseudoscalar masses and decay constants at $\beta=3.90$. }
\label{MpsFps_b390}
\end{center}
\end{table}

\begin{table}[!h]
\begin{center}
\begin{tabular}{|c|c|l|l|l|l|}
\hline 
\multicolumn{6}{|c|}{\textbf{$\beta=4.05$ $(32^{3}\times64)a^{4}$}}\tabularnewline
\hline
\hline 
$a\mu_{\ell}=a\mu_{sea}$ & $a\mu_{``s"}$ & $aM^{34}$ & $aM^{12}$ & $aF^{34}$ & $aF^{12}$\tabularnewline
\hline
\hline 
 & 0.0120 & 0.1602(8)  & 0.1931(18)  & 0.0564(3)  & 0.0558(7)\tabularnewline
\cline{2-6} 
0.0030 & 0.0150 & 0.1751(8)  & 0.2053(17)  & 0.0578(3)  & 0.0566(7)\tabularnewline
\cline{2-6} 
 & 0.0180 & 0.1889(8)  & 0.2169(16)  & 0.0591(3)  & 0.0573(7)\tabularnewline
\hline
\hline 
 & 0.0120 & 0.1739(6)  & 0.2034(11)  & 0.0600(4)  & 0.0585(8)\tabularnewline
\cline{2-6} 
0.0060 & 0.0150 & 0.1877(6)  & 0.2153(11)  & 0.0613(3)  & 0.0596(8)\tabularnewline
\cline{2-6} 
 & 0.0180 & 0.2007(6)  & 0.2266(11)  & 0.0625(3)  & 0.0605(8)\tabularnewline
\hline
\hline 
 & 0.0120 & 0.1840(5)  & 0.2127(9)  & 0.0615(4)  & 0.0604(12)\tabularnewline
\cline{2-6} 
0.0080 & 0.0150 & 0.1972(5)  & 0.2242(9)  & 0.0627(4)  & 0.0616(12)\tabularnewline
\cline{2-6} 
 & 0.0180 & 0.2097(5)  & 0.2351(9)  & 0.0638(4)  & 0.0626(12)\tabularnewline
\hline
\end{tabular}
\caption{Pseudoscalar masses and decay constants at $\beta=4.05$. }
\label{MpsFps_b405}
\end{center}
\end{table}

\newpage

\begin{table}[!h]
\begin{center}
\begin{tabular}{|c|c|l|l|l|l|}
\hline 
\multicolumn{6}{|c|}{\textbf{$\beta=4.20$ $(48^{3}\times 96)a^{4}$}}\tabularnewline
\hline
\hline 
$a\mu_{\ell}=a\mu_{sea}$ & $a\mu_{``s"}$ & $aM^{34}$ & $aM^{12}$ & $aF^{34}$ & $aF^{12}$\tabularnewline
\hline
\hline 
 & 0.0116 &  0.1277(8) &  0.1433(20) & 0.0446(3) & 0.0438(9)\tabularnewline
\cline{2-6} 
0.0020 & 0.0129 &  0.1397(8) & 0.1536(19)  & 0.0456(3) & 0.0445(9)\tabularnewline
\cline{2-6} 
 & 0.0142 & 0.1509(9) & 0.1633(19) & 0.0465(4) &  0.0452(9)\tabularnewline
\hline
\multicolumn{6}{|c|}{\textbf{$\beta=4.20$ $(32^{3}\times 64)a^{4}$}}\tabularnewline
\hline
 & 0.0116 & 0.1522(11)  & 0.1682(20)  & 0.0483(5) & 0.0476(8)\tabularnewline
\cline{2-6} 
0.0065 & 0.0129 & 0.1628(10)  & 0.1777(18)  & 0.0494(5) & 0.0486(7)\tabularnewline
\cline{2-6} 
 & 0.0142 & 0.1729(10)  & 0.1868(17)  & 0.0503(5)  & 0.0495(7)\tabularnewline
\hline
\end{tabular}
\caption{ Pseudoscalar masses and decay constants at $\beta=4.20$.}
\label{MpsFps_b420}
\end{center}
\end{table}

\newpage


\begin{table}[!ht]
\begin{center}
\begin{tabular}{|c|c|l|l|l|l|}
\hline 
\multicolumn{6}{|c|}{\textbf{$\beta=3.80$ $(24^{3}\times48)a^{4}$}}\tabularnewline
\hline
\hline 
$a\mu_{\ell}=a\mu_{sea}$ & $a\mu_{``s"}$ & $ -R_2^{(b)} $ & $ R_3^{(b)} $ & $ R_4^{(b)} $ & $ R_5^{(b)} $\tabularnewline
\hline
\hline 
 & 0.0165 & 13.14(7) & 3.19(2) & 24.29(12) & 8.18(4)\tabularnewline
\cline{2-6} 
0.0080 & 0.0200 & 11.92(6) & 2.89(1) & 21.93(10) & 7.43(3)\tabularnewline
\cline{2-6} 
 & 0.0250 & 10.58(4) & 2.56(1) & 19.34(8) & 6.61(3)\tabularnewline
\hline
\hline 
 & 0.0165 & 12.16(6) & 2.95(1) & 21.89(10) & 7.41(3)\tabularnewline
\cline{2-6} 
0.0110 & 0.0200 & 11.15(5) & 2.70(1) & 19.99(8) & 6.81(3)\tabularnewline
\cline{2-6} 
 & 0.0250 & 9.99(4) & 2.41(1) & 17.84(7) & 6.13(2)\tabularnewline
\hline
\end{tabular}
\caption{$ R_i^{(b)}$ for $i=2, \ldots, 5$, as obtained from 
Eq.~(\ref{OioverO1}),
at each combination of the quark mass pair 
$(a\mu_\ell, a\mu_{``s"})$ and at $\beta=3.80$. }
\label{Rbare_b380}
\end{center}
\end{table}

\begin{table}[!h]
\begin{center}
\begin{tabular}{|c|c|l|l|l|l|}
\hline 
\multicolumn{6}{|c|}{\textbf{$\beta=3.90$ $(24^{3}\times48)a^{4}$}}\tabularnewline
\hline
\hline 
$a\mu_{\ell}=a\mu_{sea}$ & $a\mu_{``s"}$ & $ -R_2^{(b)} $ & $ R_3^{(b)} $ & $ R_4^{(b)} $ & $ R_5^{(b)} $\tabularnewline
\hline
\hline 
 & 0.0150 & 16.42(6) & 4.15(1) & 32.51(12) & 10.79(4)\tabularnewline
\cline{2-6} 
0.0040 & 0.0220 & 13.11(4) & 3.20(1) & 24.99(7) & 8.41(2)\tabularnewline
\cline{2-6} 
 & 0.0270 & 11.38(3) & 2.77(1) & 21.57(6) & 7.32(2)\tabularnewline
\hline
\hline 
 & 0.0150 & 15.37(8) & 3.77(2) & 29.02(14) & 9.68(4)\tabularnewline
\cline{2-6} 
0.0064 & 0.0220 & 12.21(5) & 2.98(1) & 22.89(9) & 7.75(3)\tabularnewline
\cline{2-6} 
 & 0.0270 & 10.70(4) & 2.60(1) & 19.97(7) & 6.82(2)\tabularnewline
\hline
\hline 
 & 0.0150 & 14.10(5) & 3.44(1) & 26.56(9) & 8.92(3)\tabularnewline
\cline{2-6} 
0.0085 & 0.0220 & 11.41(3) & 2.77(1) & 21.32(6) & 7.25(2)\tabularnewline
\cline{2-6} 
 & 0.0270 & 10.09(3) & 2.45(1) & 18.75(5) & 6.44(1)\tabularnewline
\hline
\hline 
 & 0.0150 & 13.52(5) & 3.31(1) & 25.30(9) & 8.53(3)\tabularnewline
\cline{2-6} 
0.0100 & 0.0220 & 11.07(4) & 2.70(1) & 20.55(6) & 7.02(2)\tabularnewline
\cline{2-6} 
 & 0.0270 & 9.85(3) & 2.39(1) & 18.28(5) & 6.26(2)\tabularnewline
\hline 
\multicolumn{6}{|c|}{\textbf{$\beta=3.90$ $(32^{3}\times64)a^{4}$}}\tabularnewline
\hline 
 & 0.0150 & 17.18(8) & 4.22(2) & 33.45(14) & 11.09(5)\tabularnewline
\cline{2-6} 
0.0030 & 0.0220 & 13.27(5) & 3.24(1) & 25.66(12) & 8.63(3)\tabularnewline
\cline{2-6} 
 & 0.0270 & 11.47(5) & 2.80(1) & 22.09(9) & 7.50(3)\tabularnewline
\hline
\hline 
 & 0.0150 & 16.32(9) & 4.10(2) & 32.22(18) & 10.68(6)\tabularnewline
\cline{2-6} 
0.0040 & 0.0220 & 13.01(6) & 3.17(1) & 24.88(12) & 8.34(4)\tabularnewline
\cline{2-6} 
 & 0.0270 & 11.31(5) & 2.74(1) & 21.42(9) & 7.27(3)\tabularnewline
\hline
\end{tabular}
\caption{$ R_i^{(b)}$ for $i=2, \ldots, 5$, as obtained from 
Eq.~(\ref{OioverO1}), at each combination of the quark mass pair 
$(a\mu_\ell, a\mu_{``s"})$ and at $\beta=3.90$. }
\label{Rbare_b390}
\end{center}
\end{table}

\begin{table}[!h]
\begin{center}
\begin{tabular}{|c|c|l|l|l|l|}
\hline 
\multicolumn{6}{|c|}{\textbf{$\beta=4.05$ $(32^{3}\times64)a^{4}$}}\tabularnewline
\hline
\hline 
$a\mu_{\ell}=a\mu_{sea}$ & $a\mu_{``s"}$ & $ -R_2^{(b)} $ & $ R_3^{(b)} $ & $ R_4^{(b)} $ & $ R_5^{(b)} $\tabularnewline
\hline
\hline 
 & 0.0120 & 20.47(12) & 5.07(3) & 40.49(22) & 13.32(7)\tabularnewline
\cline{2-6} 
0.0030 & 0.0150 & 17.35(9) & 4.29(2) & 34.29(17) & 11.36(6)\tabularnewline
\cline{2-6} 
 & 0.0180 & 15.10(7) & 3.72(2) & 29.78(13) & 9.93(4)\tabularnewline
\hline
\hline 
 & 0.0120 & 16.61(9) & 4.11(2) & 32.49(15) & 10.79(5)\tabularnewline
\cline{2-6} 
0.0060 & 0.0150 & 14.55(7) & 3.59(2) & 28.35(11) & 9.48(4)\tabularnewline
\cline{2-6} 
 & 0.0180 & 12.97(5) & 3.20(2) & 25.19(9) & 8.48(3)\tabularnewline
\hline
\hline 
 & 0.0120 & 15.13(5) & 3.74(1) & 29.36(9) & 9.79(3)\tabularnewline
\cline{2-6} 
0.0080 & 0.0150 & 13.42(4) & 3.31(1) & 25.99(7) & 8.72(2)\tabularnewline
\cline{2-6} 
 & 0.0180 & 12.09(3) & 2.97(1) & 23.34(6) & 7.88(2)\tabularnewline
\hline
\end{tabular}
\caption{$ R_i^{(b)}$ for $i=2, \ldots, 5$, as obtained from 
Eq.~(\ref{OioverO1}), at each combination of the quark mass pair 
$(a\mu_\ell, a\mu_{``s"})$ and at $\beta=4.05$. }
\label{Rbare_b405}
\end{center}
\end{table}

\begin{table}[!ht]
\begin{center}
\begin{tabular}{|c|c|l|l|l|l|}
\hline 
\multicolumn{6}{|c|}{\textbf{$\beta=4.20$ $(48^{3}\times96)a^{4}$}}\tabularnewline
\hline
\hline 
$a\mu_{\ell}=a\mu_{sea}$ & $a\mu_{``s"}$ & $ -R_2^{(b)} $ & $ R_3^{(b)} $ & $ R_4^{(b)} $ & $ R_5^{(b)} $\tabularnewline
\hline
\hline 
       & 0.0116 & 21.56(12)  & 5.28(3) & 44.37(25) & 14.56(8)\tabularnewline
\cline{2-6} 
0.0020 & 0.0129 & 18.28(9)  & 4.56(2) & 37.52(20) & 12.43(7)\tabularnewline
\cline{2-6} 
       & 0.0142 & 15.88(8)  & 3.95(2) & 32.57(17) & 10.86(6)\tabularnewline
 \hline 
\multicolumn{6}{|c|}{\textbf{$\beta=4.20$ $(32^{3}\times64)a^{4}$}}\tabularnewline
\hline
       & 0.0116 & 17.87(25) & 4.46(6) & 34.83(33) & 11.57(11)\tabularnewline
\cline{2-6} 
0.0065 & 0.0129 & 15.52(20) & 3.87(5) & 30.31(27) & 10.13(9)\tabularnewline
\cline{2-6} 
       & 0.0142 & 13.75(16) & 3.42(4) & 26.85(23) &  9.03(7)\tabularnewline
\hline
\end{tabular}
\caption{$ R_i^{(b)}$ for $i=2, \ldots, 5$, as obtained from 
Eq.~(\ref{OioverO1}),
at each combination of the quark mass pair 
$(a\mu_\ell, a\mu_{``s"})$ and at $\beta=4.20$. }
\label{Rbare_b420}
\end{center}
\end{table}

\begin{table}[!h]
\begin{center}
\begin{tabular}{|c|c|l|l|l|l|}
\hline 
\multicolumn{6}{|c|}{\textbf{$\beta=3.80$ $(24^{3}\times48)a^{4}$}}\tabularnewline
\hline
\hline 
$a\mu_{\ell}=a\mu_{sea}$ & $a\mu_{``s"}$ & $-\xi_{2}B_{2}^{(b)}$ & $\xi_{3}B_{3}^{(b)}$ 
& $\xi_{4}B_{4}^{(b)}$ & $\xi_{5}B_{5}^{(b)}$\tabularnewline
\hline
\hline 
 & 0.0165 & 1.015(12) & 0.247(3) & 1.877(21) & 0.632(7)\tabularnewline
\cline{2-6} 
0.0080 & 0.0200 & 1.029(11) & 0.249(2) & 1.892(21) & 0.641(7)\tabularnewline
\cline{2-6} 
 & 0.0250 & 1.046(11) & 0.253(2) & 1.912(20) & 0.654(7)\tabularnewline
\hline
\hline 
 & 0.0165 & 1.038(7) & 0.252(2) & 1.868(12) & 0.632(4)\tabularnewline
\cline{2-6} 
0.0110 & 0.0200 & 1.050(7) & 0.254(2) & 1.883(11) & 0.641(4)\tabularnewline
\cline{2-6} 
 & 0.0250 & 1.065(6) & 0.257(2) & 1.902(11) & 0.654(4)\tabularnewline
\hline
\end{tabular}
\caption{Bare $\xi_i B_i^{(b)}$ for $i=2, \ldots, 5$, as obtained from 
Eq.~(\ref{bareBi}), at each combination of the quark mass pair 
$(a\mu_\ell, a\mu_{``s"})$ and at $\beta=3.80$. }
\label{Bbare_b380}
\end{center}
\end{table}

\begin{table}[!h]
\begin{center}
\begin{tabular}{|c|c|l|l|l|l|}
\hline 
\multicolumn{6}{|c|}{\textbf{$\beta=3.90$ $(24^{3}\times48)a^{4}$}}\tabularnewline
\hline
\hline 
$a\mu_{\ell}=a\mu_{sea}$ & $a\mu_{``s"}$ & $-\xi_{2}B_{2}^{(b)}$ & $\xi_{3}B_{3}^{(b)}$ & $\xi_{4}B_{4}^{(b)}$ & $\xi_{5}B_{5}^{(b)}$\tabularnewline
\hline
\hline 
 & 0.0150 & 0.961(7) & 0.236(2) & 1.848(14) & 0.613(5)\tabularnewline
\cline{2-6} 
0.0040 & 0.0220 & 0.991(7) & 0.242(2) & 1.888(13) & 0.635(4)\tabularnewline
\cline{2-6} 
 & 0.0270 & 1.009(7) & 0.245(2) & 1.911(13) & 0.649(4)\tabularnewline
\hline
\hline 
 & 0.0150 & 0.979(9) & 0.240(2) & 1.848(20) & 0.617(6)\tabularnewline
\cline{2-6} 
0.0064 & 0.0220 & 1.006(9) & 0.245(2) & 1.887(16) & 0.639(5)\tabularnewline
\cline{2-6} 
 & 0.0270 & 1.023(8) & 0.249(2) & 1.909(16) & 0.652(5)\tabularnewline
\hline
\hline 
 & 0.0150 & 0.987(8) & 0.241(2) & 1.860(16) & 0.624(5)\tabularnewline
\cline{2-6} 
0.0085 & 0.0220 & 1.014(8) & 0.246(2) & 1.894(15) & 0.644(5)\tabularnewline
\cline{2-6} 
 & 0.0270 & 1.030(8) & 0.250(2) & 1.914(15) & 0.657(5)\tabularnewline
\hline
\hline 
 & 0.0150 & 0.993(12) & 0.243(3) & 1.861(24) & 0.627(8)\tabularnewline
\cline{2-6} 
0.0100 & 0.0220 & 1.019(12) & 0.248(3) & 1.894(23) & 0.647(8)\tabularnewline
\cline{2-6} 
 & 0.0270 & 1.036(12) & 0.252(3) & 1.915(23) & 0.660(8)\tabularnewline
\hline 
\multicolumn{6}{|c|}{\textbf{$\beta=3.90$ $(32^{3}\times64)a^{4}$}}\tabularnewline
\hline 
 & 0.0150 & 0.953(5) & 0.234(1) & 1.855(10) & 0.615(3)\tabularnewline
\cline{2-6} 
0.0030 & 0.0220 & 0.982(5) & 0.240(1) & 1.898(9) & 0.638(3)\tabularnewline
\cline{2-6} 
 & 0.0270 & 1.000(5) & 0.244(1) & 1.924(9) & 0.653(3)\tabularnewline
\hline
\hline 
 & 0.0150 & 0.962(6) & 0.236(2) & 1.831(12) & 0.608(4)\tabularnewline
\cline{2-6} 
0.0040 & 0.0220 & 0.992(6) & 0.242(2) & 1.875(12) & 0.632(4)\tabularnewline
\cline{2-6} 
 & 0.0270 & 1.011(6) & 0.246(2) & 1.901(11) & 0.646(4)\tabularnewline
\hline
\end{tabular}
\caption{Bare $\xi_i B_i^{(b)}$ for $i=2, \ldots, 5$, as obtained from 
Eq.~(\ref{bareBi}), at each combination of the quark mass pair 
$(a\mu_\ell, a\mu_{``s"})$ and at $\beta=3.90$. }
\label{Bbare_b390}
\end{center}
\end{table}

\begin{table}[!h]
\begin{center}
\begin{tabular}{|c|c|l|l|l|l|}
\hline 
\multicolumn{6}{|c|}{\textbf{$\beta=4.05$ $(32^{3}\times64)a^{4}$}}\tabularnewline
\hline
\hline 
$a\mu_{\ell}=a\mu_{sea}$ & $a\mu_{``s"}$ & $-\xi_{2}B_{2}^{(b)}$ & $\xi_{3}B_{3}^{(b)}$ & $\xi_{4}B_{4}^{(b)}$ & $\xi_{5}B_{5}^{(b)}$\tabularnewline
\hline
\hline 
 & 0.0120 & 0.915(10) & 0.227(2) & 1.810(17) & 0.596(6)\tabularnewline
\cline{2-6} 
0.0030 & 0.0150 & 0.929(9) & 0.230(2) & 1.837(17) & 0.609(6)\tabularnewline
\cline{2-6} 
 & 0.0180 & 0.943(9) & 0.232(2) & 1.860(16) & 0.620(5)\tabularnewline
\hline
\hline 
 & 0.0120 & 0.931(9) & 0.230(2) & 1.820(18) & 0.605(6)\tabularnewline
\cline{2-6} 
0.0060 & 0.0150 & 0.946(9) & 0.234(2) & 1.843(17) & 0.617(6)\tabularnewline
\cline{2-6} 
 & 0.0180 & 0.960(9) & 0.237(2) & 1.864(17) & 0.628(6)\tabularnewline
\hline
\hline 
 & 0.0120 & 0.948(9) & 0.234(2) & 1.842(17) & 0.614(6)\tabularnewline
\cline{2-6} 
0.0080 & 0.0150 & 0.962(8) & 0.237(2) & 1.861(16) & 0.625(5)\tabularnewline
\cline{2-6} 
 & 0.0180 & 0.974(8) & 0.240(2) & 1.8883(16) & 0.636(5)\tabularnewline
\hline
\end{tabular}
\caption{Bare $\xi_i B_i^{(b)}$ for $i=2, \ldots, 5$, as obtained from 
Eq.~(\ref{bareBi}), at each combination of the quark mass pair 
$(a\mu_\ell, a\mu_{``s"})$ and at $\beta=4.05$. }
\label{Bbare_b405}
\end{center}
\end{table}

\begin{table}[!h]
\begin{center}
\begin{tabular}{|c|c|l|l|l|l|}
\hline 
\multicolumn{6}{|c|}{\textbf{$\beta=4.20$ $(48^{3}\times96)a^{4}$}}\tabularnewline
\hline
\hline 
$a\mu_{\ell}=a\mu_{sea}$ & $a\mu_{``s"}$ & -$\xi_{2}B_{2}^{(b)}$ & $\xi_{3}B_{3}^{(b)}$ 
& $\xi_{4}B_{4}^{(b)}$ & $\xi_{5}B_{5}^{(b)}$\tabularnewline
\hline
\hline 
          & 0.0116 & 0.878(7) & 0.219(2) & 1.801(13) & 0.592(4)\tabularnewline
\cline{2-6} 
   0.0020 & 0.0129 & 0.892(7) & 0.222(2) & 1.830(13) & 0.606(4)\tabularnewline
\cline{2-6}
          & 0.0142 & 0.905(7) & 0.225(2) & 1.855(14) & 0.618(5)\tabularnewline
\hline
\multicolumn{6}{|c|}{\textbf{$\beta=4.20$ $(32^{3}\times64)a^{4}$}}\tabularnewline

\hline 
          & 0.0116 & 0.939(20) & 0.234(5) & 1.831(38) & 0.608(12)\tabularnewline
\cline{2-6} 
0.0065    & 0.0129 & 0.950(19) & 0.237(5) & 1.856(36) & 0.628(12)\tabularnewline
\cline{2-6} 
          & 0.0142 & 0.961(18) & 0.239(5) & 1.877(35) & 0.632(12)\tabularnewline
\hline
\end{tabular}
\caption{Bare $\xi_i B_i^{(b)}$ for $i=2, \ldots, 5$, as obtained from 
Eq.~(\ref{bareBi}), at each combination of the quark mass pair 
$(a\mu_\ell, a\mu_{``s"})$ and at $\beta=4.20$. }
\label{Bbare_b420}
\end{center}
\end{table}

\clearpage
\vspace{1.2cm}

\section{  Results for $R_i$ and $B_i$}
\label{App_CL}

In this appendix we present in detail our results in the $\overline{\rm{MS}}$ scheme of Ref.~\cite{mu:4ferm-nlo}
at 2 GeV for the quantities $R_i$ and $B_i$ (c.f. Eqs.~(\ref{REN_Riratio},~\ref{REN_Riratio_V2}) and 
Eq.~(\ref{REN_RBi}) respectively).
We also give the $R_i$ results computed in the indirect way of Eq.~(\ref{ratio_indirect}). 
In Table~\ref{mq_OisuO1} we gather results obtained employing M1-type RCs and using ChPT (NLO) fit formula, polynomial 
and linear fit functions (see Eqs.~(\ref{ChPTBi}-\ref{ChPTRi}),  and $n=2$ and $n=1$ of Eq.~(\ref{Yfit}) respectively). 
In Table~\ref{mq_OisuO1_M2_MODIFIED} we show the respective results when using M2-type RCs. 
Instead of using the definition of Eq.~(\ref{REN_Riratio}), we have employed a slightly different but equivalent one which 
reads 
\begin{equation}
 \tilde{R}_i^{\prime}=\tilde{R}_i ~\frac{[G_K^{\rm{34}} G_K^{\rm{12}}]|_
{\rm{M1}}}{[G_K^{\rm{34}} G_K^{\rm{12}}]|_{\rm{M2}}}
\label{Ri_mod}
\end{equation}
where indices M1 and M2 refer to the use of the respective type of renormalisation constants and we define 
$G_K^{(12,34)}|_{(M1,M2)}\, = \, \langle 0|P^{(12,34)}|K\rangle|_{(M1,M2)}$. 
We find that the  quantity defined in Eq.~(\ref{Ri_mod}) 
has smaller $O(a^2)$ effects.

\begin{table}[!h]
\begin{center}
\begin{tabular}{|c|c|c|c|c|}
\hline 
Fit & $i$ & $R_{i}$ & $R_{i}({\rm via~ Eq.}~(\ref{ratio_indirect}))$ & $B_{i}$\tabularnewline
\hline
\multicolumn{1}{|c|}{} & 1 & 1 & 1 & 0.53(2) \tabularnewline
\cline{2-5} 
 & 2 & -13.7(3) & -15.4(2.2) & 0.52(2)\tabularnewline
\cline{2-5} 
ChPT & 3 & 4.8(2) & 5.3(8) & 0.89(5)\tabularnewline
\cline{2-5} 
 & 4 & 24.7(6) & 28.3(4.1) & 0.79(3)\tabularnewline
\cline{2-5} 
 & 5 & 5.9(3) & 6.9(1.1) & 0.58(4)\tabularnewline
\hline 
\multicolumn{5}{|c|}{}\tabularnewline
\hline
\multicolumn{1}{|c|}{} & 1 & 1 & 1 & 0.53(2) \tabularnewline
\cline{2-5} 
 & 2 & -14.3(5) & -15.5(2.3) & 0.52(2)\tabularnewline
\cline{2-5} 
Pol & 3 & 4.9(3) & 5.3(9) & 0.89(7)\tabularnewline
\cline{2-5} 
 & 4 & 24.7(8) & 27.7(4.2) & 0.77(4)\tabularnewline
\cline{2-5} 
 & 5 & 5.9(5) & 6.8(1.2) & 0.57(5)\tabularnewline
\hline 
\multicolumn{5}{|c|}{}\tabularnewline
\hline 
 & 1 & 1 & 1 & 0.53(2)\tabularnewline
\cline{2-5} 
 & 2 & -13.7(3) & -15.7(2.2) & 0.52(2)\tabularnewline
\cline{2-5} 
\multicolumn{1}{|c|}{L} & 3 & 4.8(2) & 5.4(8) & 0.90(5)\tabularnewline
\cline{2-5} 
 & 4 & 23.5(6) & 27.7(4.0) & 0.78(3)\tabularnewline
\cline{2-5} 
 & 5 & 5.8(3) & 6.7(1.1) & 0.57(4)\tabularnewline
\hline
\end{tabular}
\caption{$R_i$ (direct computation through Eq.~(\ref{REN_Riratio}) and indirect computation through 
Eq.~(\ref{ratio_indirect})) and $B_i$ results using M1-type RCs for three kinds of fit function, namely a ChPT (NLO) fit, 
a polynomial and a linear fit with respect to the light quark mass. For $i=2,3$ the ChPT (NLO) fit formula for $R_i$ 
 coincides with the linear one (we refer to  results of the 3rd column).}
\label{mq_OisuO1}
\end{center}
\end{table}

\begin{table}[!h]
\begin{center}
\begin{tabular}{|c|c|c|c|c|}
\hline 
Fit & $i$ & $R_{i}$ & $R_{i}({\rm via~ Eq.}~(\ref{ratio_indirect}))$ & $B_{i}$\tabularnewline
\hline
\multicolumn{1}{|c|}{} & 1 & 1 & 1 & 0.53(2)\tabularnewline
\cline{2-5} 
 & 2 & -13.6(2) & -14.9(2.1) & 0.50(1)\tabularnewline
\cline{2-5} 
ChPT & 3 & 4.7(1) & 5.1(7) & 0.87(3)\tabularnewline
\cline{2-5} 
 & 4 & 24.7(4) & 27.5(3.4) & 0.77(2)\tabularnewline
\cline{2-5} 
 & 5 & 5.9(2) & 6.5(9) & 0.55(2)\tabularnewline
\hline 
\multicolumn{5}{|c|}{}\tabularnewline
\hline 
\multicolumn{1}{|c|}{} & 1 & 1 & 1 & 0.53(2)\tabularnewline
\cline{2-5} 
 & 2 & -14.0(4) & -15.0(2.2) & 0.50(2)\tabularnewline
\cline{2-5} 
Pol & 3 & 4.8(2) & 5.2(8) & 0.87(5)\tabularnewline
\cline{2-5} 
 & 4 & 24.4(7) & 26.4(3.9) & 0.74(3)\tabularnewline
\cline{2-5} 
 & 5 & 5.8(3) & 6.3(1.0) & 0.53(4)\tabularnewline
\hline 
\multicolumn{5}{|c|}{}\tabularnewline
\hline 
 & 1 & 1 & 1 & 0.54(2)\tabularnewline
\cline{2-5} 
 & 2 & -13.6(2) & -15.0(2.1) & 0.51(1)\tabularnewline
\cline{2-5} 
\multicolumn{1}{|c|}{L} & 3 & 4.7(1) & 5.2(8) & 0.88(3)\tabularnewline
\cline{2-5} 
 & 4 & 23.7(4) & 26.5(3.5) & 0.75(2)\tabularnewline
\cline{2-5} 
 & 5 & 5.7(2) & 6.2(9) & 0.53(2)\tabularnewline
\hline
\end{tabular}

\caption{$R_i$ (direct computation through Eq.~(\ref{Ri_mod}) and indirect computation through 
Eq.~(\ref{ratio_indirect})) and $B_i$ results using M2-type RCs for three kinds of fit function, 
namely a ChPT (NLO) fit, a polynomial and a linear fit with respect to the light quark mass. 
For $i=2,3$ the ChPT (NLO) fit formula for $R_i$ 
coincides with the linear one (we refer to  results of the 3rd column).
The results in the lines corresponding to $i=5$ here are not reliable,
due to very large cutoff effects resulting in this case from the use of  
M2-type RCs (see text).}
\label{mq_OisuO1_M2_MODIFIED}
\end{center}
\end{table}

\begin{figure}[!ht]
\subfigure[]{\includegraphics[scale=0.55,angle=-0]{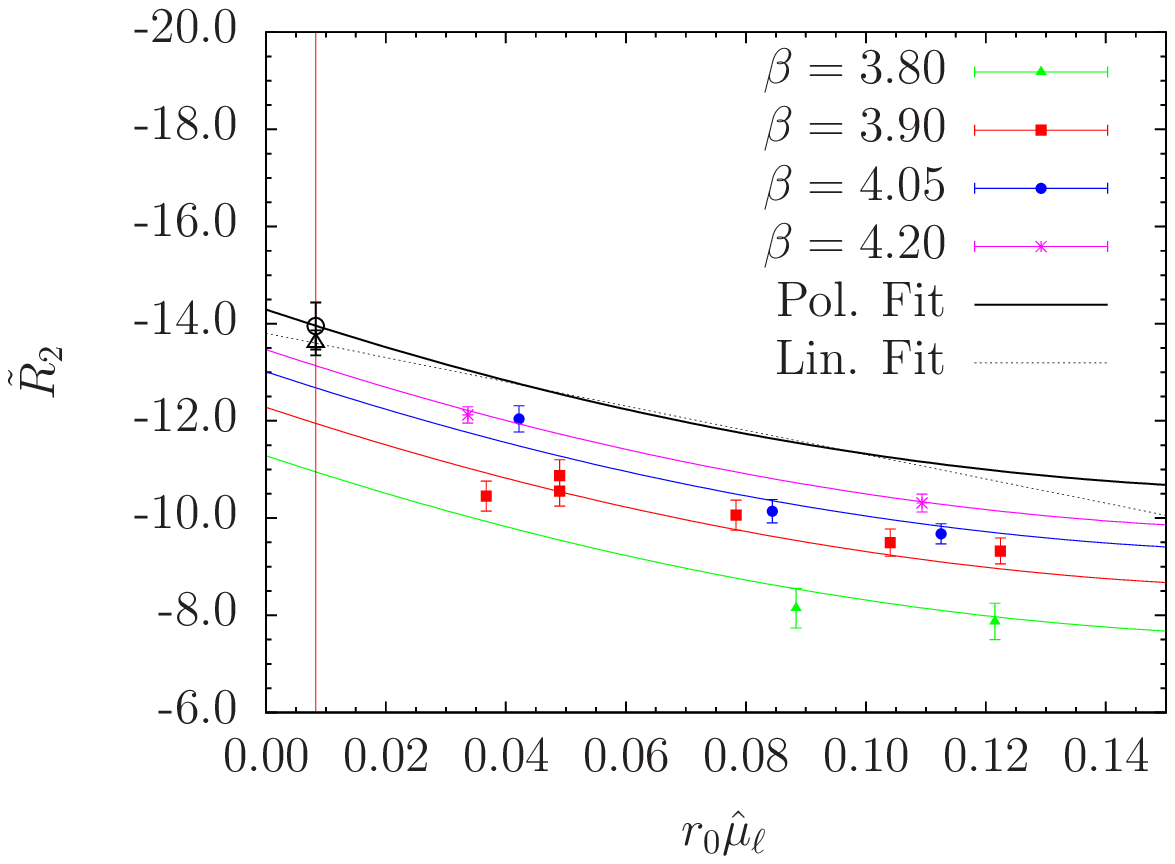}}
\subfigure[]{\includegraphics[scale=0.55,angle=-0]{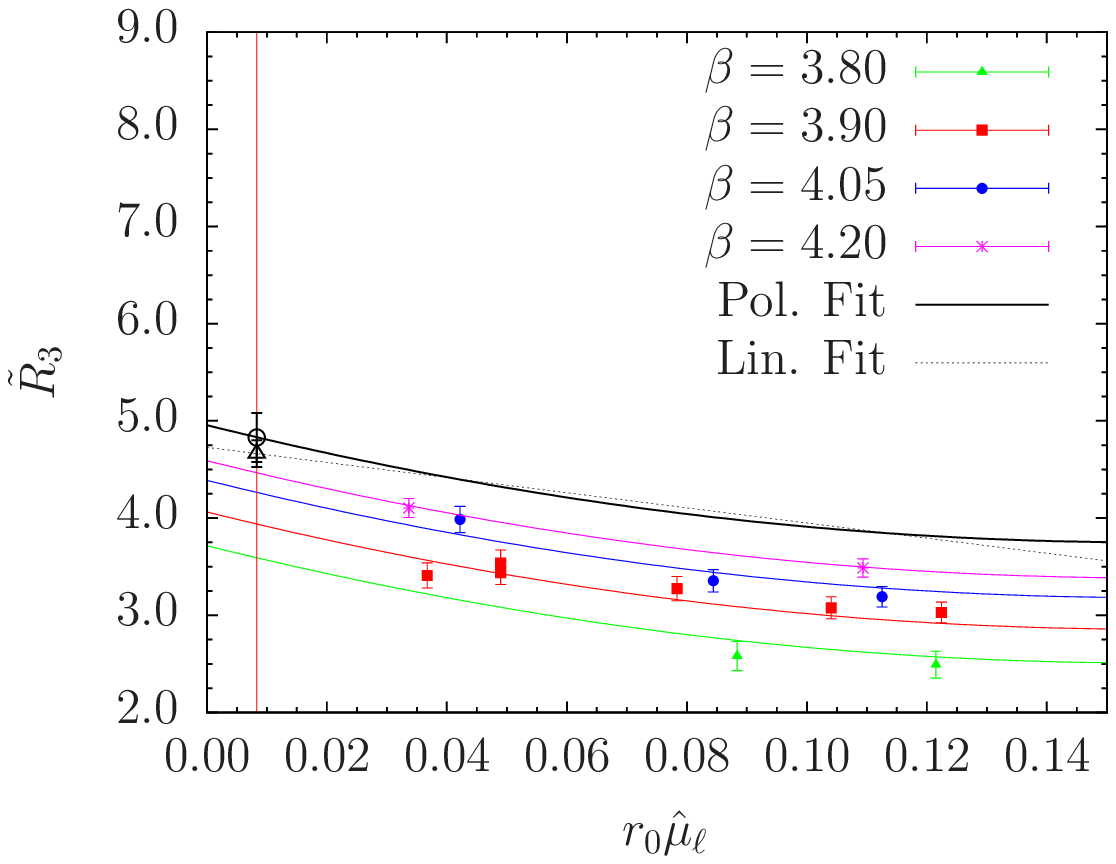}}
\hspace*{0.1cm} \subfigure[]{\includegraphics[scale=0.55,angle=-0]{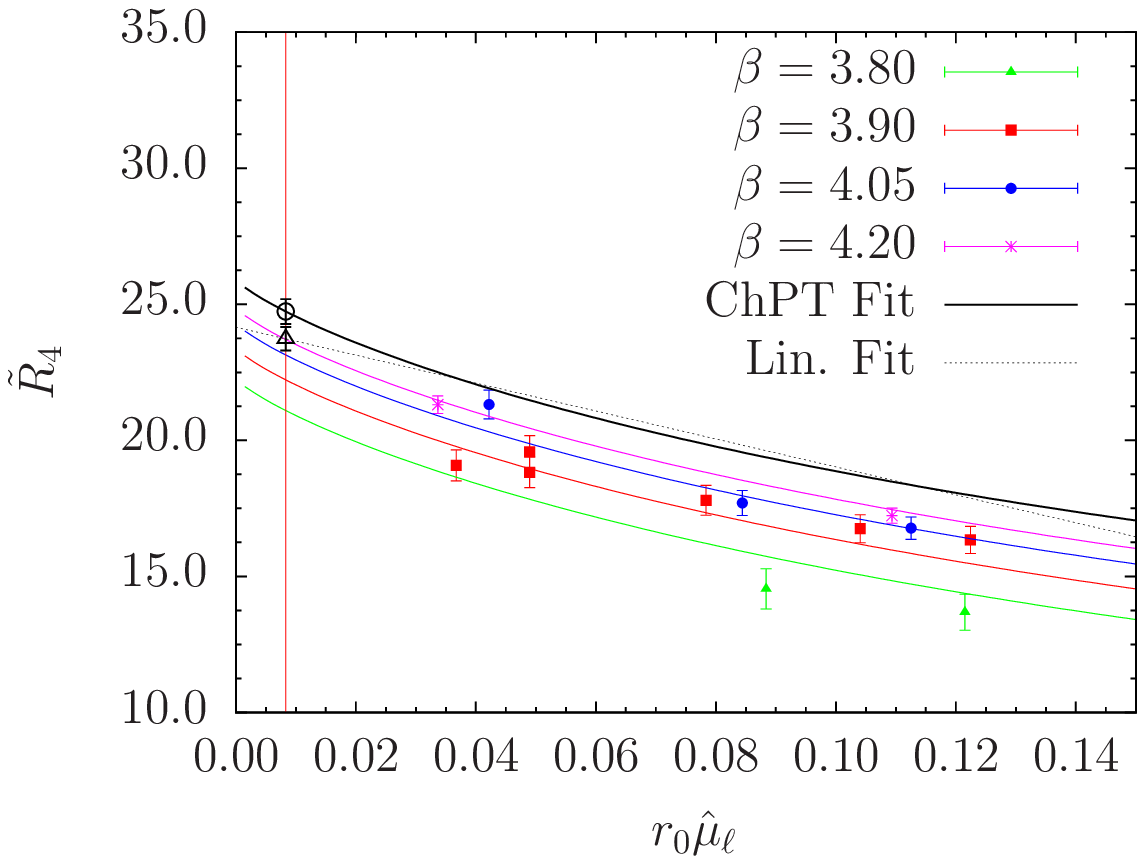}}
\hspace*{1.cm} \subfigure[]{\includegraphics[scale=0.55,angle=-0]{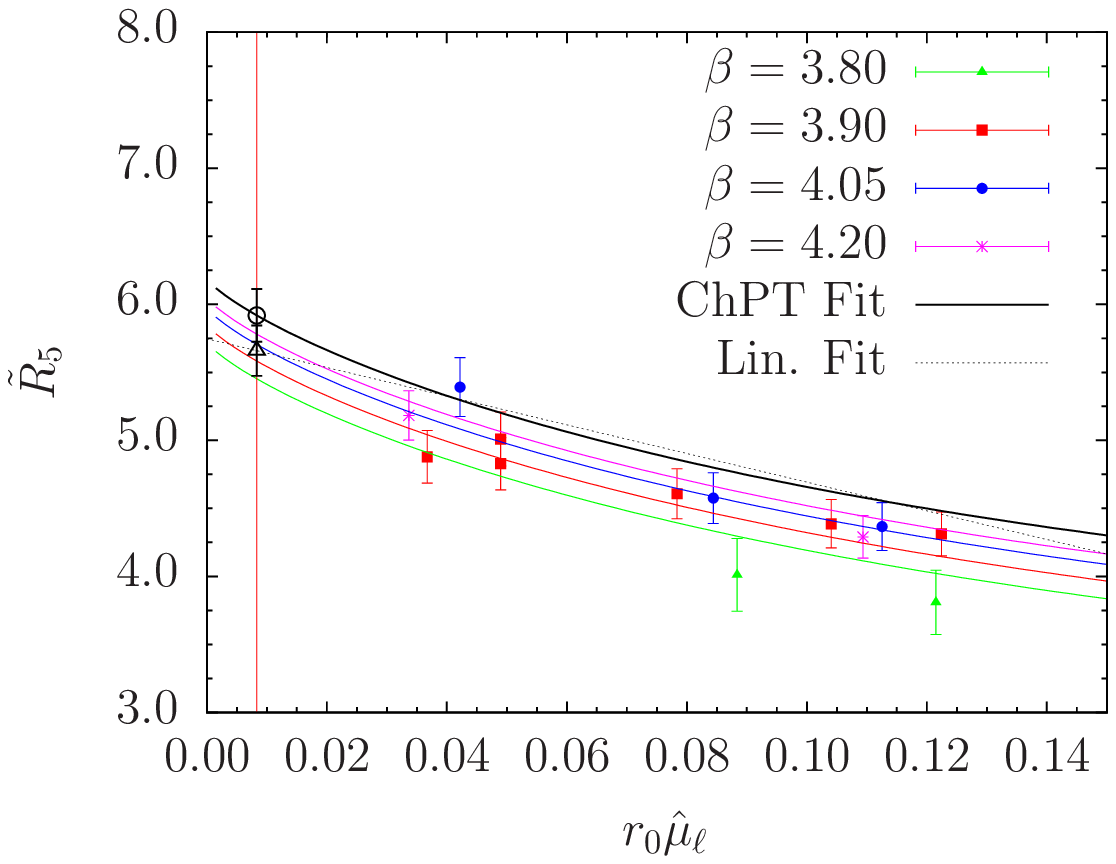}}
\vskip -0.5cm
\begin{center}
\caption{
\sl Solid lines in panels (a) and (b) show the behaviour vs. the renormalized light quark mass of the 
combined chiral and continuum fits (according to the polynomial formula~(\ref{Yfit}) with $n=2$) of the 
$\tilde{R}_i^{'}$ (see Eq.~(\ref{Ri_mod})), 
with $i=2$ and $i=3$  respectively, renormalized in the $\overline{\rm{MS}}$ scheme of Ref.~\cite{mu:4ferm-nlo} 
at 2~GeV with the M2-type RCs. The full black line is the continuum limit curve. 
In panels (c) and (d), solid lines, instead, show the combined chiral and continuum  described by NLO-ChPT, 
Eq.~(\ref{ChPTRi}) for $i=4$ and $i=5$, respectively. The full black line is the continuum limit curve. 
The dashed black line represents the continuum limit curve in 
the case of the linear fit ansatz. Black open circles and triangles stand for the results at the physical point 
corresponding to the  polynomial (panels (a) and (b)) and ChPT fit (panels (c) and (d)),   
and linear fit ansatz, respectively.
}
\label{fig:Rivsmul_M2mod}
\end{center}
\end{figure}

\begin{figure}[!ht]
\subfigure[]{\includegraphics[scale=0.55,angle=-0]{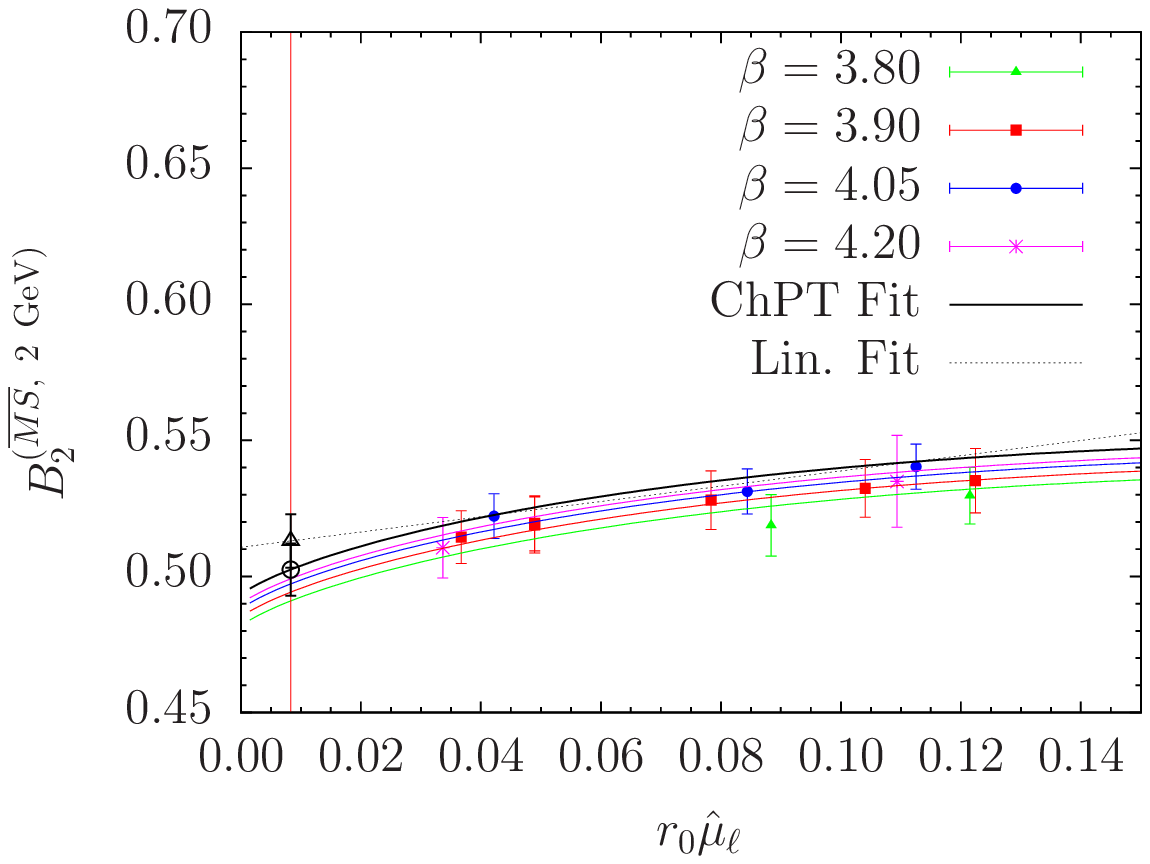}}
\subfigure[]{\includegraphics[scale=0.55,angle=-0]{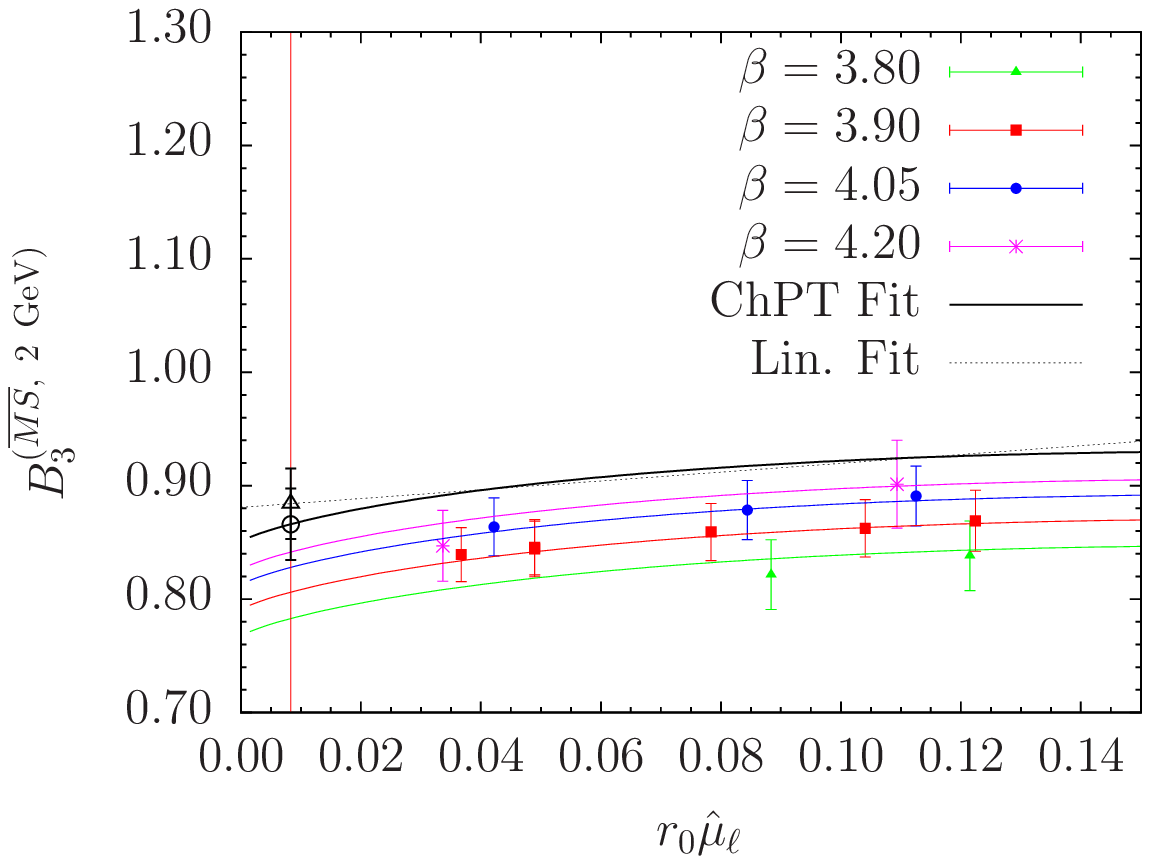}}
\hspace*{0.1cm} \subfigure[]{\includegraphics[scale=0.55,angle=-0]{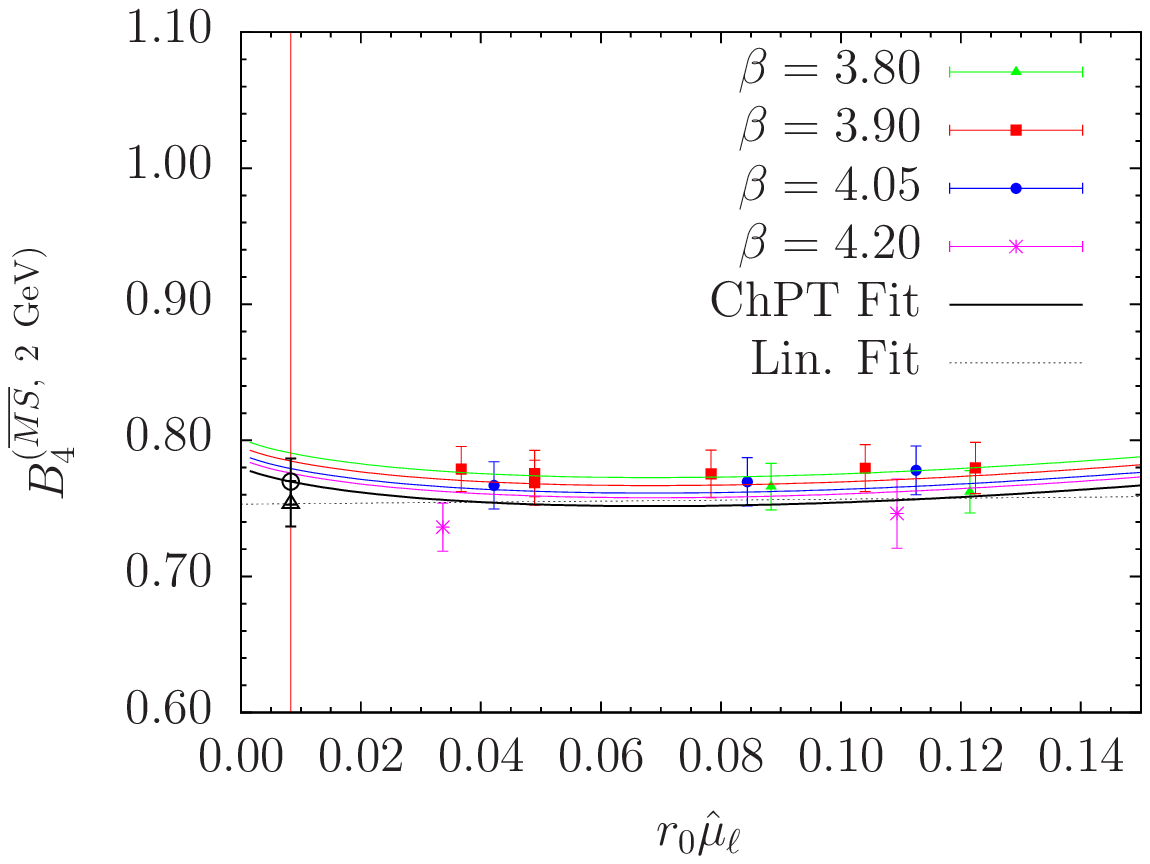}}
\hspace*{1.cm} \subfigure[]{\includegraphics[scale=0.55,angle=-0]{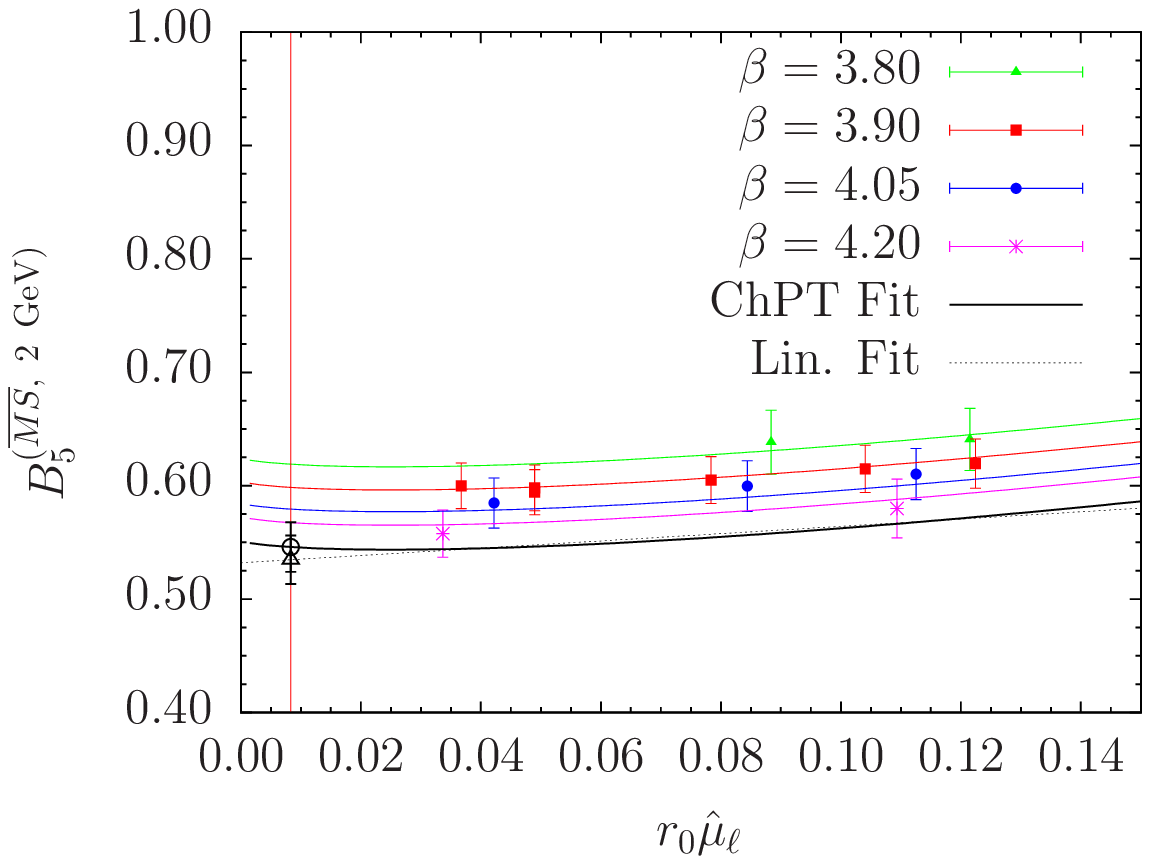}}
\vskip -0.5cm
\begin{center}
\caption{\sl Solid lines in panels (a) to (d) show the behaviour vs.\ the renormalized light quark mass of the
combined chiral and continuum fits (according to the ChPT fit formula ~(\ref{ChPTBi})) for the $B_i$ parameters
with $i=2,\ldots,5$ respectively, renormalized in the $\overline{\rm{MS}}$ scheme of 
Ref.~\cite{mu:4ferm-nlo} at 2~GeV with the M2-type RCs.
The full black line is the continuum limit curve~(\ref{Yfit}).
The dashed black line represents the continuum limit curve in
the case of the linear fit ansatz. Black open circles and triangles stand for the results at the physical point
corresponding to the  ChPT fit and linear fit ansatz, respectively.   
}
\label{fig:Bivsmul_M2}
\end{center}
\end{figure}

In Figs.~\ref{fig:Rivsmul_M2mod} and \ref{fig:Bivsmul_M2} we show the combined fit for 
the ratios, $\tilde{R}_i^{'}$ and 
bag parameters $B_i$ ($i=2, \ldots, 5$) against the light quark mass when M2-type RCs are used.

We remark that a good agreement between the continuum limit results for
the bag parameters $B_i$ and the matrix elements ratios $R_i$ obtained
using M2-type RCs and their counterparts based on M1-type RCs, as we
observe for $i=1,2,3,4$ (for $i=1$ see also Ref.~\cite{Constantinou:2010qv}),
provides a valuable check of the smallness of residual systematic errors
in the evaluation of RI-MOM RCs with the M1-method. In particular possible
systematic errors stemming in the M1-method from the inadequacy at
non-high momenta ($\tilde{p}^2$) of the perturbative operator anomalous
dimensions used in the analysis or from the removal of the leading cutoff
effects via a linear fit in $\tilde{p}^2$ are strongly reduced or absent
when using the M2-method for RCs. This is so because in this latter approach (see 
Ref.~\cite{Constantinou:2010gr}  and Appendix~\ref{App_RIMOM}) 
the RCs are extracted from Landau gauge correlators at a
rather high $\tilde{p}^2$-value  (fixed to $\sim 9$~GeV$^2$  for all
$\beta$'s) but comes at the price of generically larger lattice artifacts
on the RCs, which we partly suppress by removing the perturbatively known
O$(a^2g^2)$ contributions. For the case of $O_i$ with $i=1,2,3,4$ the
resulting cutoff effects on $R_i$ and $B_i$ (see Figs~\ref{fig:Rivsmul_M2mod} and 
\ref{fig:Bivsmul_M2}) appear to
be under control and the continuum extrapolation is reliable. On the 
contrary, the case of $B_5$ and $R_5$ (see panel d) of the figures above)
is a typical one where too large cutoff effects affecting the M2-type RCs
make unreliable the results appearing in the $i=5$-lines of 
Table~\ref{mq_OisuO1_M2_MODIFIED}.

\clearpage
Finally we give our continuum results for $B_i$ and $R_i$
in the $\overline{\rm{MS}}$  scheme of Buras {\it et al.}, defined in Ref.~\cite{mu:4ferm-nlo}, and  the RI-MOM scheme at 3~GeV,
see Tables~\ref{results_MSbar_3GeV} and~\ref{results_RIMOM_3GeV} respectively.

\begin{table}[!h]
\begin{center}
\begin{tabular}{|c|c|c|c|c|}
\hline
\multicolumn{5}{|c|}{$\overline {\rm{MS}}$ (3 GeV)}\tabularnewline
\hline
\hline
$B_{1}$ & $B_{2}$ & $B_{3}$ & $B_{4}$ & $B_{5}$\tabularnewline
\hline
\hline
0.51(2) & 0.47(2) & 0.78(4) & 0.76(3) & 0.58(3)\tabularnewline
\hline
\multicolumn{1}{c}{}  \tabularnewline
\hline
$R_{1}$ & $R_{2}$ & $R_{3}$ & $R_{4}$ & $R_{5}$\tabularnewline
\hline
\hline
1 & -15.6(5) & 5.3(3) & 28.5(9) & 7.3(4)\tabularnewline
\hline
\end{tabular}
\caption{Continuum limit results for $B_i$ and $R_i$, renormalized in the $\overline{\rm{MS}}$ scheme of Ref.~\cite{mu:4ferm-nlo}
at 3~GeV.}
\label{results_MSbar_3GeV}
\end{center}
\end{table}

\begin{table}[!h]
\begin{center}
\begin{tabular}{|c|c|c|c|c|}
\hline
\multicolumn{5}{|c|}{RI-MOM (3 GeV)}\tabularnewline
\hline
\hline
$B_{1}$ & $B_{2}$ & $B_{3}$ & $B_{4}$ & $B_{5}$\tabularnewline
\hline
\hline
0.51(2) & 0.61(2) & 1.02(5) & 0.92(4) & 0.68(5)\tabularnewline
\hline
\multicolumn{1}{c} {}\tabularnewline
\hline
$R_{1}$ & $R_{2}$ & $R_{3}$ & $R_{4}$ & $R_{5}$\tabularnewline
\hline
\hline
1 & -14.6(5) & 5.0(3) & 25.6(9) & 6.2(4)\tabularnewline
\hline
\end{tabular}
\caption{Continuum limit results for $B_i$ and $R_i$, renormalized in the RI-MOM scheme at 3~GeV.}
\label{results_RIMOM_3GeV}
\end{center}
\end{table}

\clearpage
\end{appendices}
\bibliography{lattice}

\end{document}